\RequirePackage{ifpdf}
\ifpdf
\else
\PassOptionsToPackage{dvipdfmx}{graphicx}
\PassOptionsToPackage{dvipdfmx}{hyperref}
\fi

\documentclass[12pt,a4paper]{article}
\usepackage{jheppub}
\usepackage[utf8]{inputenc}

\allowdisplaybreaks[1]
\usepackage{amsmath,bm}
\usepackage{subfigure}
\usepackage{amssymb}
\usepackage{colortbl}

\title{
Superradiant instability of black resonators and geons
}

\author[1]{Takaaki~Ishii,}
\author[2]{Keiju~Murata,}
\author[3,4]{Jorge~E.~Santos,}
\author[5]{Benson~Way}

\affiliation[1]{Department of Physics, Kyoto University, Kyoto 606-8502, Japan}
\affiliation[2]{Department of Physics, College of Humanities and Sciences, Nihon University, Tokyo 156-8550, Japan}
\affiliation[3]{DAMTP, Centre for Mathematical Sciences, University of Cambridge, Wilberforce Road, Cambridge CB3 0WA, UK}
\affiliation[4]{Institute for Advanced Study, Princeton, NJ 08540, USA}
\affiliation[5]{Departament de F\'{i}sica Qu\`{a}ntica i Astrof\'{i}sica, Institut de Ci\`{e}ncies del Cosmos\\
Universitat de Barcelona, Mart\'{i} i Franqu\`{e}s, 1, E-08028 Barcelona, Spain}
\emailAdd{ishiitk@gauge.scphys.kyoto-u.ac.jp}
\emailAdd{murata.keiju@nihon-u.ac.jp}
\emailAdd{jss55@cam.ac.uk}
\emailAdd{benson@icc.ub.edu}

\abstract{%
Black resonators and geons in global AdS are rapidly rotating, low-energy solutions with a helical Killing field.  We study the linear mode stability of equal angular momenta, five-dimensional black resonators and geons under scalar, electromagnetic, and gravitational perturbations. We find that black resonators are unstable to the superradiant instability, in agreement with previously known results.  Perhaps surprisingly, many geons appear linearly stable, despite having an ergoregion.  This apparent stability implies that geons are important long-lived, low-energy states in the dual gauge theory.  However, we do find that geons are unstable within a certain range of parameter space.  We comment on the nature of this instability and to its possible endpoints. We also report on new non-spinning oscillating geons, which we construct within a cohomogeneity two ansatz. Given the existing arguments that suggest our linear stability results may be extended nonlinearly, our findings indicate that most geons are generic and long-lived solutions.
}

\preprint{KUNS-2813}

\begin{document}
\maketitle
\section{Introduction}
\label{sec:intro}
In recent years, there has been an increased focus on gravitational dynamics in anti-de Sitter space (AdS), due in large part to the development of gauge/gravity duality \cite{Maldacena:1997re,Gubser:1998bc,Witten:1998qj}. Despite the growing attention, there is a fundamental question that remains unresolved: in classical general relativity, what are the typical low-energy configurations in global AdS?

The answer to this question is surprisingly intricate.  The timelike boundary of AdS causes gravitational dynamics to behave quite differently from that of de Sitter or Minkowski space.  With energy and momentum conserving boundary conditions, the boundary is reflective. Small-energy excitations in AdS therefore do not disperse and instead either form a small black hole or reflect off the boundary indefinitely \cite{Dafermos2006,DafermosHolzegel2006,Bizon:2011gg,Dias:2012tq,Maliborski:2013jca,Buchel:2013uba,Balasubramanian:2014cja,Craps:2014vaa,Bizon:2014bya,Craps:2014jwa,Bizon:2015pfa,Balasubramanian:2015uua,Dimitrakopoulos:2015pwa,Green:2015dsa,Choptuik:2017cyd}.  Recent results indicate that both of these scenarios are generic in the space of initial data \cite{Dias:2012tq,Dimitrakopoulos:2015pwa,Choptuik:2018ptp,Masachs:2019znp}.

Suppose a small black hole is formed. Generic configurations contain angular momentum, which implies that the black hole will be rapidly rotating.  But such a black hole is superradiant \cite{Detweiler:1980uk,Hawking:1999dp,Cardoso:2004hs} (see \cite{Brito:2015oca} for a review) and can amplify waves that approach its horizon.  These amplified waves will reflect off the boundary and interact again with the black hole, leading to a cascade of instabilities \cite{Reall:2002bh,Kunduri:2006qa,Cardoso:2006wa,Murata:2008xr,Kodama:2009rq,Dias:2011at,Dias:2013sdc,Cardoso:2013pza}.  The endpoint of this instability remains an open question, with recent studies suggesting that the endpoint is not a steady-state solution \cite{Dias:2011ss,Dias:2015rxy,Niehoff:2015oga,Chesler:2018txn}.

Suppose instead that a black hole is not formed. Such configurations tend to resemble nonlinear extensions of normal modes of AdS \cite{Maliborski:2013jca,Fodor:2015eia,Dias:2011at,Liebling:2012fv,Buchel:2013uba,Choptuik:2017cyd,Choptuik:2018ptp}.  For pure gravity, the extensions of normal modes are called geons \cite{Dias:2011ss,Horowitz:2014hja,Martinon:2017uyo,Fodor:2017spc}.  Yet, even the linear stability of geons has never been established.  They do not have a horizon, and hence the instability results from \cite{Green:2015kur} cannot be applied.  However, they still rotate rapidly enough to contain ergoregions and hence may suffer from an ergoregion instability (see e.g. \cite{Brito:2015oca} for a review). Perhaps surprisingly, if linear stability is established, the arguments presented in \cite{Dias:2012tq} indicate that nonlinear stability is ensured.

Our understanding of low-energy states in AdS thus hinges upon the endpoint of the superradiant instability of rapidly rotating black holes and the stability of geons, both of which remain open problems.

The superradiant instability and geons are intimately connected to each other via black resonators \cite{Dias:2015rxy}.  Black resonators are black holes that branch off from the onset of specific unstable superradiant modes.  They are non-stationary solutions but contain a ``helical'' Killing vector that makes them time-periodic. Though black resonators are stable to the mode that generated them, they are still rapidly rotating and hence remain unstable to other superradiant modes \cite{Dias:2011ss,Dias:2015rxy,Niehoff:2015oga,Chesler:2018txn,Green:2015kur}.

Black resonators form intermediate states in the dynamics of superradiant instabilities \cite{Niehoff:2015oga,Chesler:2018txn}. The nature of these instabilities, say in Kerr-AdS, is such that one unstable mode often dominates the dynamics at early times.  The growth of this mode causes the solution to approach a black resonator until the instabilities of the black resonator itself develop and drive the continuing dynamics.

Black resonators are also connected to geons, as their limit of zero horizon size is not AdS, but a geon.  Like black resonators, the geons also have a helical Killing field.  A small black resonator can be regarded as a small black hole placed in a geon where the black hole and geon have a matching angular frequency.

Geons and black resonators have been constructed in four dimensions \cite{Horowitz:2014hja,Dias:2015rxy,Fodor:2015eia,Fodor:2017spc}, where this helical Killing vector forms the only continuous symmetry.  This lack of symmetry poses a challenge in studying the stability of these solutions.  But recently, by going to five dimensions and exploiting the extra symmetries that emerge when both angular momenta are equal, the authors in \cite{Ishii:2018oms} have constructed black resonators and geons where the metric is cohomogeneity-1 (i.e.~the metric functions have a single variable).  Cohomogeneity-1 black resonators and geons coupled to electromagnetic waves have also been obtained in \cite{Ishii:2019wfs}.

We therefore set out in this paper to study the linear stability of the five-dimensional cohomogeneity-1 black resonators and geons obtained in \cite{Ishii:2018oms}. The simplicity of the background makes it easy to handle the perturbations as they can be expanded in terms of the spherical harmonics on $S^3$, for which the Wigner D-matrices are a convenient basis.  We consider scalar field, Maxwell, and gravitational perturbations comprehensively.  We map out the onset of superradiant instabilities in black resonators and determine the linear mode stability of geons.  We will find that many geons are stable, but there are certain ranges of parameters where they are unstable.

This paper is organized as follows. In Section~\ref{sec:review}, we review the five-dimensional cohomogeneity-1 black resonator background obtained in \cite{Ishii:2018oms}. The Wigner D-matrices are introduced in Section~\ref{sec:WignerD}. In Section~\ref{sec:computation}, we give an overview of our linear calculation, while technical details are explained in Appendices~\ref{sec:techscalar}--\ref{sec:techgrav}. Results for the study of these perturbations are shown in Section~\ref{sec:results}. In Section~\ref{multisolutions}, we briefly discuss multi-resonators and multi-geons that are expected to emerge from the instabilities of the perturbations. We then construct oscillating geon solutions in Section~\ref{sec:oscillgeons}, with technical details in Appendix~\ref{sec:techoscillgeons}. The paper is concluded with a summary and discussion in Section~\ref{sec:discussion}.

\section{Cohomogeneity-1 Black Resonators and Geons}
\label{sec:review}
We begin with a review of the construction of the cohomogeneity-1 black resonators and geons in \cite{Ishii:2018oms}.  They are obtained within pure Einstein gravity in global AdS$_5$. Throughout this paper, we set the AdS radius to unity, and the Einstein equation is given by $G_{\mu\nu}-6g_{\mu\nu}=0$.

Consider the following metric ansatz:
\begin{multline}
 ds^2=-(1+r^2)f(r)\mathrm{d}\tau^2 + \frac{\mathrm{d}r^2}{(1+r^2)g(r)}\\
+\frac{r^2}{4} \left[
\alpha(r)\sigma_1^2 + \frac{1}{\alpha(r)}\sigma_2^2 + \beta(r)(\sigma_3+2 h(r)\mathrm{d}\tau)^2
\right]\ ,
\label{metricanz}
\end{multline}
where the 1-forms $\sigma_a$ ($a=1,2,3$) are defined by
\begin{equation}
\begin{split}
  \sigma_1 &= -\sin\chi \mathrm{d}\theta + \cos\chi\sin\theta \mathrm{d}\phi\ ,\\
  \sigma_2 &= \cos\chi \mathrm{d}\theta + \sin\chi\sin\theta \mathrm{d}\phi\ ,\\
  \sigma_3 &= \mathrm{d}\chi + \cos\theta \mathrm{d}\phi  \ .
\end{split}
\label{invf}
\end{equation}
The 1-forms satisfy the Maurer-Cartan equation $\mathrm{d}\sigma_a = (1/2) \epsilon_{abc} \sigma_b \wedge \sigma_c$. The ranges of the coordinates $(\theta,\phi,\chi)$ are $0\leq \theta < \pi $, $0\leq \phi <2\pi$, and $0\leq \chi <4\pi$.  The radial coordinate $r$ will range from either the origin $r=0$ (for geons) or some horizon radius $r=r_h$ (for black holes) to asymptotic infinity $r\to\infty$.  We restrict ourselves to asymptotically global AdS$_5$ solutions where the boundary metric is conformal to $R^{(t)}\times S^3$.  The metric of AdS$_5$ can be recovered when $h$ is constant and the remaining metric functions are $f=g=\alpha=\beta=1$.

There is another closed-form solution to the Einstein equation that falls within this ansatz: the equal angular momentum Myers-Perry AdS black hole.  The Myers-Perry black hole~\cite{Myers:1986un,Hawking:1998kw,Gibbons:2004uw,Gibbons:2004js} is a rotating black hole in higher dimensions, which can be viewed as the higher dimensional generalization of the Kerr black hole solution (see \cite{Emparan:2008eg} for a review of higher dimensional black holes).  When the angular momenta of the Myers-Perry black hole are set equal, the solution can be written in cohomogeneity-1 form.  We will henceforth restrict ourselves only to the equal angular momenta family when discussing the Myers-Perry solutions, and abbreviate the Myers-Perry AdS black hole as ``MPAdS''.  Within our ansatz, MPAdS can be expressed as
\begin{equation}
\begin{split}
g(r)&=1-\frac{2\mu (1-a^2)}{r^2(1+r^2)} +\frac{2a^2\mu}{r^4(1+r^2)}\ ,\quad
\beta(r)=1+\frac{2 a^2\mu}{r^4}\ ,\\
h(r)&=\Omega-\frac{2\mu a}{r^4+2 a^2\mu}\ ,\quad
f(r)=\frac{g(r)}{\beta(r)}\ ,\qquad \alpha(r)=1\;,
\end{split}
\label{MPAdSFunctions}
\end{equation}
where the angular velocity $\Omega$ is
\begin{equation}
\Omega = \frac{2\mu a}{r_h^4+2 a^2\mu}\ ,
\label{OmegaMPAdS}
\end{equation}
and the horizon radius $r_h$ is given by the largest real root of $g(r_h)=0$. MPAdS is a two-parameter family, which can be conveniently parametrized by $r_h$ and $\Omega$.

Let us now discuss the symmetries of the ansatz \eqref{metricanz}, as well as those of some special cases.  Generically, the isometry group of the ansatz is $\mathrm R \times SU(2)$, which can be specified by the following four Killing vectors:
\begin{equation}
\begin{split}
K&=\partial_\tau\ ,\\
\xi_x &= \cos\phi\partial_\theta +
\frac{\sin\phi}{\sin\theta}\partial_\chi -
\cot\theta\sin\phi\partial_\phi\ ,\\
\xi_y &= -\sin\phi\partial_\theta +
\frac{\cos\phi}{\sin\theta}\partial_\chi -
\cot\theta\cos\phi\partial_\phi\ ,\\
\xi_z &= \partial_\phi\ .
\end{split}
\label{killing4}
\end{equation}
From these Killing vectors, we can define the angular momentum operators
\begin{equation}
L_i= i\xi_i \quad (i=x,y,z)\ ,
\label{SU2L}
\end{equation}
which satisfy the commutation relation of $SU(2)$: $[L_i,L_j]=i \epsilon_{ijk}L_k$. Each 1-form $\sigma_a$ is invariant under the $SU(2)$: $L_i \sigma_a \equiv \mathcal{L}_{L_i}\sigma_a = 0$.

An additional $SU(2)$ symmetry appears when $h$ is constant and $\alpha=\beta=1$, essentially because the 1-forms \eqref{invf} form the basis of a round $S^3$, which has $SU(2) \times SU(2)$ isometries.  The additional Killing vectors $\bar{\xi}_i$ can be expressed by
\begin{equation}
\begin{split}
\bar{\xi}_x &= -\sin \chi \partial_\theta + \frac{\cos \chi}{\sin \theta} \partial_\phi - \cot \theta \cos \chi \partial_\chi \ , \\
\bar{\xi}_y &= \cos \chi \partial_\theta + \frac{\sin \chi}{\sin \theta} \partial_\phi - \cot \theta \sin \chi \partial_\chi \ , \\
\bar{\xi}_z &= \partial_\chi \ .
\end{split}
\label{rkv}
\end{equation}
From these, one can define
\begin{equation}
R_i= i \bar{\xi}_i \ ,
\label{SU2R}
\end{equation}
which satisfy the $SU(2)$ commutation relation $[R_i,R_j]=-i \epsilon_{ijk}R_k$.

Generically, $\alpha$, $\beta$, and $h$ in the metric ansatz \eqref{metricanz} will completely break the second $SU(2)$ isometry corresponding to \eqref{rkv}, and $\bar{\xi}_i$ will no longer be Killing vectors.  However, when $\alpha(r)=1$ identically as in the case of MPAdS, this $SU(2)$ is broken only down to the $U(1)$ subgroup specified by $R_z = i \partial_\chi$. This implies that the spacetime~\eqref{metricanz} recovers this $U(1)$ symmetry, and together with the $SU(2)$ symmetries given by $\xi_i$, the isometry group enhances to $\mathrm R \times U(2)$.

The $U(1)$-generator $R_z$ commutes with the other symmetry generators (\ref{killing4}): $[L_i,R_z]=[K,R_z]=0$. This operator generates ``rotation'' of $\sigma_1$ and $\sigma_2$,
\begin{equation}
R_z\sigma_\pm = \pm \sigma_\pm\ ,\quad R_z \sigma_3 =0\ ,
\label{Wcharge}
\end{equation}
where we defined
\begin{equation}
\sigma_\pm = \frac{1}{2}(\sigma_1 \mp i  \sigma_2)\ .
\label{sigma_pm}
\end{equation}
Eq.\,\eqref{Wcharge} indicates that $\sigma_\pm$ and $\sigma_3$ have $U(1)$-charges $\pm 1$ and $0$, respectively. Note that, if $\alpha(r) \neq 1$, $R_z$ is not a symmetry generator of the spacetime~\eqref{metricanz}.

The spacetime~(\ref{metricanz}) also has a discrete isometry:
\begin{equation}
 (\tau,\phi,\chi)\to -(\tau,\phi,\chi)\ .
\label{parity}
\end{equation}
By this isometry, the invariant 1-forms transform as
$(\sigma_\pm,\sigma_3)\to (-\sigma_\mp,-\sigma_3)$.

Let us now describe the construction of geons and black resonators in our ansatz \eqref{metricanz}.  For the boundary conditions, we require regularity at the origin (for geons), or the horizon (for black resonators).  For black resonators, we also take $h=0$ at the horizon. The solutions must also be asymptotically AdS, so as $r \to \infty$ we require
\begin{equation}
 f,g,\alpha,\beta\to 1\ ,\quad h\to \Omega\ ,\quad (r\to \infty)\ ,
\label{AsymAdS}
\end{equation}
where $\Omega$ is a constant representing the angular velocity of the spacetime.

The asymptotic behavior \eqref{AsymAdS} corresponds to a rotating frame at infinity.  We can introduce another frame that is non-rotating at infinity $(t,\psi)$ via $\mathrm{d}t=\mathrm{d}\tau$ and $\mathrm{d}\psi=\mathrm{d}\chi+2\Omega \mathrm{d}\tau$. The function $h$ is then shifted as $\bar{h} = h - \Omega$, whose asymptotic behavior is $\bar{h} \to 0 \ (r \to \infty)$. In this frame, the Killing vector $K$ can be rewritten as
\begin{equation}
K=\frac{\partial}{\partial t} + \Omega \frac{\partial}{\partial(\psi/2)}\ .
\end{equation}

In Ref.\,\cite{Ishii:2018oms}, black resonators and geons were numerically constructed under these boundary conditions with the ansatz~(\ref{metricanz}) as solutions possessing a \emph{helical Killing vector}. It was found that they always satisfy $\Omega>1$. The Killing vector $K$ becomes spacelike near infinity:
\begin{equation}
K^\mu K_\mu\to -(1-\Omega^2)r^2 \ ,\quad (r\to \infty)\ .
\end{equation}
In the frame that is non-rotating at infinity, the Killing vector $K$ is helical.  There is no continuous time translation symmetry, but what remains is a discrete one: $t\sim t+\pi/(2\Omega)$. Strictly speaking, the lack of a Killing vector that is timelike everywhere at infinity implies that the spacetime is non-stationary.  Nevertheless, for black resonators, this Killing field serves as a horizon generator, and the entropy of the black hole does not change.  In this sense, black resonators are steady-state solutions.  The zero horizon size limit of black resonators is a geon, which also has this helical Killing field.

\begin{figure}
  \centering
\subfigure[The mass of geons.]
 {\includegraphics[scale=0.45]{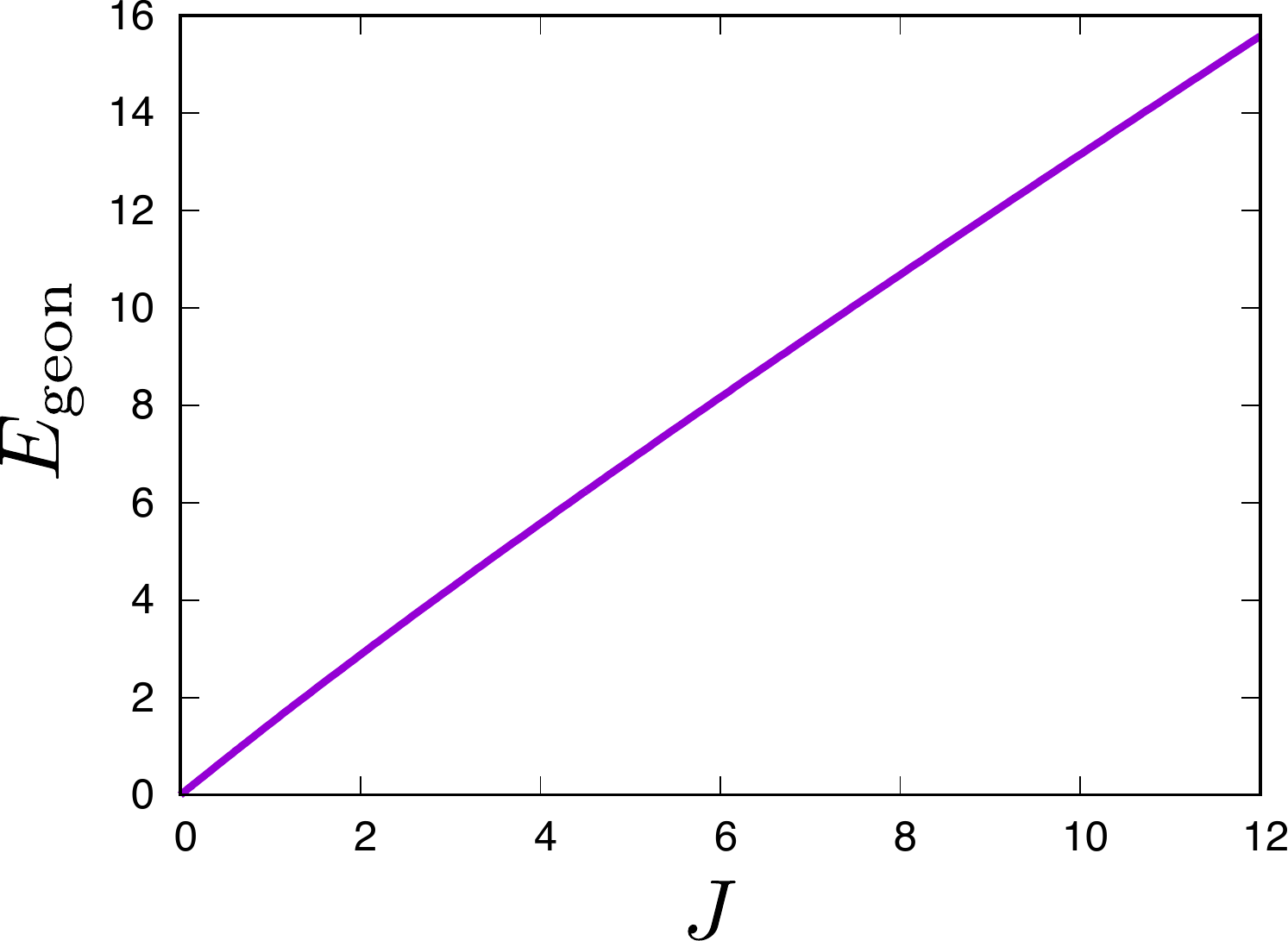}
  }
  \subfigure[Phase diagram of black resonators]
 {\includegraphics[scale=0.45]{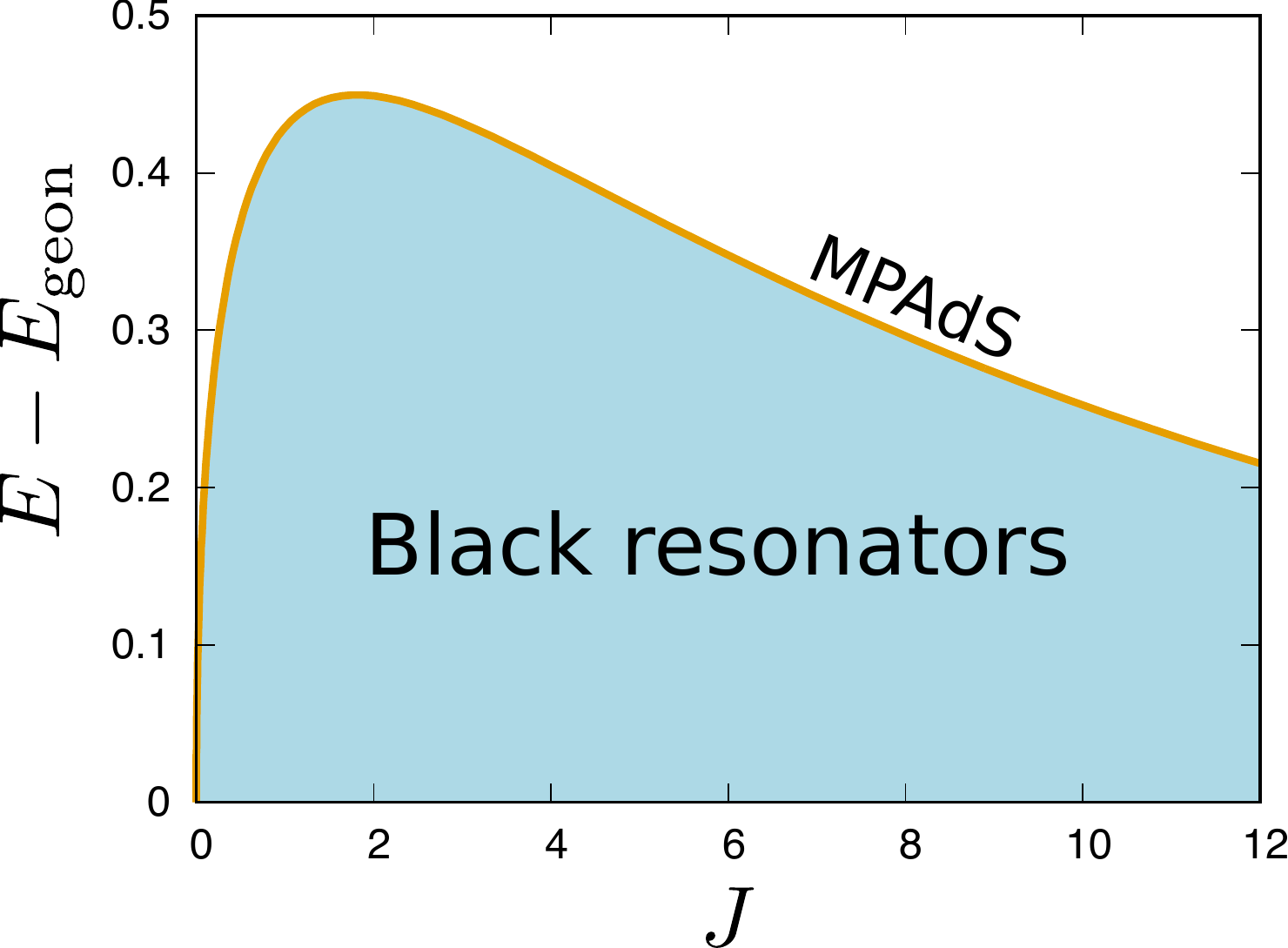}
  }
 \caption{
(a): The mass of geons $E_\textrm{geon}$ as a function of angular momentum $J$.
(b): Phase diagram of cohomogeneity-1 black resonators.
In the vertical axis, $E-E_\textrm{geon}$ is used for visibility.
The black resonators exist in the filled region.
The horizontal axis represents the geon limit.
The orange curve indicates the onset of the superradiant instability of MPAdS, where the black resonators branch off.
}
\label{BRregion}
\end{figure}

The phase diagram of the five-dimensional black resonator and geon solutions within this ansatz is shown in Fig.\,\ref{BRregion}.\footnote{In addition, there are an infinite number of black resonators and geons labeled by a radial overtone number $n$, i.e.~the number of the nodes in the $r$-direction. We focus only on the fundamental $n=0$ which has the lowest number of nodes.} We denote by $E$ and $J$ the mass and angular momentum, respectively.\footnote{For notational simplicity, we set the five dimensional gravitational constant $G_5=1$.  One can easily restore it as $E\to G_5 E$ and $J\to G_5 J$.}
The mass of geons is denoted by $E_\textrm{geon}$, in particular. In the left figure, $E_\textrm{geon}$ is shown as a function of the angular momentum.
In the right figure, the difference between the mass of the black resonator and that of the geon with the same angular momentum, $E-E_\textrm{geon}$, is used as the vertical axis for visibility.
Black resonator solutions exist in the filled region, which is bounded by the horizontal axis and the upper orange curve, where the latter gives the onset of the superradiant instability in MPAdS for the mode that generates the black resonator.
MPAdS (not shown) also exists slightly below this orange curve until extremality is reached.  When the MPAdS competes thermodynamically with the black resonators in the microcanonical ensemble, the black resonators dominate (i.e.~black resonators have more entropy than MPAdS for the same energy and angular momentum).

\section{Wigner D-matrices}
\label{sec:WignerD}

We will study the stability of the cohomogeneity-1 black resonators and geons against scalar field, Maxwell, and gravitational perturbations. These can be expanded in terms of the Wigner D-matrix $D^j_{mk}(\theta,\phi,\chi)$. (For an introduction to the Wigner D-matrix, see e.g.~Ref.\,\cite{Sakurai:2011zz}. See also Refs.\,\cite{Hu:1974hh,Murata:2007gv,Kimura:2007cr,Murata:2008yx,Murata:2008xr} for its application to the scalar field, Maxwell, and gravitational perturbations in general relativity.)

The Wigner D-matrix has three quantum numbers $(j,m,k)$: the total angular momentum $j$, and the $U(1)$ charges $m$ and $k$ associated to $L_z$ and $R_z$, respectively. The ranges of the quantum numbers are
\begin{equation}
\begin{split}
 j&= 0, \, 1/2, \, 1,\, 3/2, \ldots \ , \\
m&= -j, \, -j+1,\ldots, \, j \ , \\
k&= -j, \, -j+1, \ldots, \, j \ .
\end{split}
\label{jmkrange}
\end{equation}
The Wigner D-matrix vanishes if $(m,k)$ are outside these ranges. It is the eigenfunction of $L^2=\sum_{i=x,y,z}L_i^2$, $L_z$, and $R_z$:
\begin{equation}
L^2 D^j_{mk} = j(j+1) D^j_{mk}\ ,\quad
L_z D^j_{mk} = m D^j_{mk}\ ,\quad
R_z D^j_{mk} = k D^j_{mk}\ .
\end{equation}
The subscripts of $D^j_{mk}$ are shifted by $L_{\pm} = L_x \pm i L_y$ and $R_{\pm} = R_y \pm i R_x$ as
\begin{equation}
 \begin{split}
L_+ D^j_{mk} &= \varepsilon_{m+1} D^j_{(m+1)k}\ , \quad
L_- D^j_{mk} = \varepsilon_{m} D^j_{(m-1)k}\ , \\
R_+ D^j_{mk} &= \epsilon_{k+1} D^j_{m(k+1)}\ , \quad \;
R_- D^j_{mk} = \epsilon_{k} D^j_{m(k-1)}\ ,
 \end{split}
\end{equation}
where $\varepsilon_m=\sqrt{(j+m)(j-m+1)}$ and $\epsilon_k=\sqrt{(j+k)(j-k+1)}$.

Because of the $SU(2)$ symmetry of the spacetime~(\ref{metricanz}), the modes with different $(j,m)$ are trivially decoupled when we decompose the perturbations by the Wigner D-matrices.  Therefore, in the following, we suppress the indices $(j,m)$ and use $D_k$ for $D^j_{mk}$.  (Though, remember that the wavenumber $j$ is still important in that it limits the range of $k$ and appears in various formulae involving $\epsilon_k$.) Unlike $(j,m)$, modes with different $k$ are coupled in backgrounds where $\alpha(r) \neq1$ because in that situation $\partial_\chi$ is not an isometry.  In the special case that $\alpha(r)=1$ as in MPAdS, the extra $U(1)$ isometry allows modes with different $k$ to be decoupled.

We end this section by showing useful formulae for the derivatives of the Wigner D-matrix. 
Solving 
$R_z D_k = k D_k$, $R_+ D_k = \epsilon_{k+1} D_{k+1}$, and $R_- D_k = \epsilon_{k} D_{k-1}$,
we obtain
\begin{equation}
\begin{split}
\partial_\theta D_k &= -\frac{i}{2}
(\epsilon_k e^{-i\chi} D_{k-1}
+\epsilon_{k+1} e^{i\chi} D_{k+1})\ ,\\
\partial_\phi D_k &=
-i k \cos\theta D_k
+\frac{1}{2}\epsilon_k \sin\theta e^{-i\chi} D_{k-1}
-\frac{1}{2}\epsilon_{k+1} \sin\theta e^{i\chi} D_{k+1} \ ,\\
\partial_\chi D_k &= -ik D_k\ .
\end{split}
\label{Wignerfomulae}
\end{equation}
See also Ref.\,\cite{Hu:1974hh} for a derivation of these formulae.

\section{Overview of Linear Calculation}
\label{sec:computation}
In this section, we give an overview of our computation for linear perturbations.  Here, we mostly focus on details that are necessary for understanding the results.  More technical aspects of our computation can be found in the appendices.
\subsection{Massless Scalar Field}
For the massless scalar field, we wish to solve the Klein-Gordon equation on the background of a black resonator or a geon, $\Box \Phi=0$. To find the onset of instability, however, it is convenient to consider the eigenvalue equation of the Klein-Gordon operator given by\footnote{Note that $\lambda$ is nothing but the squared mass of a massive scalar field.}
\begin{equation}
\Box \Phi = \lambda \Phi\ ,
\label{Kleineigen}
\end{equation}
and search for solutions to the above equation with zero eigenvalues $\lambda=0$.

We decompose the scalar field by using the Wigner D-matrices as
\begin{equation}
 \Phi(\tau,r,\theta,\phi,\chi)
= e^{-i\omega \tau}\sum_{|k|\leq j} \phi_k (r) D_k(\theta,\phi,\chi)
\ ,
\label{Phi_expand}
\end{equation}
where the indices of and summation over $(j,m)$ are suppressed. Substituting this into (\ref{Kleineigen}), rewriting the derivatives of the Wigner D-matrices by using the formulae~(\ref{Wignerfomulae}), and rearranging the index $k$, we obtain for each $k$ an equation of the form
\begin{equation}
L_k \phi_k+c_{k-1} \phi_{k-2} +c_{k+1} \phi_{k+2} =0\ ,
\label{phieq}
\end{equation}
where $L_k$ is a background-dependent differential operator with only radial derivatives $\partial_r$, and $c_k$ is a function depending on the radial coordinate and background metric. Their explicit expressions can be found in Appendix \ref{subsec:scalareom}. $\phi_k$ vanishes if $k$ is outside the range given in \eqref{jmkrange}.

In (\ref{phieq}), we find that the mode coupling is ``double-stepping''-- the modes with $k$ and $k\pm2$ are coupled.  We note that if $\alpha=1$ identically as in the case of MPAdS, $c_k=0$ and the individual $k$ modes decouple.  Otherwise, as in the case for the black resonators and geons, fixed-$j$ perturbations reduce to a coupled system of scalar fields $\phi_k(r)$.

Note that because of the double-stepping, there is a difference when $j$ is a half-integer or an integer.  For half-integer $j$, the two following systems are decoupled from each other:
\begin{align}\label{halfintjscalar}
\bm{v} &= \{\phi_j,\phi_{j-2},\phi_{j-4},\cdots,\phi_{-j+1}\} \ , \\
\tilde{\bm{v}} &=\{\phi_{j-1},\phi_{j-3},\phi_{j-5},\cdots,\phi_{-j}\} \ .
\end{align}
Both of these have the same equations of motion, so it suffices to consider just one of them.

For integer $j$,
\begin{align}\label{intjscalar}
\bm{v} &= \{\phi_j,\phi_{j-2},\phi_{j-4},\cdots,\phi_{-j}\} \ , \\
\tilde{\bm{v}} &=\{\phi_{j-1},\phi_{j-3},\phi_{j-5},\cdots,\phi_{-j+1}\}
\end{align}
are decoupled systems with different equations of motion.  On solutions with $\omega=0$, integer $j$ modes can be further decoupled into odd and even perturbations under the isometry \eqref{parity}:
\begin{equation}\label{parityscalar}
\bm v_\pm=\{\phi_k\pm\phi_{-k}|k\geq0\}\;,
\end{equation}
with the plus and minus signs corresponding to even and odd parity, respectively.  Note that if $\omega\neq0$, the isometry \eqref{parity} is broken, and if $j$ is a half integer, $k$ and $-k$ modes are decoupled from each other.  This is why the decoupling \eqref{parityscalar} only applies to $\omega=0$ and integer $j$.

To solve the coupled equations (\ref{phieq}), we need to impose boundary conditions. We require regularity at the horizon for black resonators or at the origin for geons.
 At asymptotic infinity, we choose no-source boundary conditions, where the slowest fall-off of the scalar field vanishes.  With these boundary conditions, a fixed background and fixed $j$ defines an ODE system for $\phi_k$ with two unknown constants $\omega$ and $\lambda$.

Unfortunately, typical growth rates $\mathrm{Im}(\omega)$ for unstable modes of black resonators are extremely small (often smaller than the 64-bit machine epsilon $\sim 10^{-16}$).  For this reason, we only compute the onset of the instability for the black resonators where $\mathrm{Im}(\omega)=0$.

In Appendix \ref{subsec:findonset}, we demonstrate that $\mathrm{Im}(\omega)=0$ implies also $\mathrm{Re}(\omega)=0$.  We can therefore set $\omega=0$, and solve the resulting ODE system with $\lambda$ in an eigenvalue. Since $\omega=0$, the equations are time independent, and we can impose a regularity condition on the horizon, which is given by the $\omega\to 0$ limit of an ingoing wave condition.  We vary parameters of the background until we find that $\lambda=0$, which corresponds to the onset of an instability.  We carry out this computation numerically using a shooting method.

For geons, there is no horizon, and the lack of dissipation implies that the frequency $\omega$ is purely real or purely imaginary.  Therefore, for geons, we set $\lambda=0$ and compute $\omega$ directly.  We do so using pseudospectral methods with Chebyshev polynomials.

\subsection{Maxwell Field}
We consider the Maxwell perturbation in the Lorenz gauge. The equations of motion and the gauge condition are given by
\begin{align}
\Box A_\mu + 4A_{\mu}&=\lambda A_\mu\ ,\label{maxwell_eigen}\\
\nabla^\mu A_{\mu}&=0\ , \label{LorenzA}
\end{align}
where we have introduced the eigenvalue $\lambda$ for the operator $\Box+4$.  We seek solutions where $\lambda=0$.  Because the Lorenz gauge condition does not fix the gauge freedom completely, we will need to check that our solutions are not gauge modes. This technical point is discussed further in Appendix \ref{comgauge}.

Let us decompose \eqref{maxwell_eigen} and \eqref{LorenzA} by using the Wigner D-matrices. First, we define the basis 1-forms
\begin{equation}
 (e^\tau,e^r,e^\pm,e^3)=(d\tau, dr, \sigma_\pm, \sigma_3+2 hd\tau)\ ,
\label{orthbasis}
\end{equation}
and write the Maxwell field as
\begin{equation}
A=A_a e^a\ ,
\label{A_def0}
\end{equation}
where $a=\tau,r,+,-,3$. The  components of the Maxwell field $A_a$ can be decomposed by the Wigner D-matrices as
\begin{equation}
A_A=e^{-i\omega \tau} \sum_{|k|\leq j} A^k_A(r) D_k\ ,\quad
A_\pm=e^{-i\omega \tau} \sum_{|k\mp 1|\leq j} A^k_\pm(r) D_{k\mp 1}\ ,
\label{A_Dexpand}
\end{equation}
where $A=\tau,r,3$. Notice that the index $k$ of the Wigner D-matrices is shifted in the expansion of $A_\pm$. This was done so that $\sigma_\pm D_{k\mp1}$ have $U(1)$ charge $k$, matching that of $D_k$. Because of the shift of the index, $A^k_{\pm}$ are defined in $|k\mp 1|\leq j$, while $A_A^k$ $(A=\tau,r,3)$ are defined in $|k|\leq j$. Then, from \eqref{maxwell_eigen} and \eqref{LorenzA}, we obtain coupled differential equations for $\bm{A}_k\equiv (A_A^k, A_\pm^k)$ of the form
\begin{align}
\bm{A}_k''&=\bm{P}[\bm{A}_k,\bm{A}_{k-2},\bm{A}_{k+2}]\ ,\label{bmAeq}\\
(A_r^k)'&=Q[\bm{A}_k,\bm{A}_{k-2},\bm{A}_{k+2}]\ ,\label{Areq}
\end{align}
where $\bm{P}$ and $Q$ are linear operators arranged in such a way that the first $r$ derivative is included in $\bm{P}$, but not in $Q$. Their explicit forms are too cumbersome, and we do not to reproduce them here.

We again find that the coupling is double-stepping ($k$ modes are coupled to $k\pm 2$), and this will again lead to different decoupled systems depending on whether $j$ is a half-integer or integer in a similar manner as \eqref{halfintjscalar} and \eqref{intjscalar}.  Integer $j$ solutions with $\omega=0$ can be further divided into even and odd perturbations in a similar manner as \eqref{parityscalar}.

In general, there are more equations than unknown linear functions.  We will only solve a subset of these equations and treat the remainder as constraint equations which we will verify after obtaining a solution.  The details of which combinations of the equations we solve and which we leave for verification afterwards, can be found in Appendix \ref{Mxw_SR_BR}.

For boundary conditions, we again require regularity at the horizon or the origin, and the slowest fall-off at infinity to vanish.

In the same way as the case of the scalar field, on black resonators we will find the onset of an instability by setting $\omega=0$ and computing $\lambda$, searching for regions of parameter space where $\lambda=0$. For geons, we will set $\lambda=0$ and compute the frequency $\omega$ directly.

\subsection{Gravitational Perturbations}
We consider gravitational perturbation using the transverse-traceless gauge. The perturbation equations are
\begin{equation}
 \Box h_{\mu\nu}+2 R_{\mu\rho\nu\sigma}h^{\rho\sigma}=\lambda h_{\mu\nu}\ ,
\label{Lich}
\end{equation}
with
\begin{equation}
 \nabla^\mu h_{\mu\nu}=0\ ,\quad h^{\mu}{}_\mu=0\ ,
\label{TTcond}
\end{equation}
where we introduced the eigenvalue $\lambda$ of the Lichnerowicz operator to find the onset of instability.  As in the Maxwell case, we will need to verify that the solutions we will find are not gauge modes.

In a similar way as in the previous sections, we expand \eqref{Lich} and \eqref{TTcond} by the Wigner D-matrices. We can decompose the metric perturbation by using the orthogonal basis~\eqref{orthbasis} as
\begin{equation}
 h_{\mu\nu}\mathrm{d}x^\mu \mathrm{d}x^\nu
=h_{ab} e^a e^b\ ,
\end{equation}
where $a,b=\tau,r,+,-,3$. The metric components $h_{ab}$ can be expanded by the Wigner D-matrices as
\begin{equation}
\begin{split}
h_{AB}&=e^{-i\omega \tau} \sum_{|k|\leq j} h^k_{AB}(r) D_k\ ,\quad
h_{+-}=e^{-i\omega \tau} \sum_{|k|\leq j} h^k_{+-}(r) D_k\ ,\\
h_{A\pm}&=e^{-i\omega \tau} \sum_{|k\mp 1|\leq j} h^k_{A\pm}(r) D_{k\mp 1}\ ,\quad
h_{\pm\pm}=e^{-i\omega \tau} \sum_{|k\mp 2|\leq j} h^k_{\pm\pm}(r) D_{k\mp 2}\ ,
\end{split}
\label{h_WignerExpansion}
\end{equation}
where $A,B=\tau,r,3$, and we shifted the index $k$ of the Wigner D-matrices in the expansion depending on the $U(1)$-charges of the corresponding 1-forms. Note that $h^k_{AB}$ and $h^k_{+-}$ are defined in $|k|\leq j$, while $h^k_{\pm\pm}$ and $h^k_{A\pm}$ in shifted domains $|k\mp 2|\leq j$ and $|k\mp 1|\leq j$, respectively. Substituting \eqref{h_WignerExpansion} into \eqref{Lich} and \eqref{TTcond}, we obtain the coupled equations for $\bm{h}_k\equiv (h^k_{++}, h^k_{\tau+}, \cdots, h^k_{--})$ schematically as
\begin{align}
\bm{h}_k''&=\bm{P}[\bm{h}_{k-4},\bm{h}_{k-2},\bm{h}_{k},\bm{h}_{k+2},\bm{h}_{k+4}]\ ,\label{Lich2}\\
(h_{ar}^k)'&=Q[\bm{h}_{k-4},\bm{h}_{k-2},\bm{h}_{k},\bm{h}_{k+2},\bm{h}_{k+4}]\ ,\label{TV2}\\
h_{+-}^k&=R[\bm{h}_{k-4},\bm{h}_{k-2},\bm{h}_{k},\bm{h}_{k+2},\bm{h}_{k+4}]\ ,\label{TL2}
\end{align}
where $\bm{P}$, $Q$ and $R$ are linear operators; $\bm{P}$ includes the first derivative terms in $r$, but $Q$ and $R$ do not contain such terms.

We again find that the mode coupling is double-stepping. Unlike the scalar and Maxwell field perturbations, $\bm{h}_{k\pm 4}$ appear in the right hand side of the above equations.  There are again different decoupled systems depending on whether $j$ is a half-integer or integer, and integer $j$ solutions with $\omega=0$ can be further divided into even and odd perturbations in a similar manner as \eqref{parityscalar}.

Like the Maxwell field, the equations of motion contain more equations than unknown functions.  We again solve only a subset of these equations and treat the remainder as constraint equations to be verified after obtaining a solution.

For boundary conditions, we require regularity at the horizon or the origin, and the slowest fall-off at infinity to vanish as in the case of the scalar and Maxwell fields.

Unlike the scalar field and Maxwell field, the growth rates of gravitational perturbations are not prohibitively small.  We therefore compute the frequency $\omega$ directly by setting $\lambda=0$ and solving for $\omega$ as an eigenvalue.  As a consistency check, we also search for the onsets of instabilities by setting $\omega=0$ and finding regions of parameter space where $\lambda=0$.  We perform both types of calculations for both black resonators and geons.

\section{Results}
\label{sec:results}
\subsection{Scalar Field}

\begin{figure}
  \centering
\subfigure[Onsets for scalar field on black resonator.]
 {\includegraphics[scale=0.45]{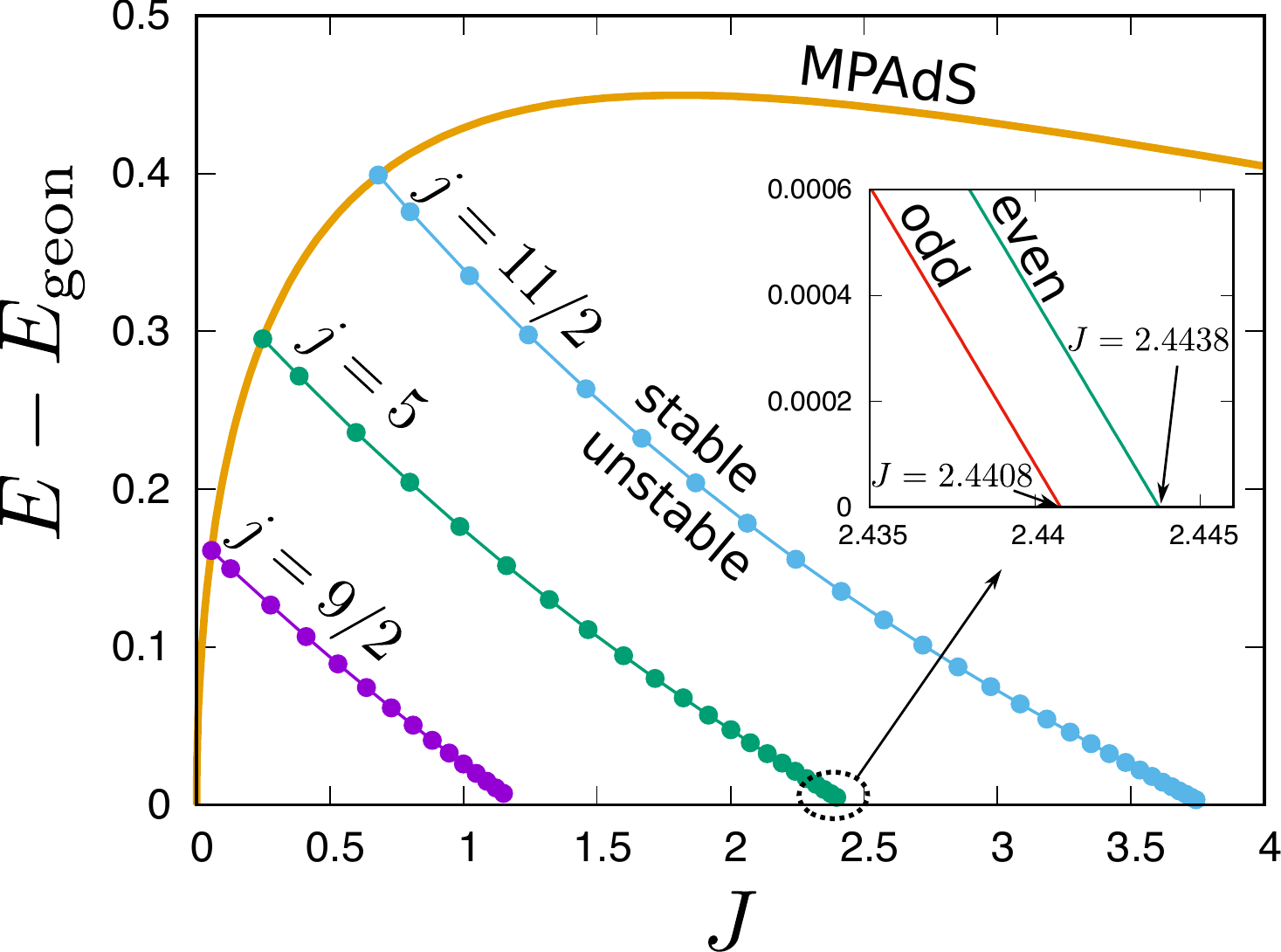}\label{scalar_J_E}
  }
  \subfigure[Scalar field frequency on geons.]
 {\includegraphics[scale=0.4]{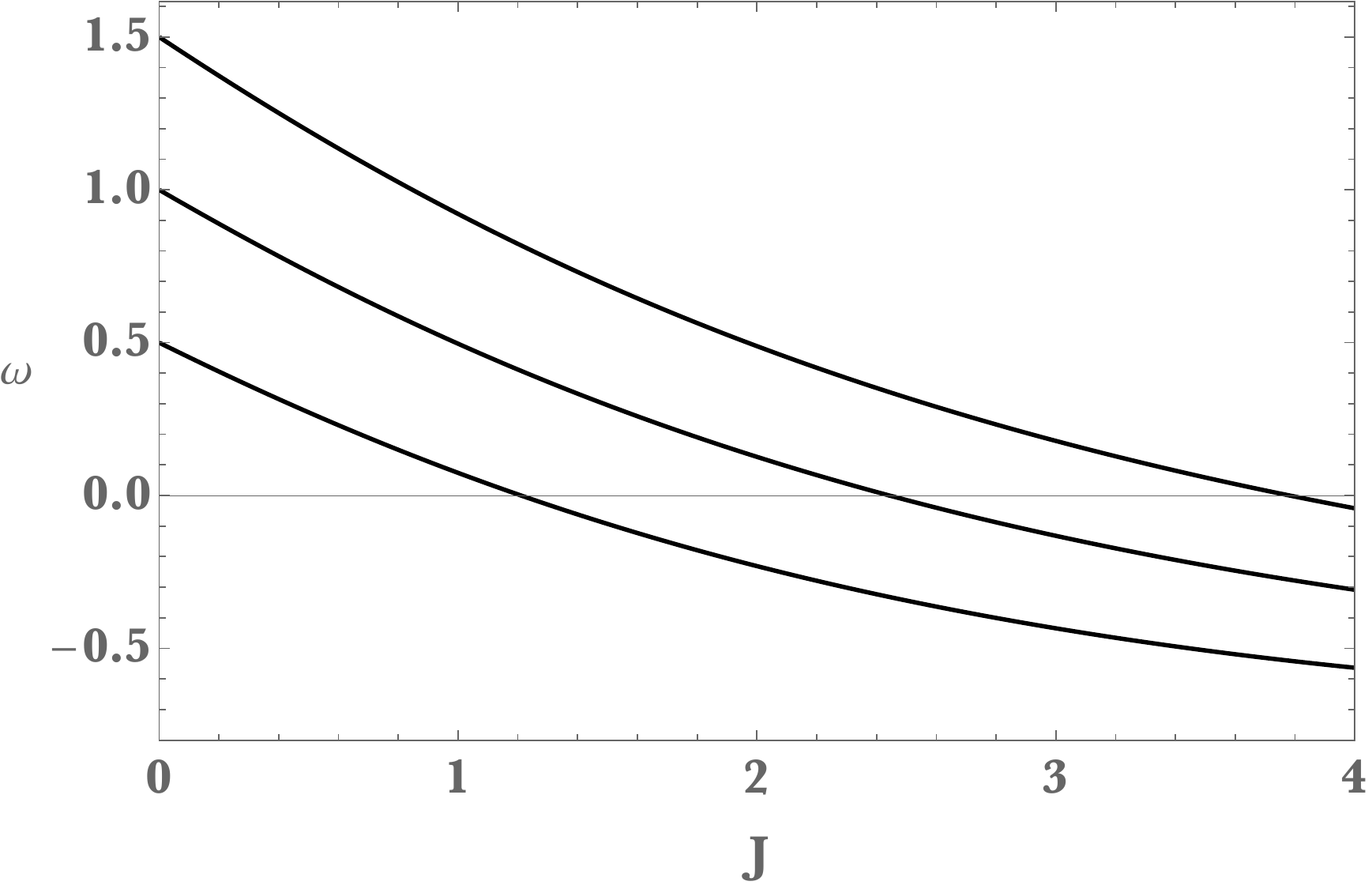}\label{scalargeon}
  }
\subfigure[Growth rate of unstable frequency on geons.]
 {\includegraphics[scale=0.4]{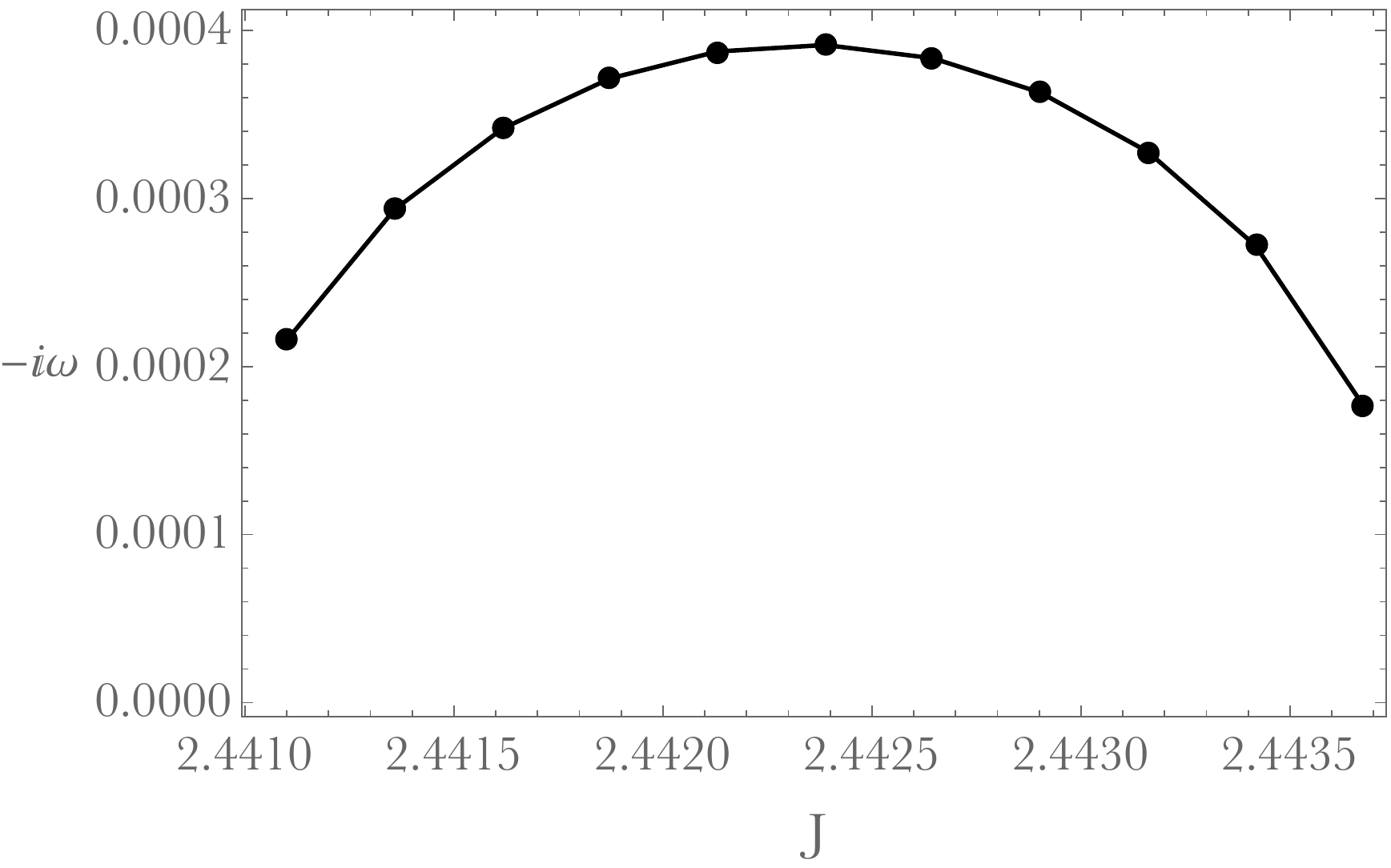}\label{scalargeon_Imomega}
  }
 \caption{(a): Phase diagram for the stability of the black resonators against a scalar field.  The colored curves correspond to the onset of an instability where $\omega=0$.  Black resonators are unstable to a given mode below each curve. Wavenumbers are for the Wigner D-matrices $D^j_{mk}$.
(b): Perturbative frequency $\omega$ for geons.
Note that the zero modes of geons where $\omega=0$ matches where the zero modes of the black resonators intersect the geon family.
(c): Growth rate of the unstable mode on the geon.
}
\label{fig:scalarmodes}
\end{figure}

We now describe the main results of our perturbation calculation, beginning with the scalar field. For a given $j$, we focus on the set of modes with a dominant $k=j$ component, which would be the mode with the highest growth rate among those with different $k$~\cite{Zouros:1979iw}. In Fig.~\ref{scalar_J_E}, we show the location of the onsets for the superradiant instability (where $\omega=0$) on black resonators for three modes with $j=9/2$, $j=5$, and $j=11/2$.  
MPAdS at branching points of black resonators are stable for $j\leq 4$, as we explain in Section~\ref{scalarMPAdS} of the Appendix. We also did not find any evidence of instability of black resonators for $j\leq 4$.
For $j=5$, there are actually two onset curves corresponding to even and odd parity modes under the discrete isometry~(\ref{parity}), but they almost coincide.  In the inset, we zoom into the region near the geon where it is easier to see the difference between the $j=5$ even and odd parity modes.

Note that the onset curves intersect the orange branching curve, where black resonators merge with MPAdS solutions.  By using the known results \cite{Kunduri:2006qa} on MPAdS (see also Appendix~\ref{scalarMPAdS} for more details), we can deduce that each mode is unstable below its onset curve, and stable above it. In other words, we see that black resonators with large angular momentum $J$ have more stable modes. Though, like MPAdS, black resonators are in general unstable to an infinite number of modes with arbitrarily high wavenumbers.

For the geon, most modes we have found for the scalar field perturbation are normal modes. That is, the modes have purely real frequency and geons are linearly stable to these modes. In Fig.~\ref{scalargeon}, we show the perturbative frequency $\omega$ on the geons for these same modes with $j=9/2$, $j=5$, and $j=11/2$, from left to right, respectively.

We note that for the $j=5$ mode, there is a small window near $J\sim 2.44$ where this mode becomes pure imaginary and unstable.\footnote{Actually, $\omega$ goes continuously from purely real, to $\omega=0$, to pure imaginary.  But one can show that if $\omega$ is an eigenfrequency, then $-\omega$ is also an eigenfrequency. There is therefore an ambiguity as to whether this $\omega$ continues to a positive imaginary mode, or negative imaginary.  We just take the positive imaginary mode because it corresponds to an instability.}  This window is imperceptibly small in Fig.~\ref{scalargeon}, but we show the growth rate in Fig.~\ref{scalargeon_Imomega}.

Note that the geons have zero modes where $\omega=0$, and that these zero modes correspond to precisely where the onset curves of the black resonators intersect the geons.  This includes both the odd and even $j=5$ modes.  Meanwhile, the half-integer zero modes of the geons are not an onset to an instability as the frequencies are all purely real.

Outside of the small window given in Fig.~\ref{scalargeon_Imomega} (and presumably other similar windows for higher integer $j$), geons are linearly stable. Continuity implies that the corresponding mode on nearby black resonators has a small growth rate $\mathrm{Im}(\omega)$.

Both the geon zero modes and the black resonator onsets correspond to $\tau$-independent perturbations. This suggests a new family of geons and black resonators with nontrivial scalar ``hair''.  We will discuss these solutions in Section \ref{multisolutions}.

\subsection{Maxwell Field}
\begin{figure}
  \centering
\subfigure[Onsets for Maxwell field on black resonators.]
 {\includegraphics[scale=0.45]{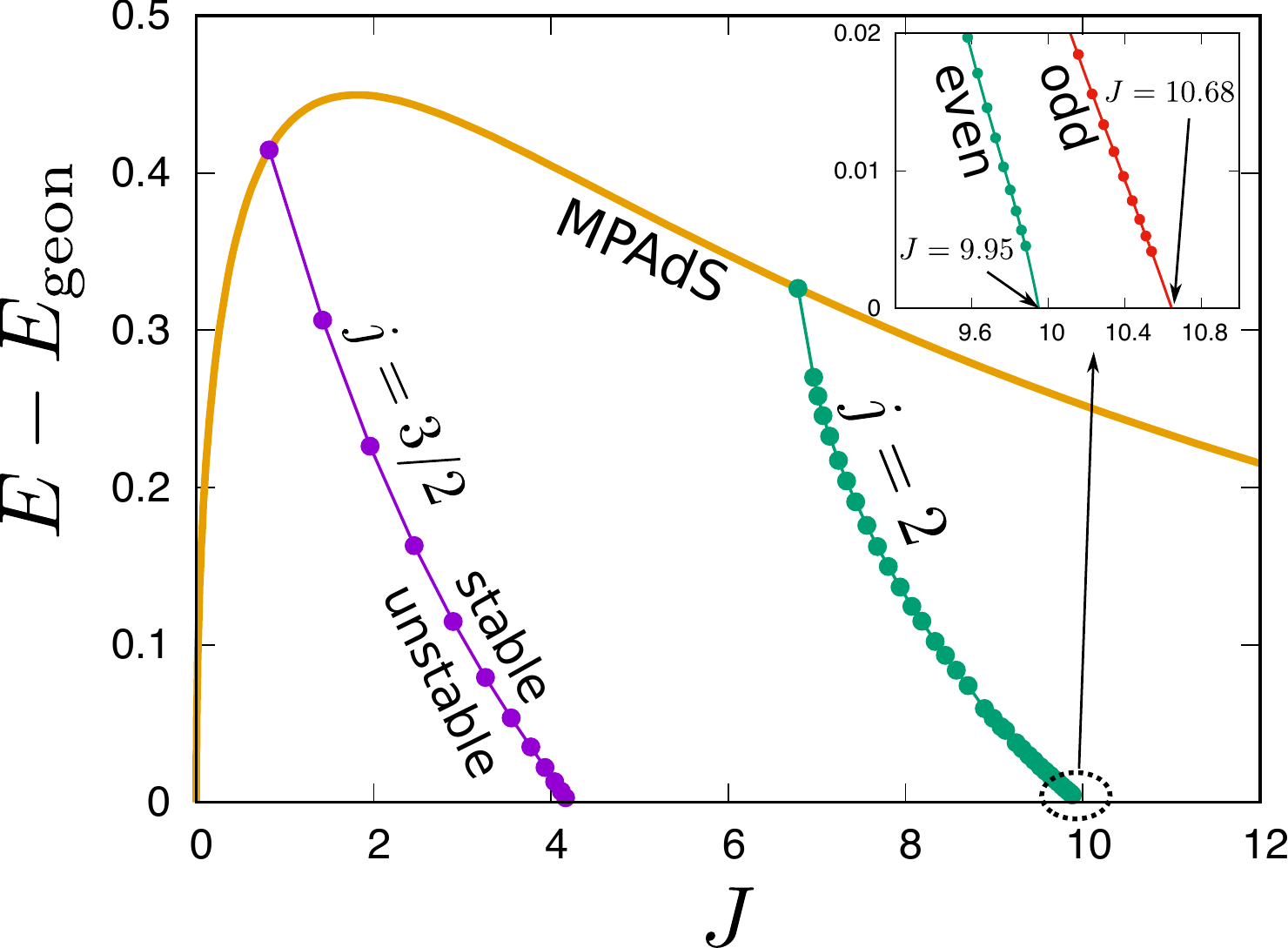}\label{maxwell_J_E}
  }
\subfigure[Maxwell field frequency on geons.]
 {\includegraphics[scale=0.4]{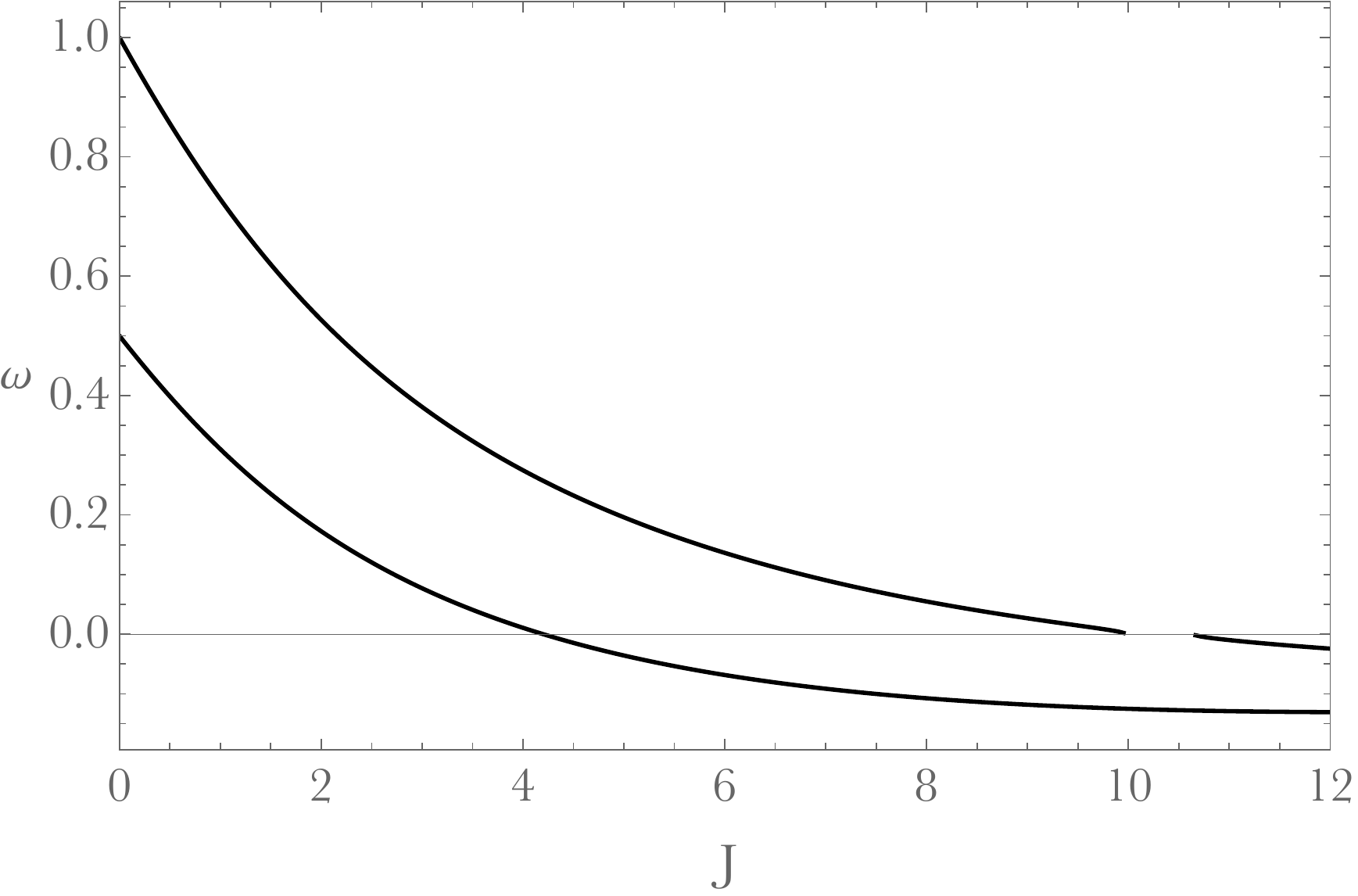}\label{maxwellgeon}
}
\subfigure[Growth rate of unstable frequency on geons.]
 {\includegraphics[scale=0.4]{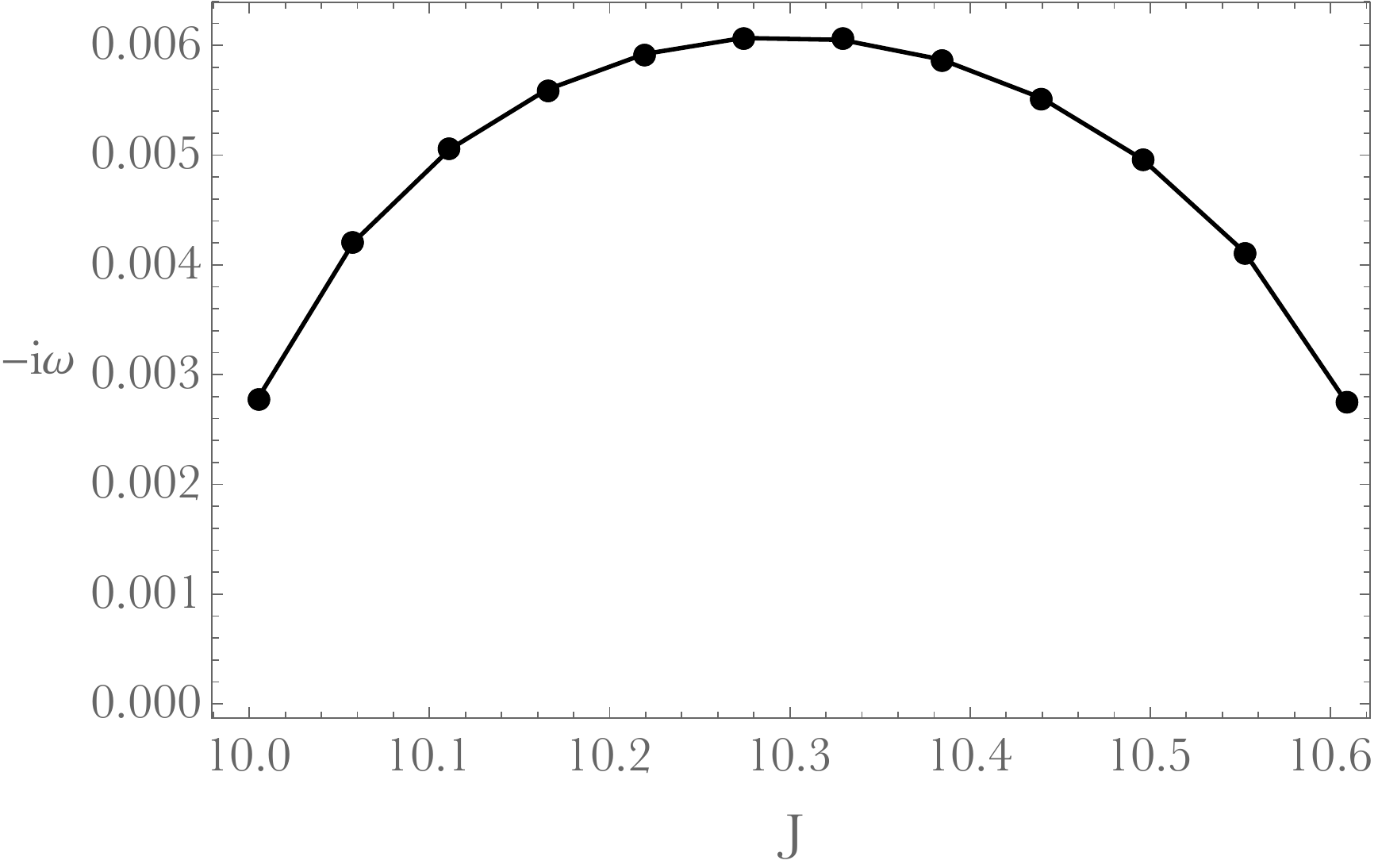}\label{maxwellgeonunstable}
}
 \caption{(a): Phase diagram for the stability of the black resonators against a Maxwell field.  The colored curves correspond to the onset of an instability where $\omega=0$.  Black resonators are unstable to a given mode in the region to the left of each curve. Wavenumbers are for Wigner D-matrices $D^j_{mk}$.  (b): Perturbative frequency $\omega$ for geons with $j=3/2$ and $j=2$.  Frequencies are purely real or purely imaginary.  Geons are unstable to the $j=2$ mode in the gap around $J\sim 10.5$, where this mode becomes purely imaginary. (c): Growth rate of the unstable mode on the geon.
 }
\label{fig:maxwell}
\end{figure}
In Fig.~\ref{fig:maxwell}, we show results for Maxwell fields for modes with $j=3/2$ and $j=2$. 
MPAdS at branching points of black resonators are stable for $j\leq 1$, as we explain in Section~\ref{Maxwell_MPAdS} of the Appendix. We also did not find any evidence of instability of black resonators for $j\leq 1$.
For $j=2$, there are two onset curves corresponding to even and odd parity modes under~(\ref{parity}); the onset of the odd perturbation is shown only in the inset in Fig.~\ref{maxwell_J_E} for visibility. Much of the behaviour is similar to that of the scalar field.  That is, black resonators with large $J$ tend to have more stable modes, although they are nevertheless still unstable to modes with arbitrarily high wavenumbers. The onset of these instabilities corresponds to a zero mode, which happens to also be a zero mode on the geon.  These zero modes are $\tau$-independent perturbations that will lead to new families of geons and black resonators with Maxwell ``hair''.

Like what was observed in the scalar field, there is a small range of parameters where the geon is unstable. Notice that in Fig.~\ref{maxwellgeon} near $J\sim 10.5$ for the $j=2$ mode, there is a small window. In this region, the mode is purely imaginary, and becomes the mode shown in Fig.~\ref{maxwellgeonunstable}.  This instability likely extends to nearby black resonators as well.

The edges of this instability just below $J=10$ and just above $J=10.6$ are zero modes for the geon.  From the inset of Fig.~\ref{maxwell_J_E}, we see that these zero modes are precisely where the $j=2$ onsets curves for even and odd party perturbations of black resonator intersect the geon.

\subsection{Gravitational Perturbations}

\begin{figure}
  \centering
\subfigure[Onset curves on black resonators]
 {\includegraphics[scale=0.45]{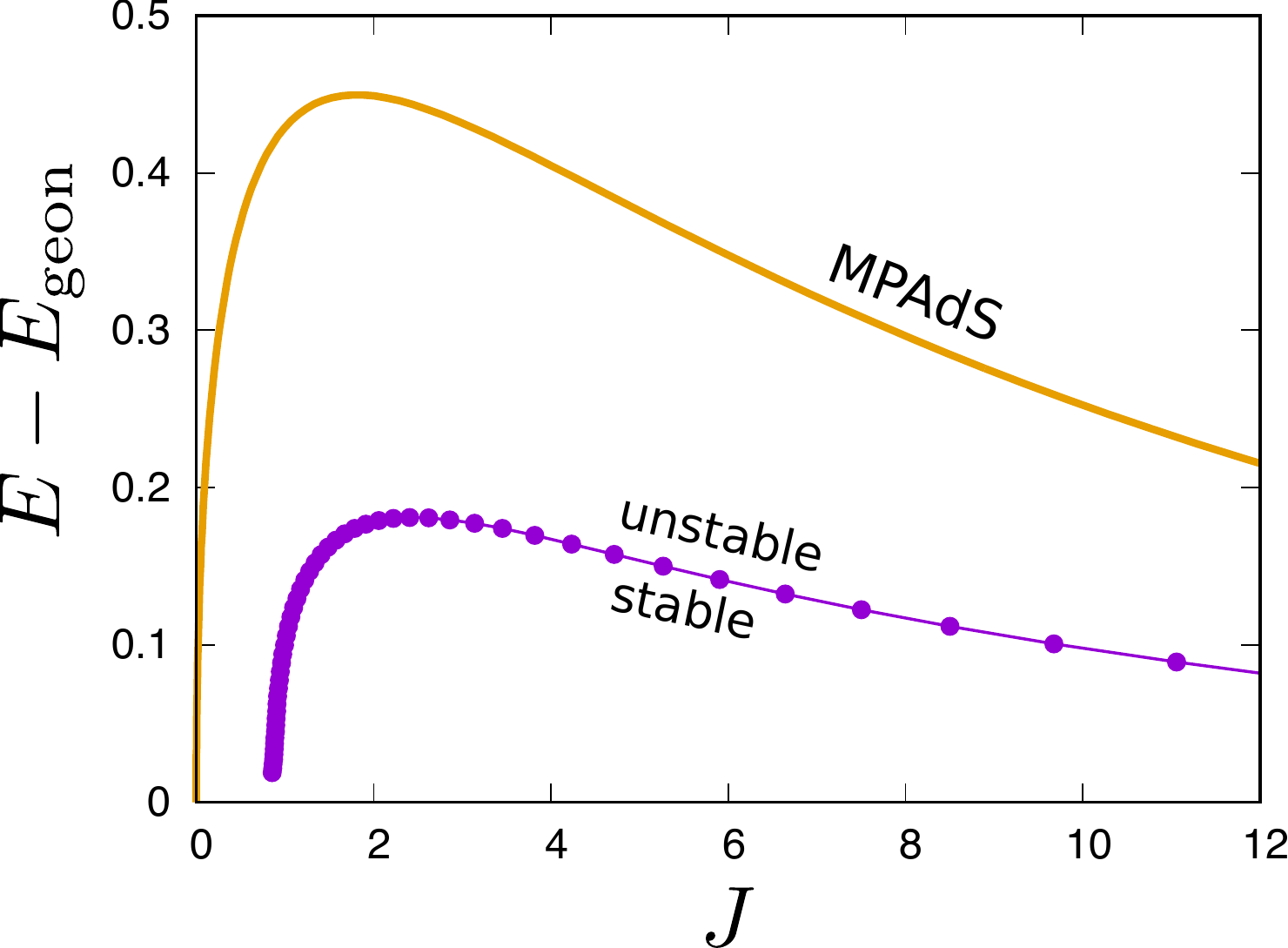}\label{J_E_grav_2j_1}
}
\subfigure[Perturbation frequency on geons]
 {\includegraphics[scale=0.4]{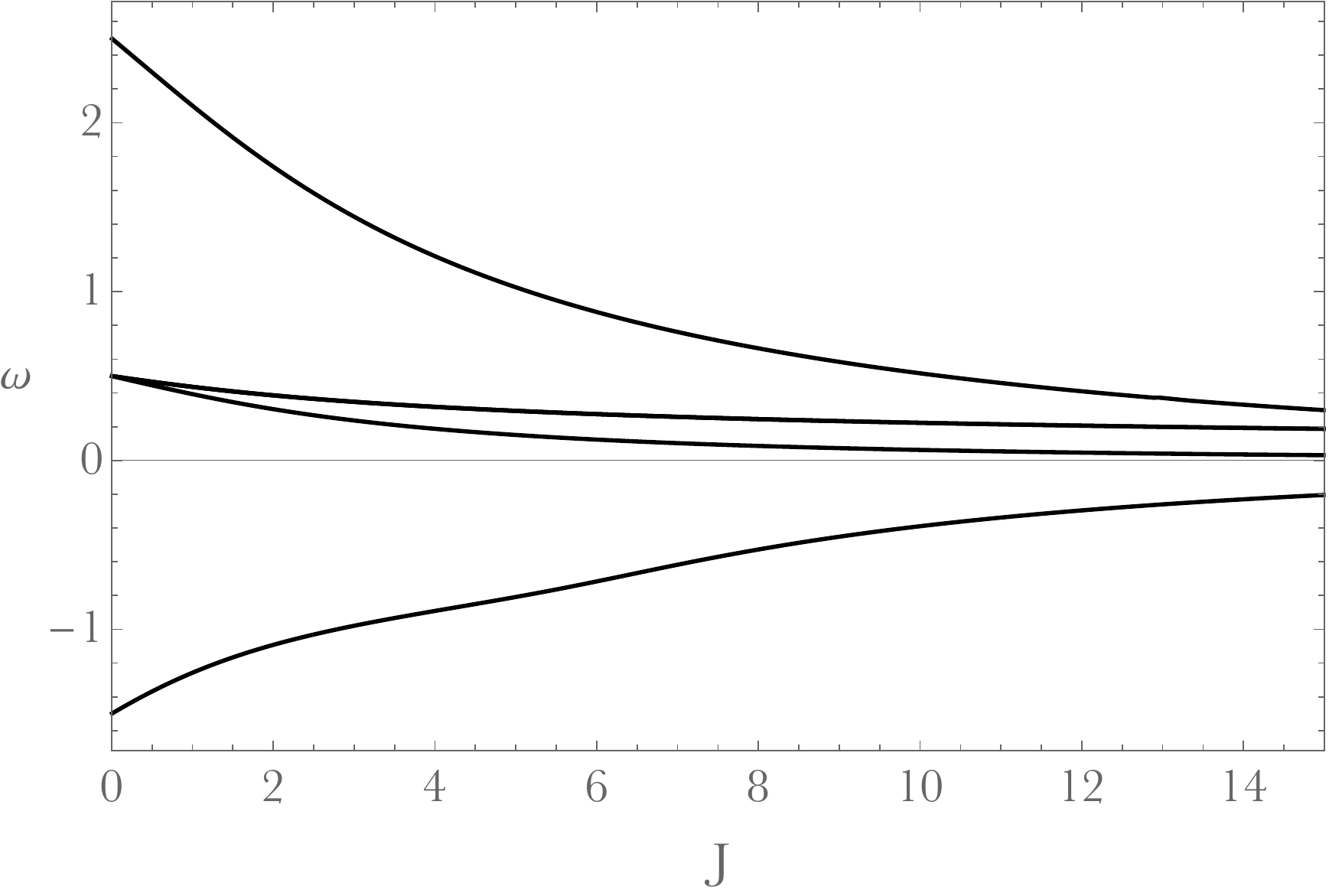}\label{gravgeonj12}
}
 \caption{Gravitational perturbations for $j=1/2$.
}
\label{gravj12}
\end{figure}

We now give results for gravitational perturbations, beginning with the $j=1/2$ modes, which we show in Fig.~\ref{gravj12}.
In Fig.~\ref{J_E_grav_2j_1}, we show the onset curve for the black resonator. The black resonators are unstable above the purple curve. Note that this onset curve does not intersect the branching points because MPAdS is already unstable to $j=1/2$ perturbations.

\begin{figure}
\begin{center}
\includegraphics[scale=0.5]{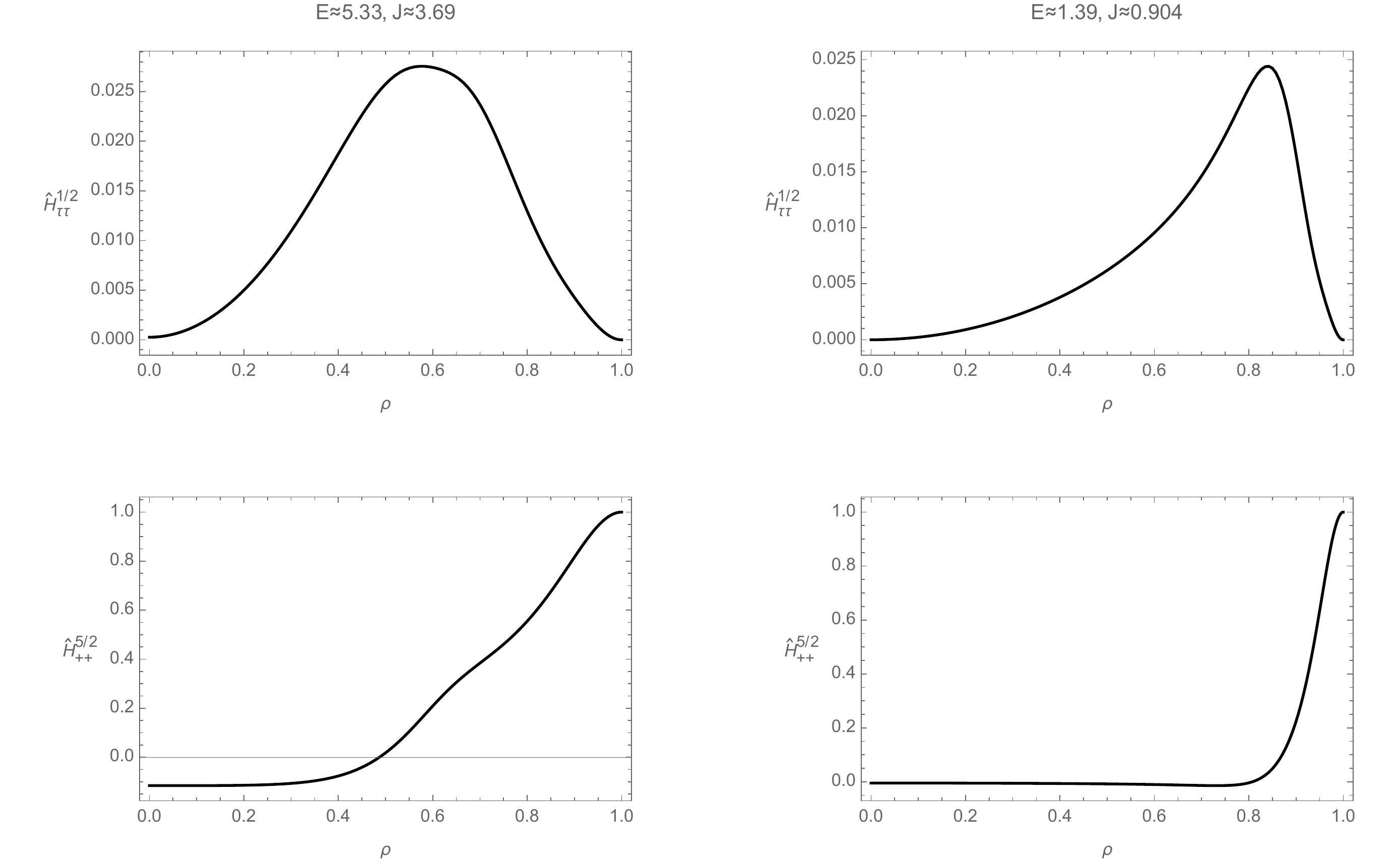}
\end{center}
\caption{Two eigenfunctions (precise definitions can be found in Appendix \ref{subsec:computeomegagrav}) for gravitational perturbations of black resonators near the onset $\omega\approx0$.  The coordinate $\rho$ is defined so that $\rho=0$ is the horizon and $\rho=1$ is the AdS boundary. Left: $E\approx5.33$, $J\approx3.69$ (farther from geon).  Right: $E\approx1.38$, $J\approx0.904$ (closer to geon).  We see that the eigenfunctions for black resonators near the geons have sharper gradients near the AdS boundary.
}
\label{fig:eifcns}
\end{figure}

In Fig.~\ref{gravgeonj12}, we plot the perturbative frequency on the geons.  Note that there are no zero modes or instabilities on the geons, even though the onset curve in Fig.~\ref{J_E_grav_2j_1} appears to intersect the geon. To explain this behaviour, in Fig.~\ref{fig:eifcns} we show two eigenfunctions for two different black resonator backgrounds.  (These eigenfunctions are defined more precisely in Appendix \ref{subsec:computeomegagrav}.) Both black resonators are close to the onset curve, but one of them is farther from the geon, and the other is closer.  We see that the eigenfunctions for near-geon black resonators have sharper gradients near the AdS boundary.  This suggests that this onset mode becomes singular in the geon limit, which would explain why zero modes can exist for black resonators but not geons.

\begin{figure}
  \centering
\subfigure[Onset curves on black resonators]
 {\includegraphics[scale=0.45]{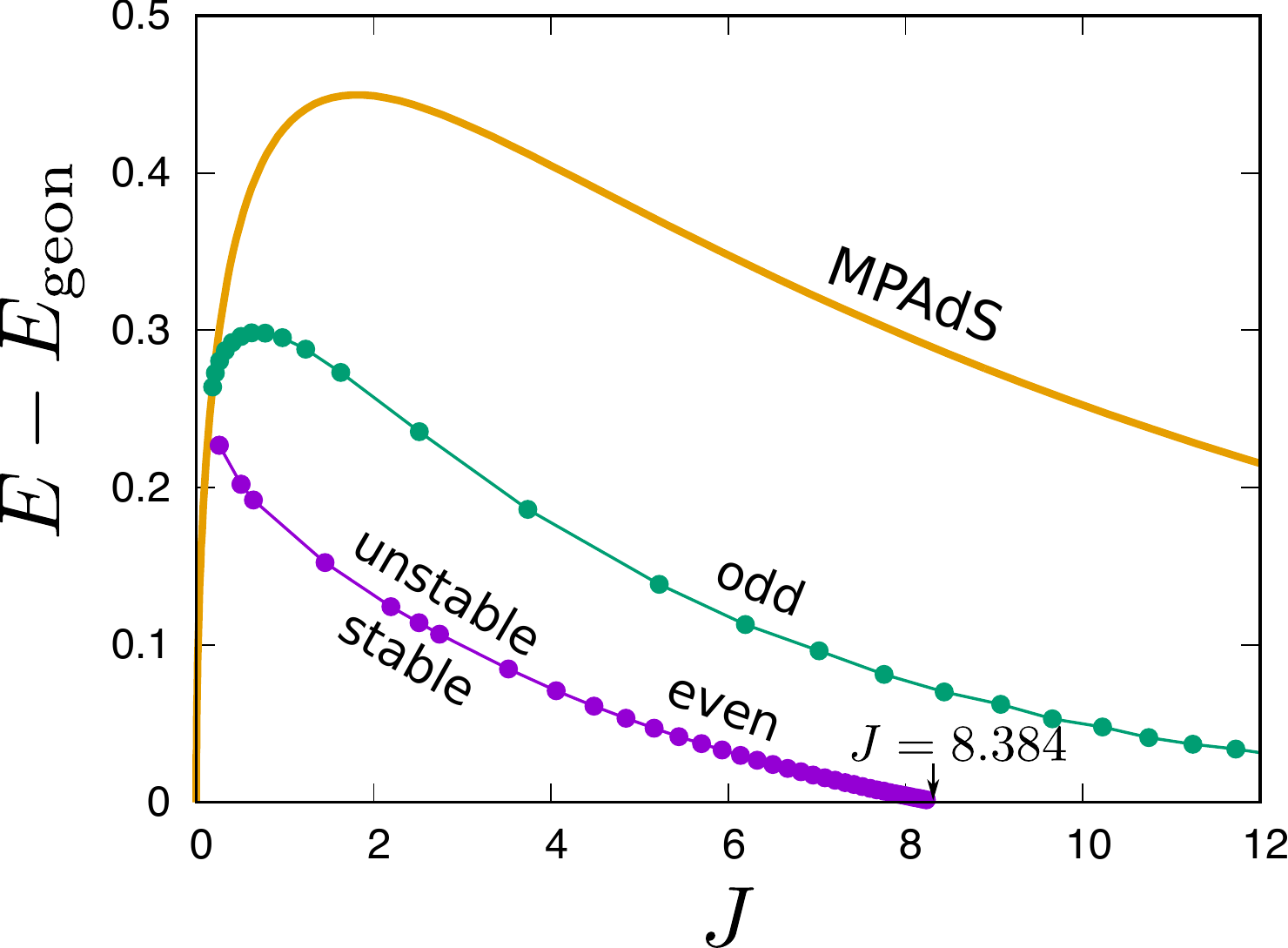}\label{J_E_grav_2j_2}
}
\subfigure[Perturbation frequency on geons.]
 {\includegraphics[scale=0.4]{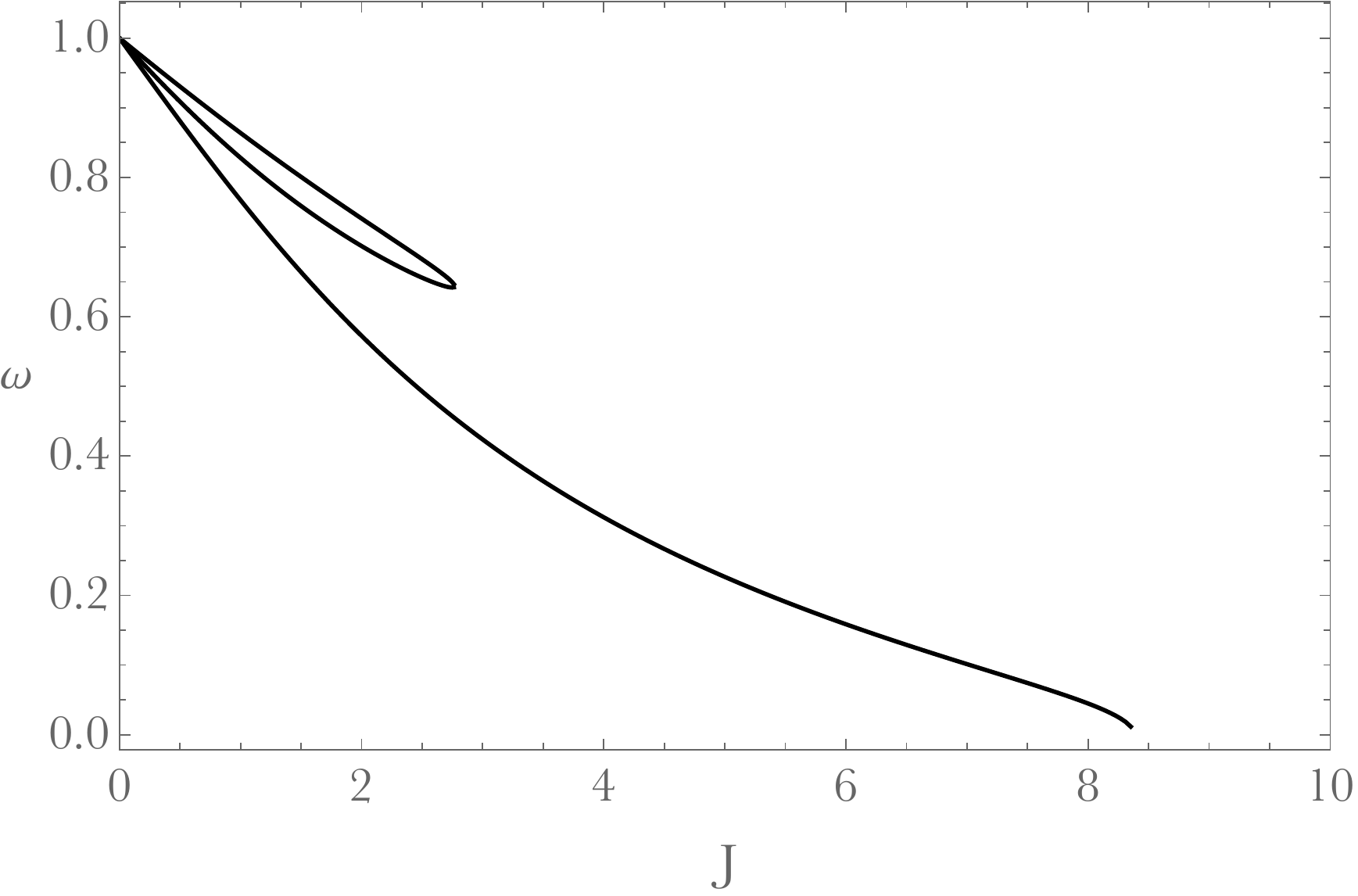}\label{gravgeonj1}
}
\subfigure[Growth rate for instability on geons.]
 {\includegraphics[scale=0.4]{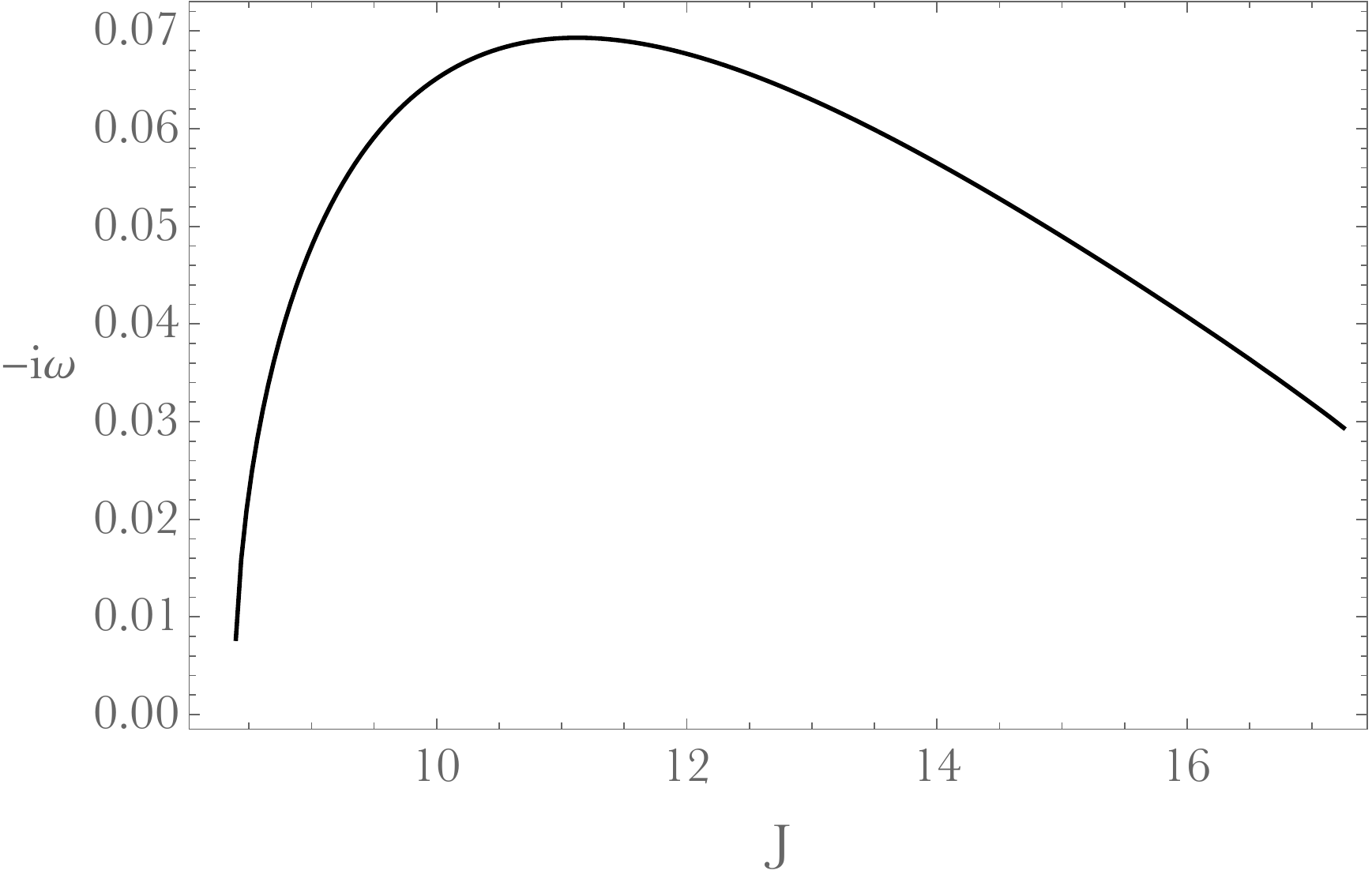}\label{gravgeonunstable}
}
 \caption{Gravitational perturbations for $j=1$.
}
\label{gravj1}
\end{figure}

Results for the $j=1$ perturbation are shown in Fig.~\ref{gravj1}.  Against the $j=1$ perturbation, MPAdS are unstable at the branching points.
The black resonators continue to be unstable in the whole region against such perturbation. (See also Figs.~\ref{lambda_grav_2j2_even} and \ref{lambda_grav_2j2_odd}.)
In addition to this mode, as shown in Fig.~\ref{J_E_grav_2j_2}, a new unstable mode appears above each of the purple and green curves.  Note that the even onset curve (in purple) intersects the geon, where in Fig.~\ref{gravgeonj1} we see that it corresponds to a zero mode of the geon.  This zero mode on the geon is the onset of an instability, the growth rate of which is shown in Fig.~\ref{gravgeonunstable}.  There is a possibility that odd onset curve (in green) also intersects the geon, where it would correspond to the location where the growth rate in Fig.~\ref{gravgeonunstable} reaches zero again. But this would occur for an angular momentum that is too high for our numerical methods to capture.  Finally, we note that the onset curve for $j=1$ perturbations intersects the $j=1/2$ perturbations though this is not shown in our figures.

\begin{figure}
\begin{center}
\includegraphics[scale=0.5]{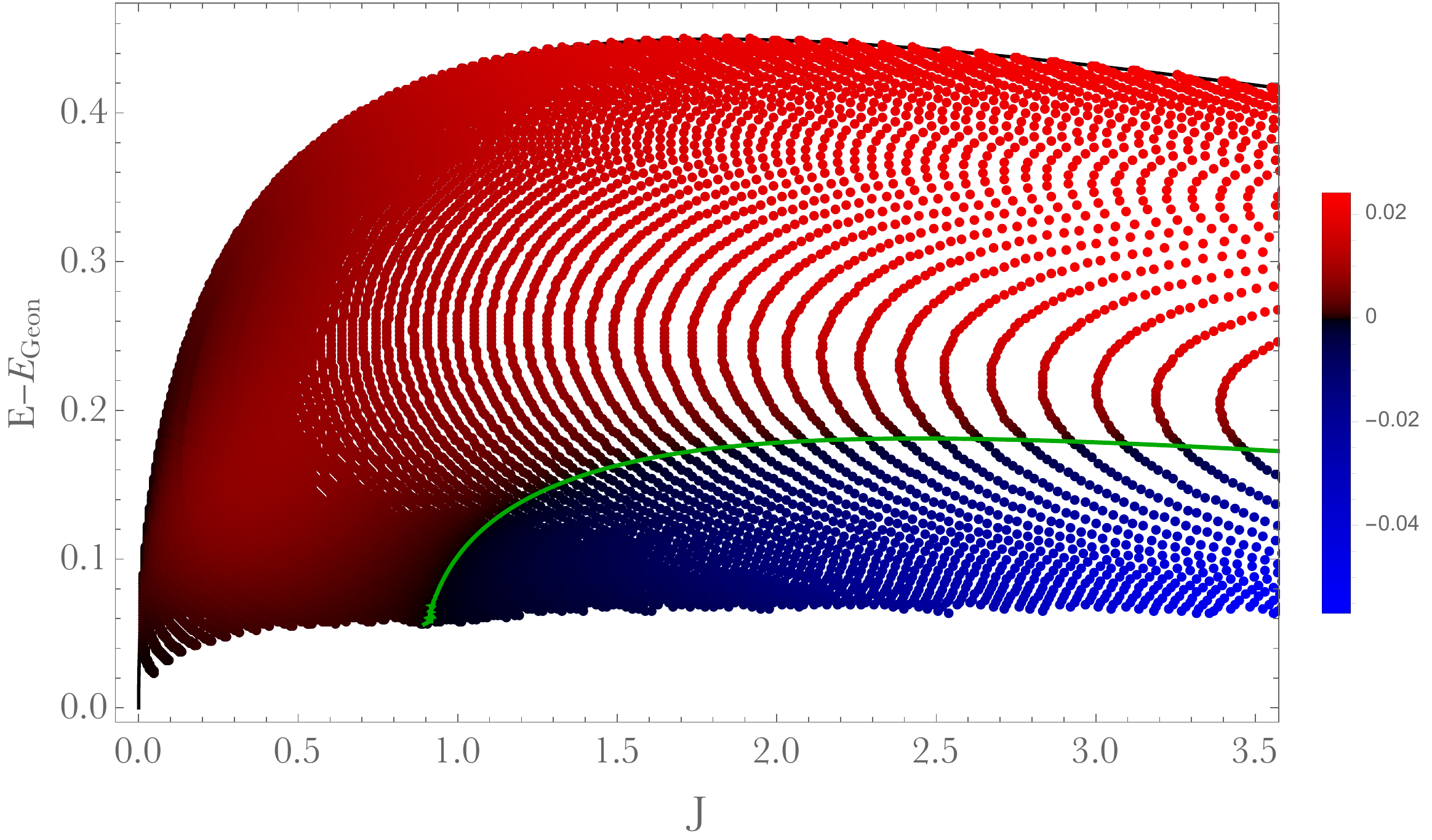}
\end{center}
\caption{Growth rate $\mathrm{Im}(\omega)$ for a $j=1/2$ gravitational perturbation of black resonators.  The green line is the onset of the instability where $\omega=0$.  The black line are the geons, which are the zero horizon size limit of black resonators.
}
\label{qnms}
\end{figure}

We expect the unstable gravitational perturbations to have a growth rate that is typically much larger than that of the scalar field and Maxwell field.  This behaviour is similar to perturbations of Kerr-AdS, where perturbing the background with higher-spin fields will yield greater growth rates in the superradiant instability \cite{Cardoso:2013pza}. This might perhaps be due to the fact that growth rates tend to decrease exponentially with increasing $j$, and gravitational perturbations tend to be unstable for smaller $j$.  Because of the larger growth rate, it is numerically feasible to compute $\omega$ directly for black resonators.  The growth rates $\mathrm{Im}(\omega)$ for the $j=1/2$ modes are shown in Fig.~\ref{qnms}.  This can be compared with Fig.~\ref{J_E_grav_2j_1}.
We also did a similar analysis for $j=0$ gravitational perturbation, which preserves $SU(2)$ symmetry, but
did not find any evidence of instability.

\section{Multi-resonators and Multi-geons}
\label{multisolutions}
At the onsets of any instability of black resonators, we find $\tau$-independent perturbations. For example, the $\tau$-independent scalar field perturbation can be written as
\begin{equation}
 \Phi(\tau,r,\theta,\phi,\chi)
= \sum_{|k|\leq j} \phi_k (r) D_k(\theta,\phi,\chi)
\ .
\end{equation}
Now, let us move to the non-rotating frame at infinity $(t,\psi)$:
\begin{equation}
 dt=d\tau\ ,\quad d\psi=d\chi+2\Omega d\tau\ .
\end{equation}
The Wigner D-matrix depends on $\chi$ as
$D_k(\theta,\phi,\chi) \propto e^{-ik\chi}$. Therefore, in the non-rotating frame, the solution of the scalar field becomes
\begin{equation}
 \Phi(t,r,\theta,\phi,\psi)
= \sum_{|k|\leq j} e^{2ik\Omega t}\phi_k (r) D_k(\theta,\phi,\psi)\ .
\end{equation}
This perturbation has eigenfrequencies
\begin{equation}
 \omega=2j\Omega, \, 2(j-1)\Omega,\cdots,\,-2j\Omega\ ,
\end{equation}
in addition to the background frequency $\omega=\pm 4\Omega$.

These perturbations will therefore generate a new solution that has deformations with multiple angular frequencies, yet preserves a helical Killing field.  These solutions can be described as either black resonators (or geons) with scalar or Maxwell hair, or as a ``double resonator'' (or ``double geon'') for gravitational deformations.  
For scalar and Maxwell perturbations, their energy momentum tensors are invariant under the change of the sign of the $\tau$-independent perturbations. We therefore only expect a single branch of solutions to emerge from these perturbations.
For a general $\tau$-independent gravitational perturbation, 
changing the sign of the perturbation leads to a physically distinct result.

But hairy black resonators and double resonators likely have their own instabilities, some of which would have onset curves.  These will lead to $\tau$-independent configurations with angular deformations from multiple different modes.  These may be described as multi-haired black resonators/geons, or perhaps as ``multi-resonators'' and ``multi-geons.''  The phase diagram of these solutions is therefore quite complicated.

In the scalar and Maxwell cases, the onset mode of the black resonators intersects a zero mode of the geon.  It is therefore natural to expect that hairy black resonators are connected to hairy geons, much like black resonators are connected to geons.  The situation is less clear for the $j=1/2$ gravitational perturbation given that there are zero modes to generate multi-resonators, but no zero modes to generate ``multi-geons.''  The $j=1$ gravitational perturbation, however, has a zero mode on the geon which will lead to a multi-geon that is presumably connected to some multi-resonators.

Because rotational superradiant instabilities are characterised by an infinite number of unstable modes, it is plausible that multi-resonators will remain unstable. Nevertheless, they are stable against several modes that are unstable in MPAdS, and these solutions may be long-lived.

Constructing these new solutions is numerically challenging due to the lack of symmetries.  Nevertheless, there is a possibility that some of these solutions could be constructed perturbatively about AdS.  We leave such investigations for future work.

\section{Oscillating Geons}
\label{sec:oscillgeons}
Because the spectrum of geons typically consists of normal modes, these perturbations generate nonlinear oscillatory solutions, such as those found in \cite{Bizon:2011gg} for gravitating scalars.  More complicated multi-oscillating solutions can be generated out of several combinations of normal modes \cite{Choptuik:2018ptp,Choptuik:2019zji,Masachs:2019znp}.  These solutions are highly asymmetric, as the helical Killing symmetry is broken in addition to many spatial symmetries.

Nevertheless, a small modification of our ansatz \eqref{metricanz} allows similar solutions to be constructed that are cohomogeneity two.  However, these oscillating geons will not have angular momentum, and are generated from perturbations about AdS rather than another geon.

Consider the metric \emph{ansatz}
\begin{multline}
\mathrm{d}s^2 = \frac{1}{y}\Bigg\{-\frac{f(\tau,y)e^{2\delta(\tau,y)}}{\Omega^2}\mathrm{d}\tau^2+\frac{\mathrm{d}y^2}{4(1-y)y f(\tau,y)}+
\\
\frac{(1-y)}{4}\left[B(\tau,y)^2 \sigma_3^2+\frac{1}{B(\tau,y)}(\sigma_1^2+\sigma_2^2)\right]\Bigg\}\,,
\label{eq:geon2}
\end{multline}
where $\Omega$ is a real parameter whose physical significance we detail below and $\{f,\delta,B\}$ are functions of $\tau$ and $y$ to be determined by imposing the Einstein equation.

To understand our choice of \emph{ansatz}, we initially set $f=B=\delta+1=1$, which brings the line element (\ref{eq:geon2}) to
\begin{subequations}
\begin{equation}
\mathrm{d}s^2 = \frac{1}{y}\Big[-\frac{\mathrm{d}\tau^2}{\Omega^2}+\frac{\mathrm{d}y^2}{4(1-y)y}+\frac{(1-y)}{4}\left(\sigma_3^2+\sigma_1^2+\sigma_2^2\right)\Big].
\label{eq:ADS2}
\end{equation}

Upon the coordinate transformation
\begin{equation}
\tau = \Omega\,t\qquad\text{and}\qquad r =\frac{\sqrt{1-y}}{\sqrt{y}}\,,
\end{equation}
\end{subequations}
one recovers the metric of global AdS. From the above considerations, it is clear that we want to consider $y\in[0,1]$, with $y=0$ marking the location of the conformal boundary and $y=1$ the AdS centre. Finally, we are interested in periodic solutions in time with frequency $\Omega$, and thus we take $\tau\in[0,2\pi]$. In fact, we can take advantage of an additional discrete symmetry to reduce the domain to $\tau\in[0,\pi]$ without loss of generality.

The ansatz can be placed in the Einstein equation and solved with the conditions of regularity at the origin and that the metric is asymptotically global AdS. The system has one more equation than unknown functions, which we treat as a constraint equation to be verified after finding a solution.  Further details can be found in Appendix \ref{sec:techoscillgeons}.

To find our solutions we employed spectral collocation methods, with a uniform cosine-type grid along the $\tau$ direction and a Chebyshev grid along the holographic direction $y$. Our findings are consistent with exponential convergence, which seems to be supported by the fact that we found no non-analytic behaviour at any of the edges of the integration domain with the gauge we used.

With a solution, we can determine the holographic stress energy tensor using \cite{deHaro:2000vlm}
\begin{equation}
\langle T_{\mu\nu}\rangle \mathrm{d}x^\mu\mathrm{d}x^\nu=\frac{1}{16 \pi}\left.\Bigg\{-3 q_2 \mathrm{d}t^2+\frac{1}{4} \left[\left(8 q_1-q_2\right) \sigma _3^2-\left(4 q_1+q_2\right) \left(\sigma _1^2+\sigma _2^2\right)\right]\Bigg\}\right|_{y=0}\,,
\end{equation}
from which one can read the total energy of the system $E$.\footnote{We will always measure the energy with respect to pure AdS, \emph{i.e.} we neglect the Casimir energy of global AdS$_5$.}

Besides solving for our solutions numerically, we can also use perturbation theory to construct these objects. This has been extensively used in the literature \cite{Bizon:2011gg,Dias:2011ss,Horowitz:2014hja,Dias:2016ewl,Martinon:2017uyo,Dias:2017tjg} in similar contexts and provides a good check of our numerical procedures. We expand all our functions about AdS in power series in a small parameter $\varepsilon$, with the $\tau$ coordinate expanded as a Fourier series.  Solving order by order, we obtain expressions for the angular frequency and energy, perturbatively about AdS:
\begin{equation}
E=\frac{9 \pi  \varepsilon ^2}{160}\;,\qquad\Omega=6-\frac{11057 \varepsilon ^2}{90090}\;.
\label{eq:perturb}
\end{equation}
The details of this perturbative calculation can be found in Appendix \ref{sec:techoscillgeons}.

\begin{figure}
\begin{center}
\includegraphics[width=0.4\textwidth]{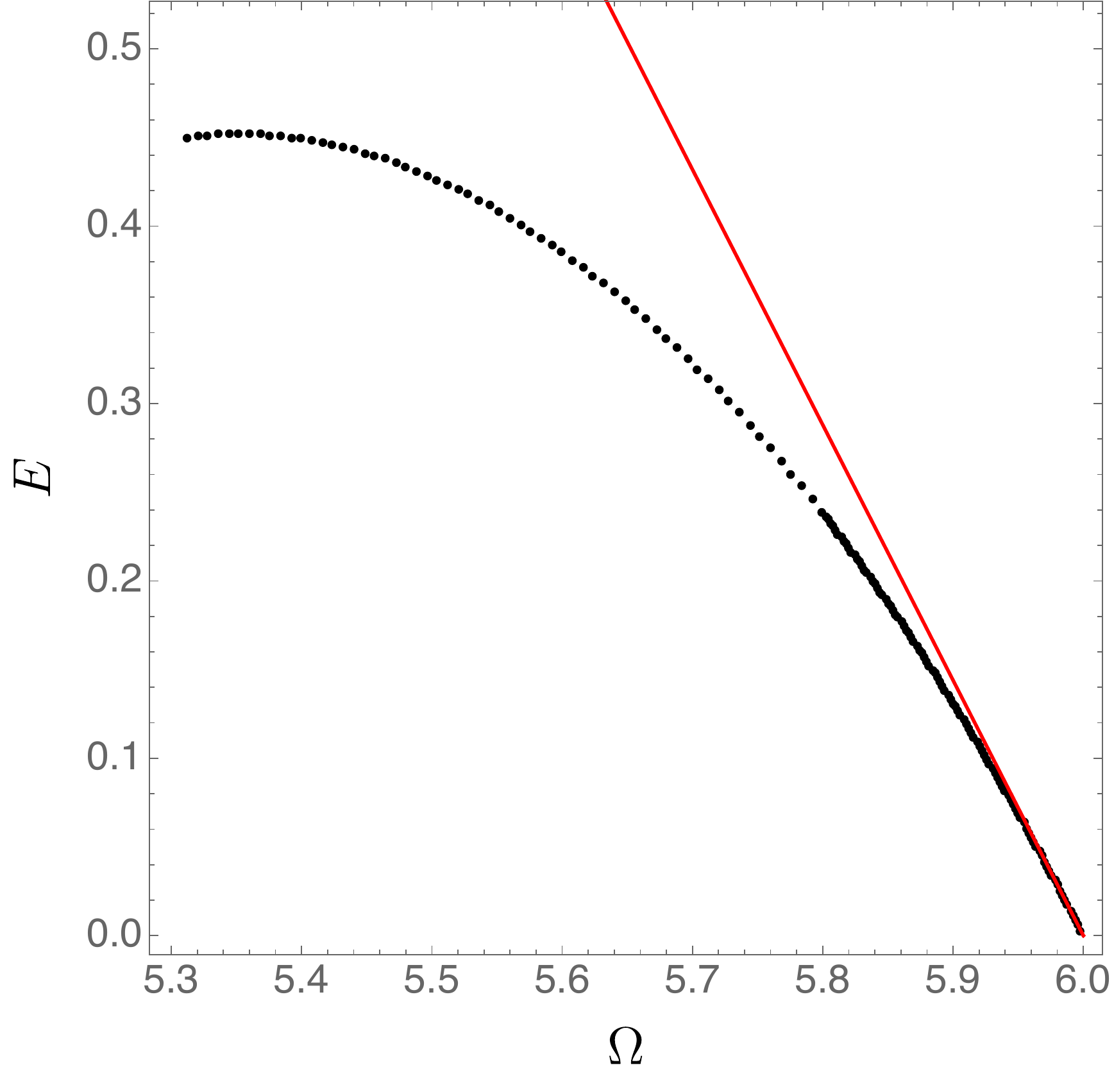}
\end{center}
\caption{Energy $E$ of the squashed geons as a function of $\Omega$ for the fundamental mode. For $\Omega\lesssim6$ the agreement between the exact numerical data (black disks) and the analytic perturbative approach (solid red line) are in good agreement.}
\label{fig:geons5D}
\end{figure}

\begin{figure}
\begin{center}
\includegraphics[width=0.4\textwidth]{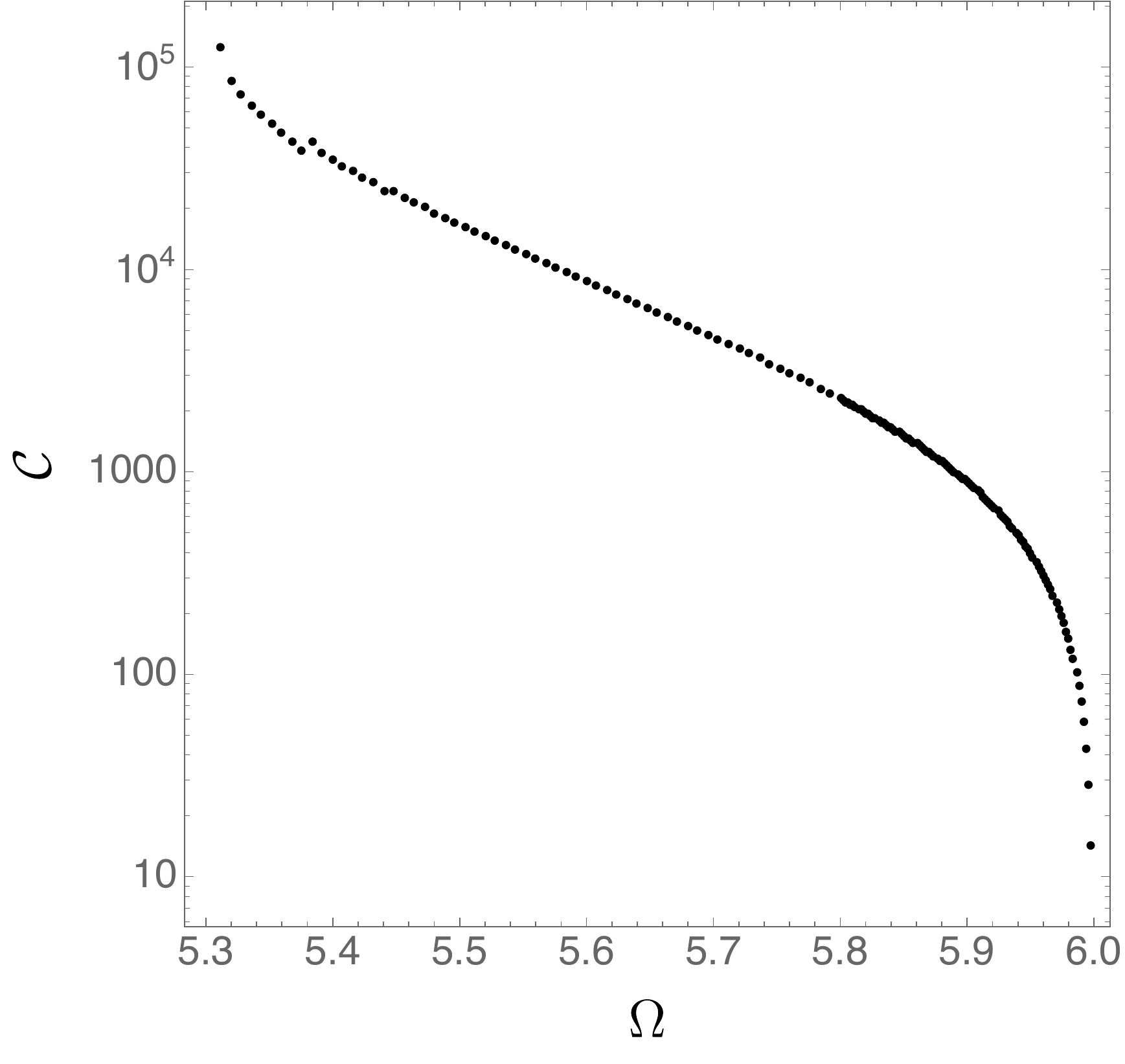}
\end{center}
\caption{$\mathcal{C}$ as a function of $\Omega$ in a log-scale. This plot shows numerical evidence that $\mathcal{C}$ is blowing up as we take the limit $\Omega\to {\Omega_{\max}}^+$.}
\label{fig:C}
\end{figure}

We are now ready to present our results and compare them with our perturbative calculations. In Fig.~\ref{fig:geons5D} we show the energy as a function of $\Omega$, with the black disks corresponding to our numerical data and the red solid line to the analytic perturbative results (\ref{eq:perturb}).
For $\Omega \lesssim 6$, we find good agreement with the perturbative and full nonlinear solutions. The energy seems to reach a maximum value $E_{\max}\simeq 0.452018$ around $\Omega=\Omega_{\max} \simeq 5.30219$. For $\Omega<\Omega_{\max}$ the energy decreases with decreasing $\Omega$. We stop finding new solutions at around $\Omega_{\star}=5.312$, and we suspect that this is due to a formation of a curvature singularity in the bulk. To inspect whether this is true we monitored
\begin{equation}
\mathcal{C}=\max_{\mathcal{M}}\;W^{abcd}W_{abcd}\,,
\end{equation}
where $W$ is the Weyl tensor. We find that the maximum of $W^{abcd}W_{abcd}$ is always located at $y=1$ and $\tau=\pi$. In Fig.~\ref{fig:C} we plot $\mathcal{C}$ as a function of $\Omega$ in a log-scale. We find evidence that $\mathcal{C}$ is blowing up as we take the limit $\Omega\to {\Omega_{\max}}^+$.

\begin{figure}
\begin{center}
\includegraphics[width=\textwidth]{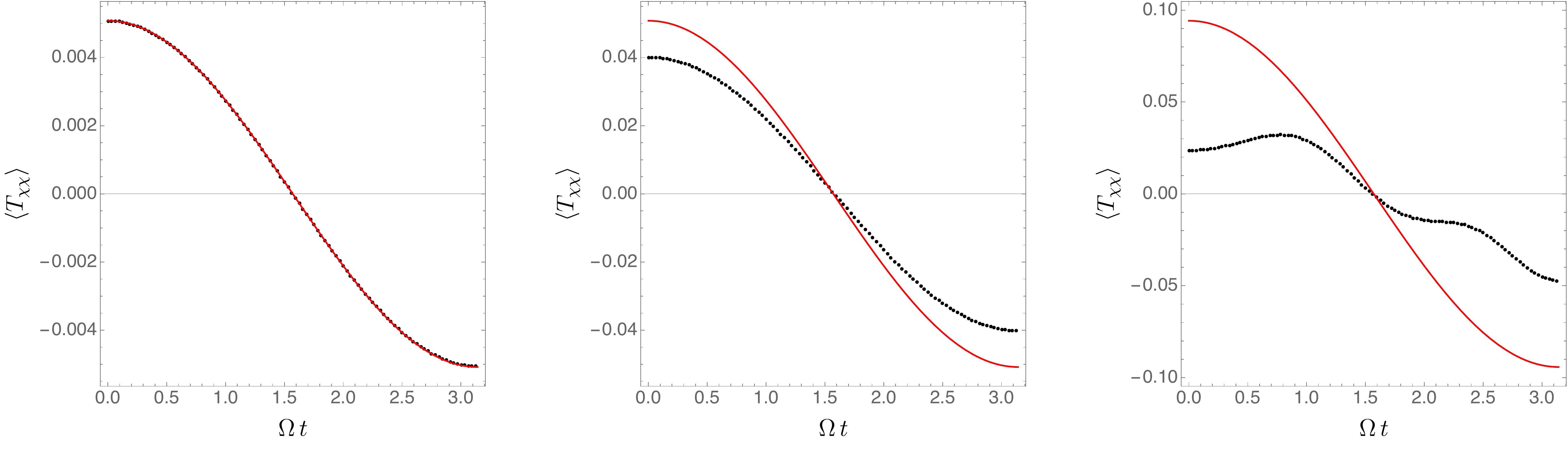}
\end{center}
\caption{$\langle T_{\chi \chi}\rangle$ as a function of $\tau \equiv \Omega\,t$ computed for several values of $\Omega$. From left to right, we have $\Omega = 5.998, 5.8, 5.312$. The black disks correspond to our numerical data and the solid red lines to our perturbative analytic approximation. For $\Omega \lesssim 6$ we find excellent agreement between the two methods to extract $\langle T_{\chi \chi}\rangle$.}
\label{fig:stress}
\end{figure}

We have also extracted $\langle T_{\chi \chi}\rangle$ for several values of $\Omega$, which we depict in Fig.~\ref{fig:stress}. From left to right, we have $\Omega = 5.998, 5.8, 5.312$. The black disks represent the exact numerical results and the solid red line the analytic perturbative results. Like it was for the energy, we find agreement between the two methods when $\Omega \lesssim 6$.

Finally let us comment on the stability of oscillating geons.  In many ways, these geons resemble boson stars and oscillaton stars \cite{Bizon:2011gg,Dias:2012tq,Maliborski:2013jca,Buchel:2013uba,Balasubramanian:2014cja,Bizon:2014bya,Balasubramanian:2015uua,Dimitrakopoulos:2015pwa}, in that they are nonlinear extensions of normal modes about AdS. There is considerable evidence for the nonlinear stability of such solutions \cite{Buchel:2013uba,Balasubramanian:2014cja,Bizon:2014bya,Balasubramanian:2015uua,Dimitrakopoulos:2015pwa}, and there are arguments that this stability should apply to geons as well.  Also like boson stars and oscillaton stars, the oscillating geons appear to have a maximum energy.  We expect that solutions past this maximum energy (i.e.~for $\Omega\lesssim 5.30219$) are unstable, as is common for such turning points \cite{poincare1885,Sorkin:1981jc,Sorkin:1982ut,Arcioni:2004ww}.

\section{Summary and Discussion}
\label{sec:discussion}
We presented a comprehensive study of the linear mode stability of certain five-dimensional black resonators and geons with equal angular momenta.  We find that black resonators are superradiantly unstable throughout moduli space to scalar, electromagnetic and gravitational perturbations.  Our findings are in agreement with those reported in \cite{Green:2015kur}.

The results pertaining to the geons are more surprising. These horizonless solutions have an ergoregion, so one might be tempted to think that they are unstable to the so-called ergoregion instability first studied by Friedman in \cite{friedman1978}, proved in \cite{Moschidis:2016zjy}  for scalar waves and reviewed in \cite{Brito:2015oca}. However, Friedman's instability only applies to asymptotically flat boundary conditions, which is not our case. Indeed, Friedman finds initial data with negative energy and then uses the fact that the Bondi mass is a decreasing function on null infinity to argue that the energy will continue to grow more negative. In AdS, with reflecting boundary conditions, the energy is conserved at infinity and thus the instability is absent. Black holes also do not have this instability because they absorb the negative energy states in the ergoregion.

In \cite{Dias:2011ss,Horowitz:2014hja}, a linear stability argument for geons was presented based on eigenvalue perturbation theory. Geons are nonlinear realizations of normal modes of AdS, and in particular, when their energy is small, it is natural that their perturbative spectrum is similar to that of AdS. Since AdS does not have any mode that is on the verge of becoming unstable, we expect the geons to be stable. However, this argument is too quick, since we cannot prove that the spectrum of linear perturbations around a geon yields a selfadjoint problem in $\omega^2$, and thus we cannot rule out the existence of an instability. Remarkably, after studying scalar and gravitational perturbations we still find that the spectrum remains real, and in fact many are linearly stable, even moderately far away from pure AdS.  Continuity implies that the growth rate of black resonators should approach zero in the small horizon limit.

However, for scalar and electromagnetic perturbations, we have found a small window of parameters where geons are unstable.  Our results also indicate that such a window is plausible for gravitational perturbations as well.  This window is bounded by the zero modes of integer $j$ perturbations with odd and even parity.

The fact that these instabilities exist for scalar, Maxwell, and gravitational perturbations implies that there is some general mechanism that drives this instability. Though at this stage, it is unclear what physical mechanism is responsible.  Such a mechanism would have to be sensitive to some symmetries of the perturbations, as the instability only seems to exist for integer $j$ modes, and not half-integer modes.

It would be natural for the instability of these geons to lead to black hole formation.  There are black resonators with the same energy and angular momentum as these geons. These are the less symmetric black resonators that are bounded in phase space by higher-mode geons and the onset of higher superradiant modes.  In the scalar and Maxwell case, there may also be hairy black holes that share the same energy and angular momentum.  All of these black hole solutions can serve as intermediate states in the evolution of these unstable geons, though the ultimate endstate is still unknown as these black holes are still unstable to superradiance.

We have also commented on the existence of black resonators with scalar or electromagnetic hair that branch from the onset curves, and also on the existence of multi-black resonators generated by the gravitational zero modes.  It is natural to expect that under time evolution, some of these instabilities will evolve towards the hairy solutions, at least as intermediate states until other instabilities grow.  We leave the construction of these hairy and multi solutions, as well as a comparison of their entropies, to future work.

For the scalar, electromagnetic, and integer $j$ gravitational perturbations, the geons also have zero modes that coincide with the zero-horizon limit of the zero modes on black resonators.  It is therefore likely that hairy and multi-geons also exist, and that they may be the zero horizon limit of hairy or multi black resonators.  However, for the $j=1/2$ (and presumable other half-integer) gravitational perturbations, the zero modes do not have a smooth zero-horizon limit.  This leads to the possible existence of black multi-resonators that likewise do not have a smooth zero-size limit. The non-smooth limit is curious in that our perturbative results suggest the singularity appears near the boundary, rather than near the origin.  This suggests that the zero-size limit is not a singularity that is covered by a horizon.

In addition to these hairy and multi-geons, the normal modes of geons suggest that oscillating geons also exist.  Though we have not constructed any such solutions branching from geons, we have constructed an oscillating geon that branches from a normal mode of AdS. By analogy with what is understood for gravitating scalars, we expect small-energy oscillating geons to be stable.  In the context of gauge/gravity duality, the existence of these oscillating geons is mysterious as they represent states that are very long lived and do not appear to thermalise.

Given that our calculations are in AdS$_5$, there are naturally implications for the famous duality between supergravity in AdS$_5\times S^5$ and $\mathcal{N}=4$ SYM.  But the presence of the $S^5$ space introduces other instabilities in black holes.  Schwarzschild-AdS$_5\times S^5$ black holes are linearly mode stable above energies $E\gtrsim0.27$\footnote{Recall we are setting the AdS radius to unity as well as the five dimensional Newton's constant.}, but unstable otherwise. A putative endpoint of this instability was found in \cite{Dias:2016eto} and can be described as a ten-dimensional spherical black hole localised on one of the poles of the $S^5$. To contrast, the Hawking-Page transition \cite{Hawking:1982dh} occurs for $E\approx 2.4$. The oscillating geons of Section~\ref{sec:oscillgeons} will compete with both the localised and unlocalised black holes with AdS$_5\times S^5$ asymptotics, since they exist for $E\lesssim0.45$. Note that since these solutions have no horizon, and thus no entropy to leading order in $N$, they will always be subdominant with respect to any black hole solution at fixed $E$. Thus, at any finite value of $N$, the new oscillating geons will tunnel to the aforementioned black hole solutions, which in turn will Hawking evaporate leaving behind a gas of gravitons. Geons and their oscillating cousins therefore do not thermalise except via tunneling which is a process suppressed by $1/N^2$. From the field theory perspective, it is far from clear how such a state can be realised.

\acknowledgments
The authors would like to thank support from Osaka City University Advanced Mathematical Institute (MEXT Joint Usage/Research Center on Mathematics and Theoretical Physics) and Nambu Yoichiro Institute of Theoretical and Experimental Physics (NITEP), and especially thank the NITEP one year anniversary conference/workshop series:~``Turbulence of all kinds'', where many fruitful discussions were had that were indispensable for completing this work. 
The work of T.~I.~was supported in part by JSPS KAKENHI Grant Number JP18H01214 and JP19K03871.
The work of K.~M.~was supported in part by JSPS KAKENHI Grant Number JP18H01214 and JP20K03976.
BW acknowledges support from ERC Advanced Grant GravBHs-692951 and MEC grant FPA2016-76005-C2-2-P. J.~E.~S. is supported in part by STFC grants PHY-1504541 and ST/P000681/1. J.~E.~S. also acknowledges support from a J. Robert Oppenheimer Visiting Professorship.

\appendix
\section{Technical Details for Scalar Field Perturbations}
\label{sec:techscalar}

\subsection{Equations of motion}
\label{subsec:scalareom}
Recall that we decompose the scalar field by using the Wigner D-matrices as \eqref{Phi_expand}, which we reproduce here:
\begin{equation}
 \Phi(\tau,r,\theta,\phi,\chi)
= e^{-i\omega \tau}\sum_{|k|\leq j} \phi_k (r) D_k(\theta,\phi,\chi)
\ .
\label{Phi_expand2}
\end{equation}

The equation of motion for the scalar field perturbations is in the form \eqref{phieq}, which we reproduce below:
\begin{equation}
L_k \phi_k+c_{k-1} \phi_{k-2} +c_{k+1} \phi_{k+2} =0\ .
\label{phieq2}
\end{equation}
We now give the operator and coefficients in full as
\begin{multline}
L_k=(1+r^2)g \frac{d^2}{dr^2}
 +\left[\frac{1+r^2}{2}\left(
\frac{f'}{f}+\frac{g'}{g}+\frac{\beta'}{\beta}
\right)+\frac{3+5r^2}{r}\right]g\frac{d}{dr}\\
-\frac{\epsilon_k^2+\epsilon_{k+1}^2}{r^2}\left(\alpha+\frac{1}{\alpha}\right)-\frac{4k^2}{r^2\beta}
+\frac{(\omega-2kh)^2}{(1+r^2)f}-\lambda\ ,
\end{multline}
and
\begin{equation}
c_k=-\frac{\epsilon_{k}\epsilon_{k+1}}{r^2}\left(\alpha-\frac{1}{\alpha}\right)\ .
\end{equation}
\subsection{Finding the onset of an instability}
\label{subsec:findonset}
As we will see shortly, an analysis of the equation \eqref{phieq} shows that only scalar field perturbations with $j\geq 9/2$ can be unstable. The growth rate of an unstable perturbation $\mathrm{Im}\,\omega>0$ is exponentially suppressed by the quantum number $j$~\cite{Kunduri:2006qa}, and for $j\geq 9/2$ it would be extremely small (smaller than the 64-bit machine epsilon which is about $10^{-16}$).\footnote{For example, for the MPAdS with $\Omega=1.5$ and $r_h=0.1$, we obtained a growth rate $\mathrm{Im}(\omega) \sim 10^{-16}$ for $j=3$. For $j\geq 9/2$, which are modes we are interested in, the growth rate is much smaller.} Therefore, direct computation of the spectrum $\omega$ would be difficult especially for a large $j$. Instead, we just focus on locating the onset of the instability.

By definition, the onset of a linear instability occurs when $\textrm{Im}\,\omega=0$.  We will now demonstrate that at an onset, we also have $\textrm{Re}\,\omega=0$, which will allow us to set $\omega=0$ and compute the eigenvalue of the Klein-Gordon operator $\lambda$.  A change in the sign of $\lambda$ will then imply a change in stability.

Let us consider the conserved current $J^\mu$:
\begin{equation}
J^\mu = \sqrt{-\textrm{det}\, g_{\mu\nu}} \left(\Phi^\ast \nabla^\mu \Phi - \Phi \nabla^\mu \Phi^\ast \right)\ ,
\end{equation}
which satisfies $\partial_\mu J^\mu=0$ by the Klein-Gordon equation. We consider the scalar field perturbation~\eqref{Phi_expand2} at the onset of instability, i.e.~$\textrm{Im}\,\omega=0$. Then, the time dependent factor $e^{-i\omega\tau}$ cancels out in $J^\mu$, and we obtain $\partial_\tau J^\mu=0$.
Using this and integrating the conservation law $\partial_\mu J^\mu=0$ with respect to the angular coordinates ($\theta,\phi$, $\chi$), we obtain the radial conservation law
\begin{equation}
 \partial_r Q = 0\ ,\quad Q\equiv \int \mathrm \mathrm{d}\theta\, \mathrm \mathrm{d}\phi\, \mathrm d\chi \,J^r\ .
\end{equation}
The explicit expression of $Q$ is given by
\begin{equation}
Q=r^3(1+r^2)\sqrt{fg\beta}\sum_{|k|\leq j}(\phi^\ast_k{}' \phi_k-\phi^\ast_k{} \phi'_k)\ .
\end{equation}
We can estimate the value of $Q$ at the infinity ($r=\infty$) and the horizon ($r=r_h$). For this, we consider (\ref{Kleineigen}). At the infinity, the quickly decaying mode behaves as
\begin{equation}
 \phi_k\sim \frac{1}{r^{2+\sqrt{4+\lambda}}}\ .
\end{equation}
This gives $Q|_{r=\infty}=0$.
Near the horizon, the background solution behaves as $f, g, h = \mathcal{O}(r-r_h)$ and $\alpha,\beta=\mathcal{O}(1)$ (see \cite{Ishii:2018oms} for details). Near the horizon, the ingoing wave solution is given by
\begin{equation}
 \phi_k \sim  (r-r_h)^{-i\omega/(2\kappa)}\ ,
\end{equation}
where
\begin{equation}
 \kappa=\frac{1}{2}(1+r^2)\sqrt{f'g'} \Big|_{r=r_h}
\end{equation}
is the surface gravity. Using these, we obtain $Q|_{r=r_h}\propto \omega$. From the radial conservation law, we conclude that $\omega=0$ at the onset of the instability.

\subsection{Scalar field superradiant instability of Myers-Perry AdS black holes}
\label{scalarMPAdS}
Since black resonators branch off from MPAdS black holes at the onset of the gravitational superradiant instability, MPAdS black holes provide good approximations to black resonators near the onset point. It would therefore be useful to compute the onset of the scalar superradiant instability in MPAdS as well. This has essentially\footnote{The calculation presented in \cite{Kunduri:2006qa} applies for any odd dimension, but no explicit results were provided for $D=5$, which is our current case of interest.} been done in \cite{Kunduri:2006qa} using a basis for $\mathrm{CP}^n$, but we repeat this calculation here using a decomposition into Wigner D-matrices.  We only need to solve the decoupled equation $L_k \phi_k=0$ because of the additional $U(1)$-symmetry.

Following the procedure we have set out in previous sections, we set $\omega=0$ and substitute \eqref{MPAdSFunctions} into $L_j \phi_j=0$. We impose a regular boundary condition at the horizon and integrate the differential equation from the horizon to the infinity. In general, the scalar field behaves as $\phi_j \sim c_1/r^{2+\sqrt{4+\lambda}} + c_2/r^{2-\sqrt{4+\lambda}}$ as $r \to \infty$, and for the solution we impose $c_2=0$, which corresponds to introducing no source of $\phi$ at the AdS boundary. To find the solution with this condition, we use the shooting method by tuning $\lambda$. We repeat this procedure for various parameters $(r_h,\Omega)$ and search the point at which the eigenvalue $\lambda$ changes its sign. Such a point marks the onset of the instability.

\begin{figure}
\begin{center}
\includegraphics[scale=0.6]{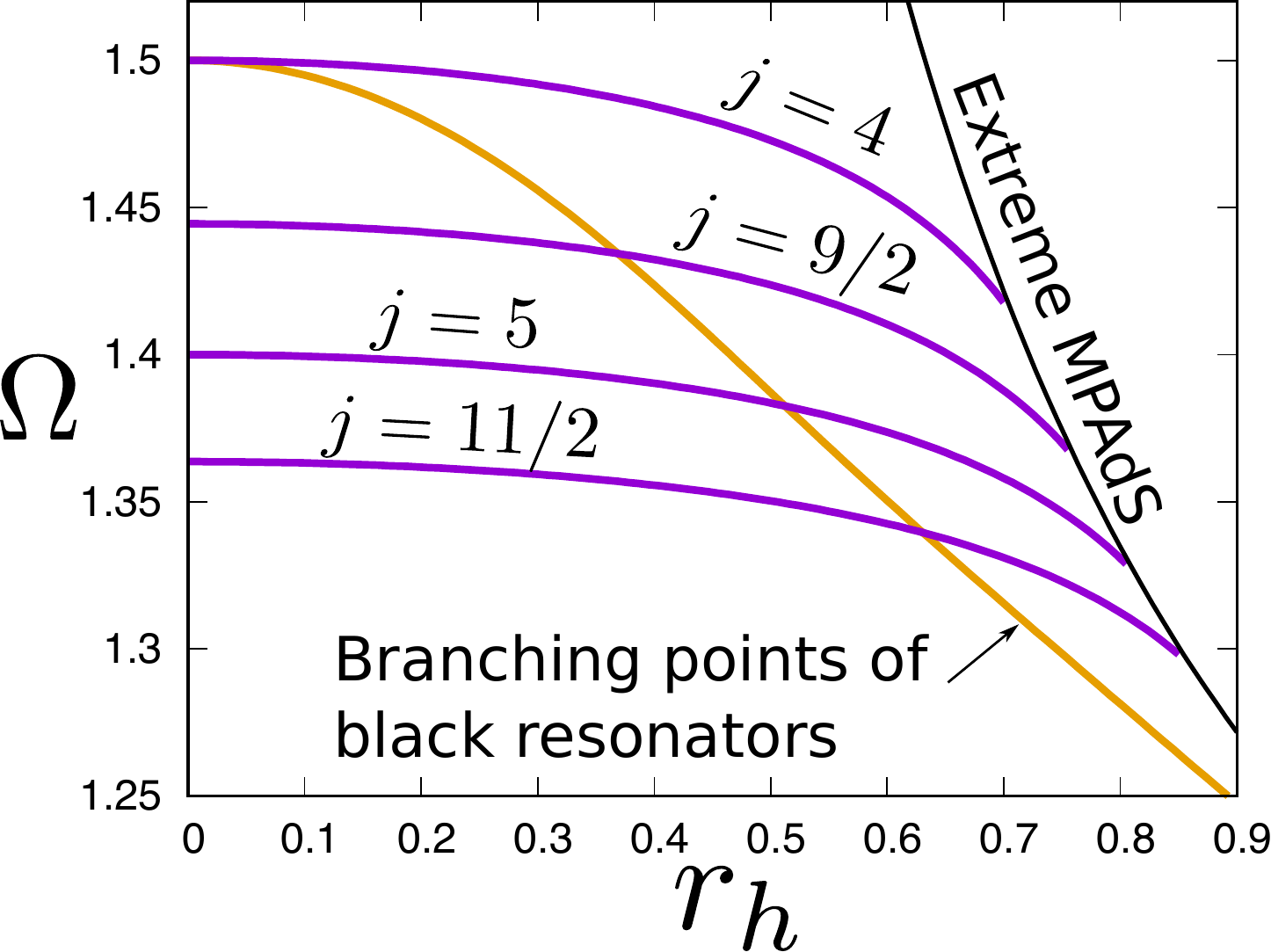}
\end{center}
 \caption{%
Onsets of the scalar field superradiant instability of MPAdS black holes. The onsets are shown in the $(r_h,\Omega)$-plane by purple curves for Wigner D-matrix's ($D^j_{mk}$) wavenumbers $j=4,9/2,5,11/2$, $k=j$, and any allowed value of $m$. Perturbative modes of MPAdS black holes are unstable in the region above the corresponding purple curve. The orange curve represents the branching points to the black resonators from MPAdS, i.e.~the onset of a gravitational superradiant instability of MPAdS.
}
\label{onset_scalar}
\end{figure}

In Fig.\,\ref{onset_scalar}, we summarize the onset of the superradiant instability of MPAdS under the scalar field perturbation with $j=4,\,9/2,\,5,\,11/2$. Above each curve, the MPAdS is unstable with respect to the corresponding mode.
The branching points to the black resonators are also plotted in the orange curve.
For example, for the $j=9/2$ perturbation, the MPAdS is unstable at the branching points in $r_h\lesssim 0.37$ but stable in $r_h\gtrsim 0.37$.  We see that the purple curves intersect with the orange curve with $r_h>0$ only for wavenumbers $j\geq 9/2$.  From this result, we can deduce that the black resonators with a small horizon radius and in the neighborhood of MPAdS are unstable to perturbations with $j\geq 9/2$.

\subsection{Classifying scalar field perturbations}
We now turn to the analysis of the stability of the black resonators. Unlike the case of MPAdS, modes with different $k$ couple. However, because of the double-stepping coupling in \eqref{phieq} and discrete isometries, we can reduce the number of coupled equations.

For half-integer $j$, we obtain two closed systems of scalar field perturbations,
\begin{align}
\bm{v} &= \{\phi_j,\phi_{j-2},\phi_{j-4},\cdots,\phi_{-j+1}\} \ , \\
\tilde{\bm{v}} &=\{\phi_{j-1},\phi_{j-3},\phi_{j-5},\cdots,\phi_{-j}\} \ .
\end{align}
These satisfy the same equations of motion, and therefore it is sufficient to solve only for $\bm{v}$.

When $j$ is an integer, $\{\phi_j$, $\phi_{j-2}$, $\cdots$, $\phi_{-j}\}$ form a closed system. At the onset of instability, where $\omega=0$,
we can further decompose this system into two parts using the discrete isometry~\eqref{parity}. Under the isometry, the Wigner D-matrix transforms as
\begin{equation}
 D_k \to D_{-k}\ ,
\label{Dtrans}
\end{equation}
and also the perturbation variable $\phi_k\to \phi_{-k}$. Therefore, under this isometry, we can divide the perturbation into the even and odd parity modes as
\begin{equation}
\bm{v} = \{\phi_{k}\pm \phi_{-k}|k=j,j-2,\cdots;k\geq 0\}\ ,
\end{equation}
where the plus and minus signs correspond to the even and odd parity, respectively. They transform as $\bm{v}\to \pm \bm{v}$ and hence are decoupled. There is also a closed system $\tilde{\bm{v}} = \{\phi_{k}\pm \phi_{-k}|k=j-1,j-3,\cdots; k\geq 0\}$. However, because we expect that the one with $k=j$ would be more unstable, we do not consider $\tilde{\bm{v}}$ in this paper.

\subsection{Scalar field superradiant instability of black resonators}
\label{sec:scalarSRBR}
We can now solve the equations for scalar field perturbations in black resonators and locate the onset to superradiance.  We opt for a shooting method from the horizon out to infinity. Near the horizon and at the onset of the instability, where $\omega=0$, the equation for the system $\bm{v}$ takes the form
\begin{equation}
\bm{v}'' + \frac{1}{r-r_h}\bm{v}'+\frac{M}{r-r_h}\bm{v} \simeq 0\ ,
\end{equation}
where $' \equiv \partial_r$, and $M$ is a constant matrix derived from \eqref{phieq}. From regularity, we obtain $\bm{v}'=-M\bm{v}$ at $r=r_h$.  The components of $\bm{v}|_{r=r_h}$ are free parameters. We set the first component of $\bm{v}$ (i.e, $\phi_j$ or $\phi_j\pm\phi_{-j}$) to 1 at $r=r_h$. This fixes the scale of the perturbation. We also have a parameter given by the eigenvalue $\lambda$.  Therefore, in total, we have $\textrm{dim}\,\bm{v}$ tuning parameters when we integrate the equations of motion from the horizon to infinity.

At the infinity, the scalar field behaves as
\begin{equation}
\bm{v} \simeq \frac{\bm{c}_1}{r^{2+\sqrt{4+\lambda}}}+\frac{\bm{c}_2}{r^{2-\sqrt{4+\lambda}}}\ ,\quad (r\to \infty)\ .
\end{equation}
We require that the coefficients of the slowly decaying solution to vanish: $\bm{c}_2=0$. This gives $\textrm{dim}\,\bm{v}$ conditions. In the shooting method, hence, we tune the $\textrm{dim}\,\bm{v}$ parameters so that $\bm{c}_2=0$ is satisfied.

\begin{figure}
\begin{center}
\includegraphics[scale=0.5]{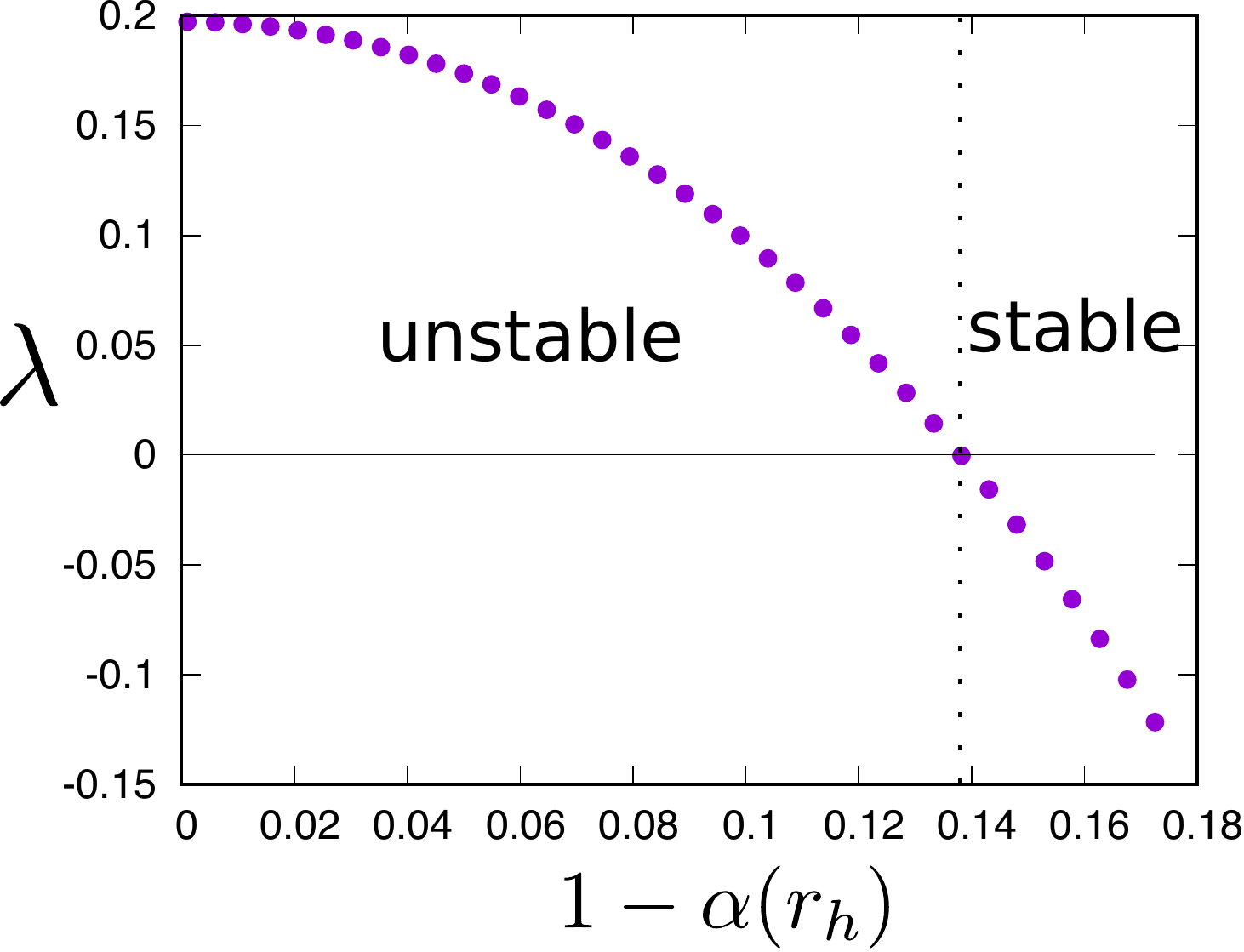}
\end{center}
\caption{Eigenvalue of the Klein-Gordon operator for $j=9/2$ and $r_h=0.35$.}
\label{lambda_KG}
\end{figure}

Solving the perturbation equation \eqref{phieq2}, we track how the eigenvalue of the Klein-Gordon operator $\lambda$ changes as we vary the parameters of the black resonator background $(r_h,\alpha(r_h))$. Fig.\,\ref{lambda_KG} displays an example for the scalar field perturbation with $j=9/2$. For this computation, we fix the horizon radius of the black resonator as $r_h=0.35$ and vary the horizon deformation parameter $\alpha(r_h)$. (Note that $\alpha(r_h)=1$ corresponds to MPAdS.) We find that the eigenvalue crosses zero around $1-\alpha(r_h)\simeq 0.138$. We also know that the MPAdS at the branching point ($\alpha=1$) is unstable for $r_h=0.35$ and $j=9/2$ from Fig.\,\ref{onset_scalar}. Therefore, in this region of parameter space, black resonators with $1-\alpha(r_h)< 0.138$ are unstable against the $j=9/2$ scalar field perturbation, and stable otherwise.\footnote{This identification of stable and unstable domains is consistent with the intuition derived from $\lambda$ being the squared mass of a massive scalar field.}

We repeat the above procedure for various $(r_h,\alpha(r_h))$, searching for the zero of $\lambda$. These parameters can be easily converted into the mass and angular momentum of the black resonators, $E$ and $J$ (see Ref.\,\cite{Ishii:2018oms}). Results of this computation are presented in Section~\ref{sec:results}.

\subsection{Scalar field perturbations of geons}
Unlike black resonators, geons do not have a horizon.  Their perturbations do not have dissipation, and therefore generate modes which are either purely real or purely imaginary.

For geons, we can compute the perturbative frequency $\omega$ directly rather than search for zero modes.  For this calculation, we will work with a different radial coordinate $\rho$ defined by $r=\rho\sqrt{2-\rho^2}/(1-\rho^2)$, so that $\rho\in[0,1]$.  We also perform a field redefinition from $\phi_k$ to $\varphi_k$ given by
\begin{equation}
\phi_k=\rho^{2j}(2-\rho^2)^j(1-\rho^2)^4\varphi_k\;.
\end{equation}
With this field redefinition, the boundary conditions are guaranteed to be satisfied if $\varphi_k$ is finite on the domain.  The basic structure of the equations, including the double-stepping coupling between modes with $k$ and $k\pm 2$, is the same as that of the black resonator.

Now we set $\lambda=0$ in the equations of motion, with the geon as a background solution.  We can then reduce the linear perturbation equations and boundary conditions to a matrix eigenvalue problem (with eigenvalue $\omega$) using pseudospectral methods.  The matrix eigenvalue problem spectrum is solved by QZ decomposition. We track individual modes in parameter space using a Newton-Raphson algorithm. See e.g.~\cite{Dias:2015nua} for an introduction to the various methods used here.  Again, we present the results of this calculation in Section~\ref{sec:results}.

\section{Technical Details for Maxwell Perturbations}
\label{sec:techmaxwell}

\subsection{Equations of motion}
Recall that we decompose the Maxwell field into Wigner D-matrices according to \eqref{A_Dexpand}, which we reproduce here:
\begin{equation}
A_A=e^{-i\omega \tau} \sum_{|k|\leq j} A^k_A(r) D_k\ ,\quad
A_\pm=e^{-i\omega \tau} \sum_{|k\mp 1|\leq j} A^k_\pm(r) D_{k\mp 1}\ ,
\label{A_Dexpand2}
\end{equation}
where $A=\tau,r,3$. The equations of motion take the form (\ref{bmAeq}-\ref{Areq})
\begin{align}
\bm{A}_k''&=\bm{P}[\bm{A}_k,\bm{A}_{k-2},\bm{A}_{k+2}]\ ,\\
(A_r^k)'&=Q[\bm{A}_k,\bm{A}_{k-2},\bm{A}_{k+2}]\ .
\end{align}
Their explicit forms are too cumbersome, and we decided not to reproduce them here.

\subsection{Maxwell field superradiant instability of Myers-Perry AdS black holes}
\label{Maxwell_MPAdS}
Again, in the limit of MPAdS, i.e.~$\alpha(r)\to 1$, the $U(1)$-symmetry generated by $R_z$ is recovered, and the modes with different $U(1)$ charges decouple.  Here, we focus on $k=j+1$, which has the largest $k$ for a fixed $j$, and then there is a single perturbation variable $A_+^{k=j+1}$.  In the same way as the scalar field perturbation of MPAdS in Section~\ref{scalarMPAdS}, we search the values of $(r_h,\Omega)$ where $\lambda=0$.

\begin{figure}
\begin{center}
\includegraphics[scale=0.6]{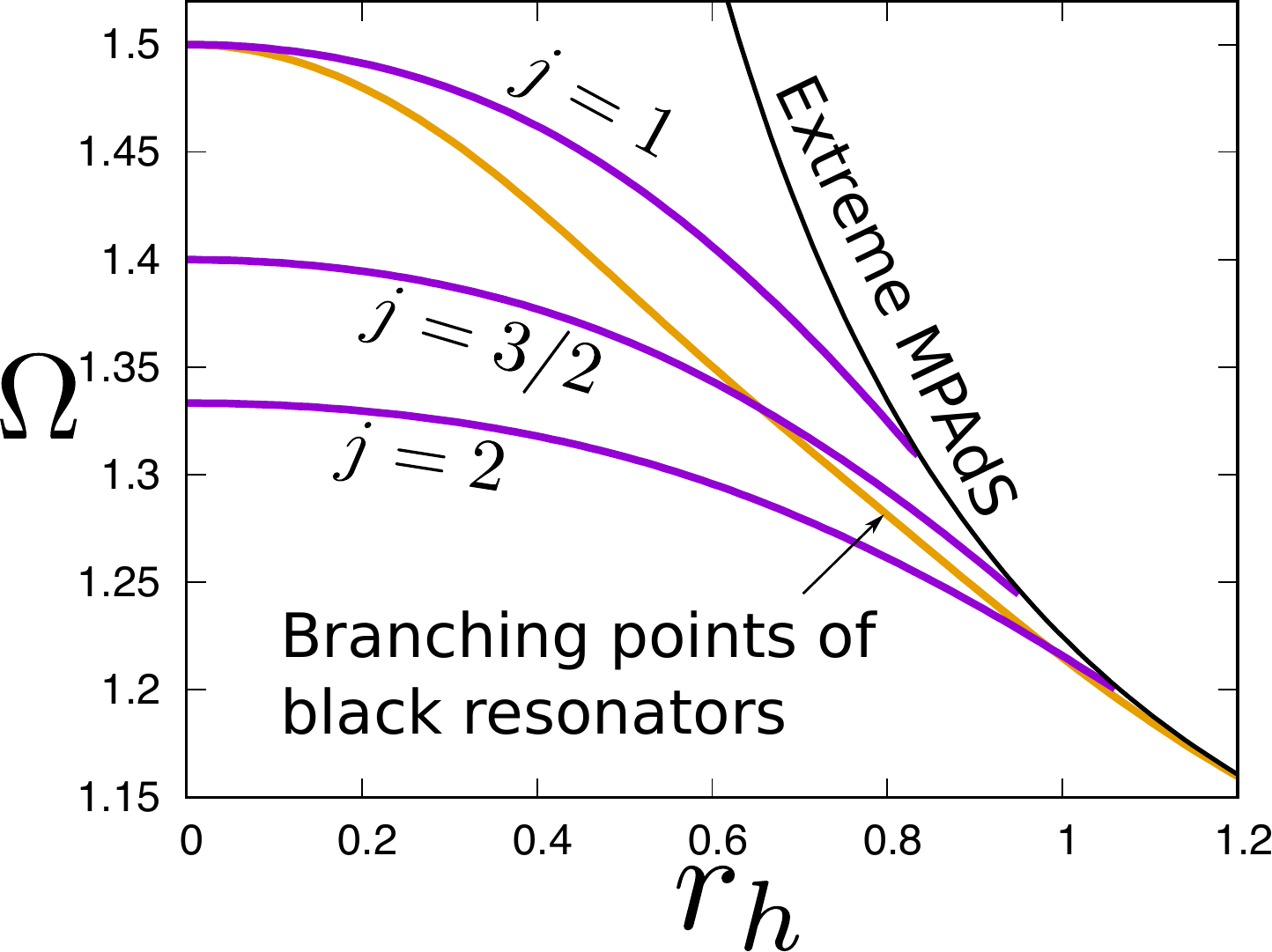}
\end{center}
 \caption{%
Onsets of the Maxwell superradiant instability of MPAdS black holes for perturbations with Wigner D-matrix's ($D^j_{mk}$) wavenumbers $j=1,3/2,2$, $k=j+1$, and any $m$. These are shown in the $(r_h,\Omega)$-plane in purple curves. The orange curve represents the branching points to black resonators from MPAdS.
}
\label{onset_maxwell_MPAdS}
\end{figure}

Results are shown in Fig.\,\ref{onset_maxwell_MPAdS} for $j=1,\, 3/2,\, 2$. In the region above each curve, MPAdS is unstable to the corresponding mode. The branching points to black resonators are also shown in the orange curve. We find that the MPAdS at the branching points change the stability against the perturbations with $j \ge 3/2$. Therefore, the black resonators with a small horizon radius and small deformation near the MPAdS should also be unstable.

\subsection{Classifying Maxwell perturbations}
We turn to the Maxwell perturbations in the black resonator background. First, we classify the coupling of the perturbations.

In the same way as the case of the scalar field perturbations, we can reduce the number of coupled variables by making use of the double-stepping coupling and the discrete isometry~\eqref{parity}. Let us consider the case of $j=3/2$ for example. All perturbation variables can be listed as
\begin{equation}
 A^{k=5/2,\, 3/2,\, 1/2,\, -1/2}_+\ ,\quad
 A^{k=3/2,\, 1/2,\, -1/2,\, -3/2}_A\ ,\quad
 A^{k=1/2,\, -1/2,\, -3/2,\, -5/2}_-\ .
\end{equation}
Among them, because of the double-stepping coupling, we find a closed system with
\begin{equation}
 A^{k=5/2,\, 1/2}_+\ ,\quad
 A^{k=1/2,\, -3/2}_A\ ,\quad
 A^{k=1/2,\, -3/2}_-\ .
 \label{system_j32maxwell}
\end{equation}
To analyze the stability, we will focus on this closed system including the variable with the highest $k$: $A^{k=j+1}_{+}$. The coupled variables for this case is summarized in Table~\ref{j32maxwell_table}. We can proceed with the same analysis for $j=2$, and the result is summarized in Table~\ref{j2maxwell_table}.

\begin{table}
\begin{center}
\begin{tabular}{|c|c|c|c|c|c|c|} \hline
$k$      &-5/2      &-3/2      &-1/2      &1/2       &3/2       &5/2       \\ \hline\hline
$A_{+}$  &\cellcolor[gray]{0.8}&\cellcolor[gray]{0.8}&          &\checkmark&          &\checkmark\\ \hline
$A_{A}$  &\cellcolor[gray]{0.8}&\checkmark&          &\checkmark&          &\cellcolor[gray]{0.8} \\ \hline
$A_{-}$  &          &\checkmark&          &\checkmark&\cellcolor[gray]{0.8}&\cellcolor[gray]{0.8} \\ \hline
\end{tabular}
\end{center}
\caption{Coupled variables for the Maxwell perturbation with $j=3/2$. The variables in \eqref{system_j32maxwell} are marked with $\checkmark$. Variables do not exist in gray cells. Other variables which are not here considered also exist in blank cells.}
\label{j32maxwell_table}
\end{table}
\begin{table}
\begin{center}
\begin{tabular}{|c|c|c|c|c|c|c|c|} \hline
$k$      &-3        &-2        &-1        &0         &1         &2         &3         \\ \hline\hline
$A_{+}$  &\cellcolor[gray]{0.8}&\cellcolor[gray]{0.8}&\checkmark&          &\checkmark&          &\checkmark\\ \hline
$A_{A}$  &\cellcolor[gray]{0.8}&          &\checkmark&          &\checkmark&          &\cellcolor[gray]{0.8} \\ \hline
$A_{-}$  &\checkmark&          &\checkmark&          &\checkmark&\cellcolor[gray]{0.8}&\cellcolor[gray]{0.8} \\ \hline
\end{tabular}
\end{center}
\caption{Coupled variables for the Maxwell perturbation with $j=2$. The closed system including $k=j+1$ is marked with $\checkmark$.}
\label{j2maxwell_table}
\end{table}

For integer $j$, we can further decompose the perturbation when $\omega=0$. By the discrete isometry~\eqref{parity}, the orthogonal basis transforms as
\begin{equation}
 (e^\tau,e^r,e^\pm,e^3)\to (-e^\tau,e^r,-e^\mp,-e^3)\ .
\label{etrans}
\end{equation}
Using this and (\ref{Dtrans}), we find that the transformation of the perturbation variables is
\begin{equation}
 (A_\tau^k,A_r^k,A_\pm^k,A_3^k)\to (-A_\tau^{-k},A_r^{-k},-A_\mp^{-k},-A_3^{-k})\ .
\end{equation}
Hence, we can group the variables into the even and odd parity modes as
\begin{equation}
\begin{split}
\textrm{Even: }&(A_\tau^k-A_\tau^{-k}, A_r^k+A_r^{-k}, A_3^k-A_3^{-k}, A_\pm^k-A_\mp^{-k})\ ,\\
\textrm{Odd: }&(A_\tau^k+A_\tau^{-k}, A_r^k-A_r^{-k}, A_3^k+A_3^{-k}, A_\pm^k+A_\mp^{-k})\ .
\end{split}
\end{equation}
They decouple in the perturbation equations.
\subsection{Maxwell field superradiant instability of black resonators}
\label{Mxw_SR_BR}
We again resort to a shooting method to solve the equations. To adjust the asymptotic behavior at the horizon and infinity, we define rescaled variables $(a_A^k, a_\pm^k)$ by
\begin{equation}
 (A_\tau^k,A_r^k,A_3^k,A_\pm^k) = \left(\frac{F(r)}{r}a_\tau^k,\frac{i}{r^3}a_r^k,\frac{1}{r}a_3^k,\frac{i}{r}a_\pm^k\right)\ ,
\end{equation}
where $F(r)\equiv1-(r_h/r)^6$. We also multiplied $i=\sqrt{-1}$ to $a_r^k$ and $a_\pm^k$ so that the coefficients in the perturbation equations are real.

We introduce the variables $\bm{v}$ and $\bm{w}$ which will be used in numerical calculations as follows. In $\bm{w}$, we collect the components whose indices include $r$. The others are packaged in $\bm{v}$. For example, for $j=3/2$, $\bm{v}$ and $\bm{w}$ are given by
\begin{equation}
\begin{split}
\bm{v}&=(
a_+^{5/2},\,
a_+^{1/2},\,
a_\tau^{1/2},\,
a_3^{1/2},\,
a_-^{1/2},\,
a_\tau^{-3/2},\,
a_3^{-3/2},\,
a_-^{-3/2}
)\ ,\\
\bm{w}&=(
a_r^{1/2},\,
a_r^{-3/2})\ .
\end{split}
\end{equation}
For $j=2$, they become
\begin{equation}
\begin{split}
\bm{v}&=(a_+^3\mp a_-^{-3}, a_+^1\mp a_-^{-1}, a_\tau^1\mp a_\tau^{-1}, a_3^1\mp a_3^{-1}, a_-^1\mp a_+^{-1})\ ,\\
\bm{w}&=(a_r^1\pm a_r^{-1})\ ,
\end{split}
\end{equation}
where the upper and lower signs correspond to the even and odd parity modes.

From (\ref{bmAeq}) and (\ref{Areq}), we obtain second and first order differential equations for $\bm{v}$ and $\bm{w}$, respectively:
\begin{align}
\bm{v}''&=N_1 \bm{v}'+N_2\bm{v}+N_3\bm{w}\label{veq}\ ,\\
\bm{w}'&=N_4\bm{v}+N_5\bm{w}\ ,\label{weq}
\end{align}
where $N_{1,\cdots,5}$ denote matrices whose entries are $r$-dependent functions. We can also obtain equations including $\bm{w}''$ from (\ref{bmAeq}) as
\begin{equation}
\bm{w}''=N_6 \bm{v}'+N_7\bm{v}+N_8\bm{w}\ ,
\label{weq2nd}
\end{equation}
where we eliminated $\bm{w}'$ by using \eqref{weq}.

We do not integrate this equation but use it as constraints to check numerical accuracy. Differentiating \eqref{weq} by $r$ and eliminating $\bm{w}''$ by using \eqref{weq2nd}, we obtain
\begin{equation}
\bm{C}=(N_4-N_6)\bm{v}'+(N_4'+N_5N_4-N_7)\bm{v}+(N_5^2+N_5'-N_7)\bm{w}=0\ ,\label{Const}
\end{equation}
where we again used \eqref{weq} to eliminate $\bm{w}'$. This constraint is conserved under the ``$r$-evolution'' by \eqref{veq} and \eqref{weq} as $\bm{C}'=N_9 \bm{C}$. This constraint equation is automatically satisfied if we impose a regular boundary condition at the horizon.

Near the horizon, we can expand the perturbations as
\begin{equation}
 \bm{v}=\sum_{m=0}^\infty \bm{v}_m(r-r_h)^m \ ,\quad
 \bm{w}=\sum_{m=0}^\infty \bm{w}_m(r-r_h)^m \ .
\end{equation}
Substituting the above series into (\ref{veq}) and (\ref{weq}), we find that $\bm{v}_{m\geq 1}$ and $\bm{w}_{m\geq 0}$ are determined by $\bm{v}_0$. We can impose the first entry of $\bm{v}_0$ to be $1$, and this fixes the scale of the perturbation:
\begin{equation}
\begin{split}
a_+^{j+1}|_{r=r_h}&=1\ ,\quad(j\textrm{: half-integer})\ ,\\
a_+^{j+1}\mp a_-^{-j-1}|_{r=r_h}&=1\ ,\quad(j\textrm{: integer})\ ,
\end{split}
\label{fixscaleA}
\end{equation}
where the upper and lower signs represent the even and odd parity. We are then left with $(\textrm{dim}\,\bm{v} - 1)$ free parameters in $\bm{v}_0$, but we also have $\lambda$ as a free parameter. Therefore, in total, we have $\textrm{dim}\,\bm{v}$ tuning parameters when we integrate \eqref{veq} and \eqref{weq} from the horizon to infinity.

Near the AdS boundary, the asymptotic solution of $\bm{v}$ takes the form
\begin{equation}
 \bm{v}\sim  \frac{\bm{c}_1}{r^{\sqrt{1+\lambda}}}+\bm{c}_2 r^{\sqrt{1+\lambda}}\ ,
\end{equation}
and we impose $\bm{c}_2=0$. One can then check that $\bm{w}$ also decays as $1/r^{\sqrt{4+\lambda}}$ when $\bm{c}_2=0$, and therefore it is sufficient to impose $\bm{c}_2=0$ only. In the shooting method, we tune the $\textrm{dim}\,\bm{v}$ tuning parameters so that $\bm{c}_2=0$ is satisfied.  Results are shown in Section~\ref{sec:results}.

\subsection{Maxwell field perturbations of geons}
\label{Mxw_SR_geon}
For the geon, we compute the frequency $\omega$ directly.  Like the scalar field, we work with the modified radial coordinate $\rho$ defined by $r=\rho\sqrt{2-\rho^2}/(1-\rho^2)$, so that $\rho\in[0,1]$. We work with rescaled variables $a^k$ defined by
\begin{equation}
A^k=\rho^{2j}(2-\rho^2)^j(1-\rho^2)^2\left[a^k_\tau\,\mathrm d\tau+i\rho^{-1}(1-\rho^2)a^k_\rho\,\mathrm d\rho+a^k_3(\sigma_3+2h\mathrm d\tau)+ia^k_-\sigma_-+ia^k_+\sigma_+\right]\;.
\end{equation}
Just as we have done for the scalar field, the boundary conditions for $A^k$ are guaranteed to be satisfied if $a^k$ is finite.  The basic structure of the equations of motion is the same as that of the black resonator.  We again set $\lambda=0$ and solve for $\omega$ using the same numerical methods as that of the scalar field.

\subsection{Comment on gauge modes}
\label{comgauge}
The Lorenz gauge condition~\eqref{LorenzA} does not completely fix the gauge freedom. However, we can show that the normal modes found at the onset of instability cannot be gauge modes. The gauge transformation of the Maxwell field is written as $\delta A_\mu = \partial_\mu \Lambda$ where $\Lambda$ is a scalar function. It can be expanded by the Wigner D-matrices as $\Lambda = \sum_{|k|\leq j} \Lambda_k D_k$. In particular, the gauge transformation of $A_{\pm}^k$ is given by
\begin{equation}
 \delta A_+^k = \epsilon_{k} \Lambda_k\ ,\quad
 \delta A_-^k = \epsilon_{k+1} \Lambda_k\ .
\end{equation}
Because $\epsilon_{j+1}=\epsilon_{-j}=0$, we obtain $\delta A^{k=\pm(j+1)}_\pm = 0$. This is consistent with the fact that $\Lambda_k$ is defined in $|k|\leq j$. Thus we find that $A^{k=\pm(j+1)}_\pm$ are gauge invariant variables.  It follows that, if $A_+^{k=j+1}$ and $A_-^{k=-j-1}$ are non-zero, the perturbation cannot be a pure gauge mode. In our numerical calculations, we impose \eqref{fixscaleA}, and therefore the perturbation cannot be a gauge mode.

\section{Technical Details for Gravitational Perturbations}
\label{sec:techgrav}

\subsection{Equations of motion}

Recall that we decompose the metric perturbation into Wigner D-matrices according to \eqref{h_WignerExpansion}, which we reproduce again here:
\begin{equation}
\begin{split}
h_{AB}&=e^{-i\omega \tau} \sum_{|k|\leq j} h^k_{AB}(r) D_k\ ,\quad
h_{+-}=e^{-i\omega \tau} \sum_{|k|\leq j} h^k_{+-}(r) D_k\ ,\\
h_{A\pm}&=e^{-i\omega \tau} \sum_{|k\mp 1|\leq j} h^k_{A\pm}(r) D_{k\mp 1}\ ,\quad
h_{\pm\pm}=e^{-i\omega \tau} \sum_{|k\mp 2|\leq j} h^k_{\pm\pm}(r) D_{k\mp 2}\ ,
\end{split}
\label{h_WignerExpansion2}
\end{equation}
where $A,B=t,r,3$.  The equations of motion we obtain are in the form
\begin{align}
\bm{h}_k''&=\bm{P}[\bm{h}_{k-4},\bm{h}_{k-2},\bm{h}_{k},\bm{h}_{k+2},\bm{h}_{k+4}]\ ,\\
(h_{ar}^k)'&=Q[\bm{h}_{k-4},\bm{h}_{k-2},\bm{h}_{k},\bm{h}_{k+2},\bm{h}_{k+4}]\ ,\\
h_{+-}^k&=R[\bm{h}_{k-4},\bm{h}_{k-2},\bm{h}_{k},\bm{h}_{k+2},\bm{h}_{k+4}]\ ,
\end{align}
where $\bm{P}$, $Q$ and $R$ are linear operators. In $\bm{P}$, we include the first derivative terms by $r$, but $Q$ and $R$ do not contain such terms. Again, we do not reproduce the lengthy expressions of $\bm{P}$, $Q$ and $R$.

\subsection{Gravitational superradiant instability of Myers-Perry AdS black holes}
\label{grav_MPAdS}
\begin{figure}
\begin{center}
\includegraphics[scale=0.6]{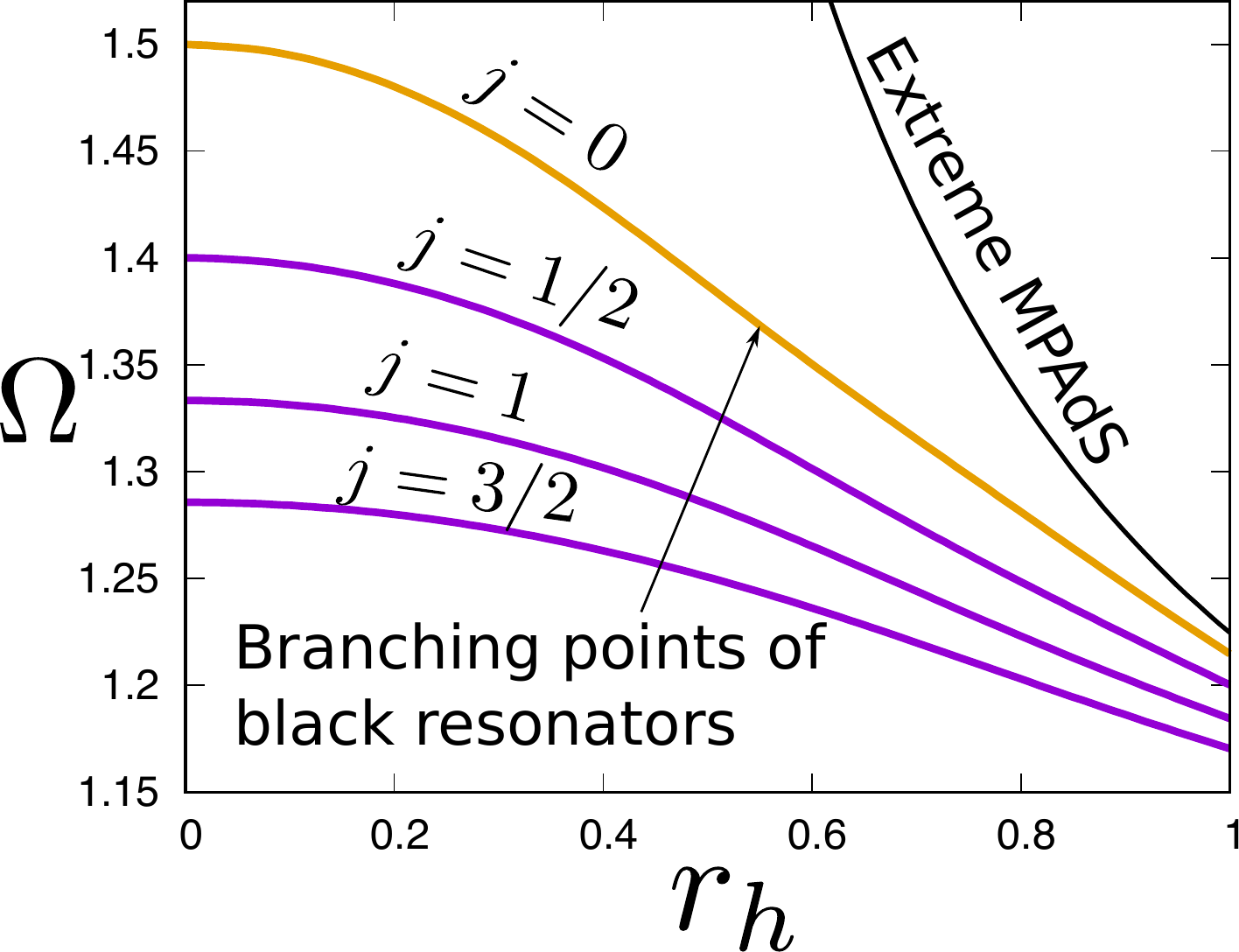}
\end{center}
 \caption{Onsets of the gravitational superradiant instability of MPAdS. The onsets for $j=1/2,1,3/2$ are shown in purple curves in the $(r_h,\Omega)$-plane. The orange curve represents the branching points to cohomogeneity-1 black resonators, which is generated from MPAdS by the $j=0$ gravitational perturbation.
}
\label{onset_grav_MPAdS}
\end{figure}

In the limit of MPAdS, i.e, $\alpha\to 1$, we have a decoupled equation for $h_{++}^{k=j+2}$. As employed in Section~\ref{scalarMPAdS}, we search the zero of the eigenvalue $\lambda$ for the onset of the gravitational superradiant instability of MPAdS. Results are shown in Fig.\,\ref{onset_grav_MPAdS} for $j=1/2,\,1,\,3/2$. (Recall that the orange branching curve is also the $j=0$ onset curve.) Although this result has already been obtained in \cite{Murata:2008xr}, we show it again to make this paper self-contained. Above each curve, MPAdS is unstable to the corresponding mode. This indicates that, on top of the branching points, MPAdS is always unstable to ($j\geq 1/2$)-modes. This is consistent with the observation in Fig.\,\ref{J_E_grav_2j_1} that an onset curve for the $j\geq 1/2$ perturbation does not intersect with the branching point curve.

\subsection{Classifying gravitational perturbations}
Let us consider the gravitational perturbation in the black resonator background.

First, we explain the classification of the perturbation using $j=1/2$ as an example. All perturbation variables for this case can be listed as
\begin{equation}
 h^{k=5/2,3/2}_{++}, \quad h^{k=3/2,1/2}_{A+}, \quad h^{k=\pm 1/2}_{AB}, \quad h^{k=\pm 1/2}_{+-},
\quad h^{k=-1/2,-3/2}_{A-}, \quad h^{k=-3/2,-5/2}_{--}\ .
\end{equation}
Because of the double-stepping coupling, we can extract a closed system of
\begin{equation}
h_{++}^{k=5/2}\ ,\quad
h_{A+}^{k=1/2}\ ,\quad
h_{AB}^{k=1/2}\ ,\quad
h_{+-}^{k=1/2}\ ,\quad
h_{A-}^{k=-3/2}\ ,\quad
h_{--}^{k=-3/2}\ .
\label{system_j12grav}
\end{equation}
The coupled variables for $j=1/2$ are summarized in Table~\ref{j12table}, where we focus only on the closed system including $h_{++}^{k=j+2}$. A similar closed system obtained for $j=1$ is shown in Table~\ref{j1table}.

\begin{table}
\begin{center}
\begin{tabular}{|c|c|c|c|c|c|c|} \hline
$k$      &-5/2      &-3/2      &-1/2      &1/2       &3/2       &5/2       \\ \hline\hline
$h_{++}$ &\cellcolor[gray]{0.8}&\cellcolor[gray]{0.8}&\cellcolor[gray]{0.8}&\cellcolor[gray]{0.8}&          &\checkmark\\ \hline
$h_{A+}$ &\cellcolor[gray]{0.8}&\cellcolor[gray]{0.8}&\cellcolor[gray]{0.8}&\checkmark&          &\cellcolor[gray]{0.8} \\ \hline
$h_{AB}$ &\cellcolor[gray]{0.8}&\cellcolor[gray]{0.8}&          &\checkmark&\cellcolor[gray]{0.8}&\cellcolor[gray]{0.8} \\ \hline
$h_{+-}$ &\cellcolor[gray]{0.8}&\cellcolor[gray]{0.8}&          &\checkmark&\cellcolor[gray]{0.8}&\cellcolor[gray]{0.8} \\ \hline
$h_{A-}$ &\cellcolor[gray]{0.8}&\checkmark&          &\cellcolor[gray]{0.8}&\cellcolor[gray]{0.8}&\cellcolor[gray]{0.8} \\ \hline
$h_{--}$ &          &\checkmark&\cellcolor[gray]{0.8}&\cellcolor[gray]{0.8}&\cellcolor[gray]{0.8}&\cellcolor[gray]{0.8} \\ \hline
\end{tabular}
\end{center}
\caption{Coupled variables of the gravitational perturbation for $j=1/2$. The variables in (\ref{system_j12grav}) are marked with $\checkmark$.}
\label{j12table}
\end{table}

\begin{table}
\begin{center}
\begin{tabular}{|c|c|c|c|c|c|c|c|} \hline
$k$      &-3        &-2        &-1        &0         &1         &2         &3         \\ \hline\hline
$h_{++}$ &\cellcolor[gray]{0.8}&\cellcolor[gray]{0.8}&\cellcolor[gray]{0.8}&\cellcolor[gray]{0.8}&\checkmark&          &\checkmark\\ \hline
$h_{A+}$ &\cellcolor[gray]{0.8}&\cellcolor[gray]{0.8}&\cellcolor[gray]{0.8}&          &\checkmark&          &\cellcolor[gray]{0.8} \\ \hline
$h_{AB}$ &\cellcolor[gray]{0.8}&\cellcolor[gray]{0.8}&\checkmark&          &\checkmark&\cellcolor[gray]{0.8}&\cellcolor[gray]{0.8} \\ \hline
$h_{+-}$ &\cellcolor[gray]{0.8}&\cellcolor[gray]{0.8}&\checkmark&          &\checkmark&\cellcolor[gray]{0.8}&\cellcolor[gray]{0.8} \\ \hline
$h_{A-}$ &\cellcolor[gray]{0.8}&          &\checkmark&          &\cellcolor[gray]{0.8}&\cellcolor[gray]{0.8}&\cellcolor[gray]{0.8} \\ \hline
$h_{--}$ &\checkmark&          &\checkmark&\cellcolor[gray]{0.8}&\cellcolor[gray]{0.8}&\cellcolor[gray]{0.8}&\cellcolor[gray]{0.8} \\ \hline
\end{tabular}
\end{center}
\caption{Coupled variables of the gravitational perturbation for $j=1$. The closed system including $k=j+2$ is marked with $\checkmark$.}
\label{j1table}
\end{table}

When $j$ is an integer, we can further decompose the perturbation into the even and odd parity modes by the discrete isometry~\eqref{parity} at the onset of the instability $\omega=0$. By using \eqref{Dtrans} and \eqref{etrans}, the transformation of the gravitational perturbation variables by the isometry is found as
\begin{multline}
 (h_{\tau\tau}^k,\,
h_{\tau r}^k,\,
h_{\tau 3}^k,\,
h_{r r}^k,\,
h_{r 3}^k,\,
h_{33}^k,\,
h_{+-}^k,\,
h_{\tau \pm}^k,\,
h_{r \pm}^k,\,
h_{3 \pm}^k,\,
h_{\pm\pm}^k)\\
\to
(h_{\tau\tau}^{-k},\,
-h_{\tau r}^{-k},\,
h_{\tau 3}^{-k},\,
h_{r r}^{-k},\,
-h_{r 3}^{-k},\,
h_{33}^{-k},\,
h_{+-}^{-k},\,
h_{\tau \mp}^{-k},\,
-h_{r \mp}^{-k},\,
h_{3 \mp}^{-k},\,
h_{\mp\mp}^{-k})\ .
\end{multline}
Therefore we can group the variables into the even and odd parity modes, which are decoupled, as
\begin{equation}
\begin{split}
\textrm{Even: }&(
h_{\tau\tau}^k+h_{\tau\tau}^{-k},\,
h_{\tau r}^k-h_{\tau r}^{-k},\,
h_{\tau 3}^k+h_{\tau 3}^{-k},\,
h_{r r}^k+h_{r r}^{-k},\,
h_{r 3}^k-h_{r 3}^{-k},\\
&\hspace*{0.5cm}h_{33}^k+h_{33}^{-k},\,
h_{+-}^k+h_{+-}^{-k},\,
h_{\tau \pm}^k+h_{\tau \mp}^{-k},\,
h_{r \pm}^k-h_{r \mp}^{-k},\,
h_{3 \pm}^k+h_{3 \mp}^{-k},\,
h_{\pm\pm}^k+h_{\mp\mp}^{-k}
)\ ,\\
\textrm{Odd: }&(
h_{\tau\tau}^k-h_{\tau\tau}^{-k},\,
h_{\tau r}^k+h_{\tau r}^{-k},\,
h_{\tau 3}^k-h_{\tau 3}^{-k},\,
h_{r r}^k-h_{r r}^{-k},\,
h_{r 3}^k+h_{r 3}^{-k},\\
&\hspace*{0.5cm}h_{33}^k-h_{33}^{-k},\,
h_{+-}^k-h_{+-}^{-k},\,
h_{\tau \pm}^k-h_{\tau \mp}^{-k},\,
h_{r \pm}^k+h_{r \mp}^{-k},\,
h_{3 \pm}^k-h_{3 \mp}^{-k},\,
h_{\pm\pm}^k-h_{\mp\mp}^{-k}
)\ .
\end{split}
\end{equation}
\subsection{Gravitational superradiant instability of black resonators}
To adjust the asymptotic behavior at the horizon and infinity, we introduce rescaled perturbation variables $H^k_{ab}$ as
\begin{align}
&\bigg(
h^k_{\tau\tau},\
h^k_{\tau r},\
h^k_{\tau 3},\
h^k_{rr},\
h^k_{r3},\
h^k_{33},\
h^k_{+-},\
h^k_{\tau\pm},\
h^k_{r\pm},\
h^k_{\pm 3},\
h^k_{\pm\pm}
\bigg) \nonumber\\
&=\bigg(
F H^k_{\tau\tau},\
ir^{-2} F H^k_{\tau r},\
F H^k_{\tau 3},\
r^{-4} F^{-1} H^k_{rr},\
ir^{-2} H^k_{r3},\
H^k_{33}, \nonumber\\
&\hspace{5cm}H^k_{+-},\
iFH^k_{\tau\pm},\
r^{-2} H^k_{r\pm},\
iH^k_{\pm 3},\
H^k_{\pm\pm}
\bigg)\ .
\end{align}
In the same way as the case of the Maxwell perturbation, we introduce the variable $\bm{v}$ and $\bm{w}$. In $\bm{w}$, we assemble the components of the perturbations whose indices include $r$, and the others are put in $\bm{v}$. However, we can eliminate $H_{+-}^k$ by using the traceless condition~\eqref{TL2}.
For example, for $j=1/2$ mode, $\bm{v}$ and $\bm{w}$ become
\begin{equation}
\begin{split}
\bm{v}&=
(
H_{++}^{k=5/2},
H_{\tau +}^{k=1/2},
H_{+3}^{k=1/2},
H_{\tau\tau}^{k=1/2},
H_{\tau 3}^{k=1/2},\\
&\hspace{4cm}
H_{33}^{k=1/2},
H_{t-}^{k=-3/2},
H_{3-}^{k=-3/2},
H_{--}^{k=-3/2})\ ,\\
\bm{w}&=
(
H_{r +}^{k=1/2},
H_{\tau r}^{k=1/2},
H_{rr}^{k=1/2},
H_{r3}^{k=1/2},
H_{r-}^{k=-3/2})\ .
\end{split}
\end{equation}
For $j=1$, they become
\begin{equation}
\begin{split}
\bm{v}&=
(
H_{++}^{k=3}\pm H_{--}^{k=-3},
H_{++}^{k=1}\pm H_{--}^{k=-1},
H_{\tau +}^{k=1} \pm H_{\tau -}^{k=-1},
H_{+3}^{k=1}\pm H_{-3}^{k=-1},\\
&\hspace{4cm}
H_{\tau\tau}^{k=1}\pm H_{\tau\tau}^{k=-1},
H_{\tau 3}^{k=1}\pm H_{\tau 3}^{k=-1},
H_{33}^{k=1}\pm H_{33}^{k=-1})\ ,\\
\bm{w}&=
(
H_{r +}^{k=1}\mp H_{r -}^{k=-1},
H_{\tau r}^{k=1}\mp H_{\tau r}^{k=-1},
H_{rr}^{k=1}\pm H_{rr}^{k=-1},
H_{r3}^{k=1}\mp H_{r3}^{k=-1})\ ,
\end{split}
\end{equation}
where the upper and lower signs correspond to the even and odd parity modes.
Then, we follow the same procedure as that for the Maxwell field in Appendix~\ref{Mxw_SR_BR}. To fix the scale of the perturbation, we impose
\begin{equation}
\begin{split}
H_{++}^{j+2}|_{r=r_h}&=1\ ,\quad(j\textrm{: half-integer})\ ,\\
H_{++}^{j+2}\pm H_{--}^{-j-2}|_{r=r_h}&=1\ ,\quad(j\textrm{: integer})\ .
\end{split}
\label{fixscaleH}
\end{equation}
Near the AdS boundary, the asymptotic solution of $\bm{v}$ is given by
\begin{equation}
\bm{v}\sim  \frac{\bm{c}_1}{r^{\sqrt{4+\lambda}}}+\bm{c}_2 r^{\sqrt{4+\lambda}}\ .
\end{equation}
We impose $\bm{c}_2=0$. In the shooting method, we tune the $\textrm{dim}\,\bm{v}$ parameters including the eigenvalue $\lambda$ at the horizon so that $\bm{c}_2=0$ is satisfied at the infinity.

Because perturbation equations are linear, there is a linear relation between free parameters at horizon $\bm{v}|_{r=r_h}$ and
coefficients of the growing mode at infinity $\bm{c}_2$ as
$\bm{c}_2=A(\lambda) \, \bm{v}|_{r=r_h}$ where $A(\lambda)$ is a $\lambda$-dependent matrix.
Computing $\bm{c_2}$ for $\bm{v}|_{r=r_h}=(1,0,\cdots,0)^T, (0,1,0,\cdots,0)^T, \cdots, (0,\cdots,0,1)^T$, we can explicitly construct the matrix $A(\lambda)$ for a given $\lambda$.
If $\textrm{det}\,A(\lambda)=0$, there is an appropriate $\bm{v}|_{r=r_h}$ such that $\bm{c}_2=0$ is satisfied, i.e, roots of $\textrm{det}\,A(\lambda)$ are eigenvalues.
We sometimes use this technique when it is difficult to track the eigenvalue by the shooing method.

We can also compute the spectrum for $\omega$ directly, by setting $\lambda=0$ and solving the resulting eigenvalue problem.  Unlike the scalar and Maxwell fields, the growth rates for the gravitational perturbations are not prohibitively small for numerics.

\begin{figure}
  \centering
\subfigure[$j=1/2$]
 {\includegraphics[scale=0.32]{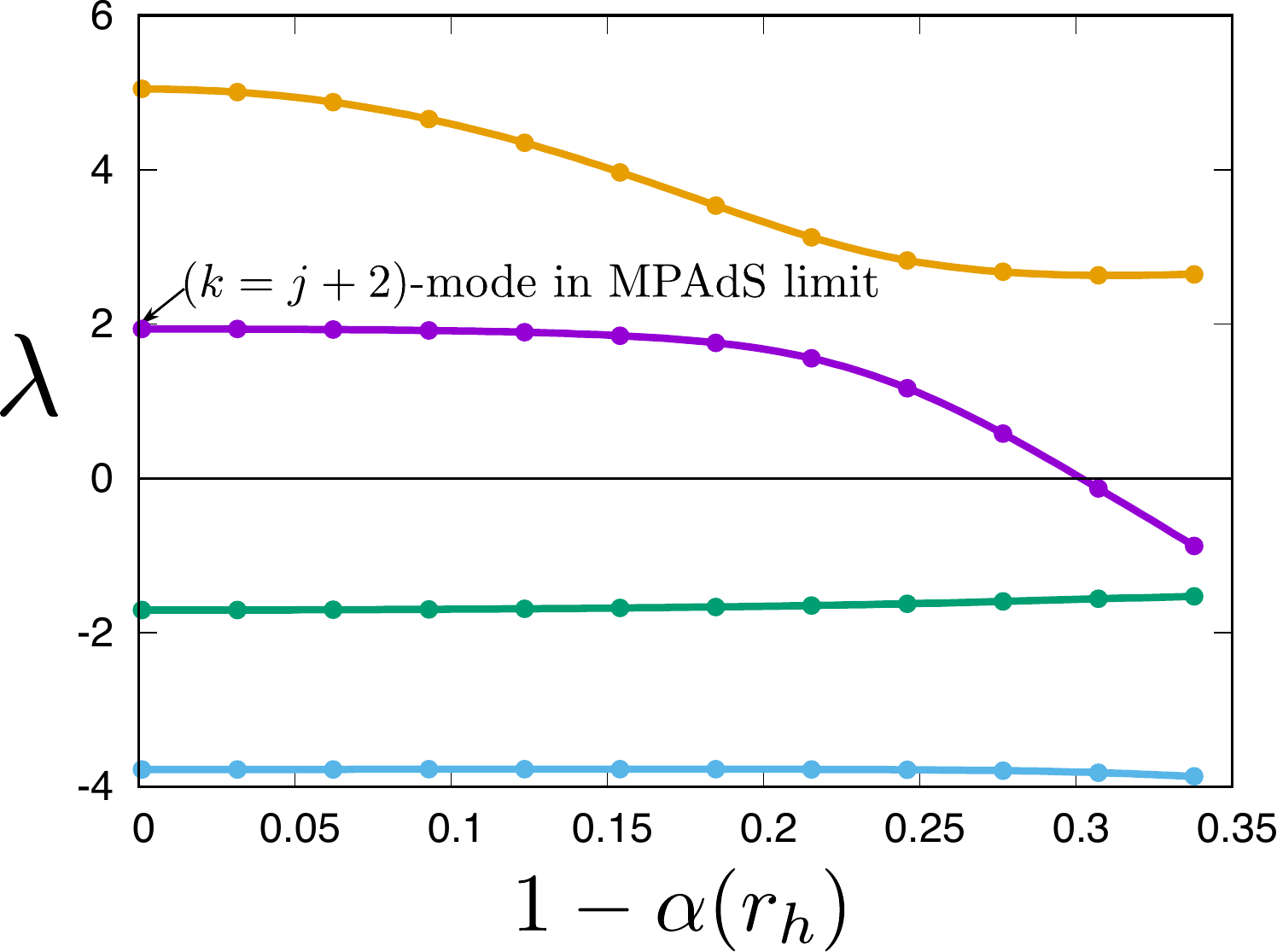}\label{lambda_grav_2j1}
  }
\subfigure[$j=1$ : even parity]
 {\includegraphics[scale=0.32]{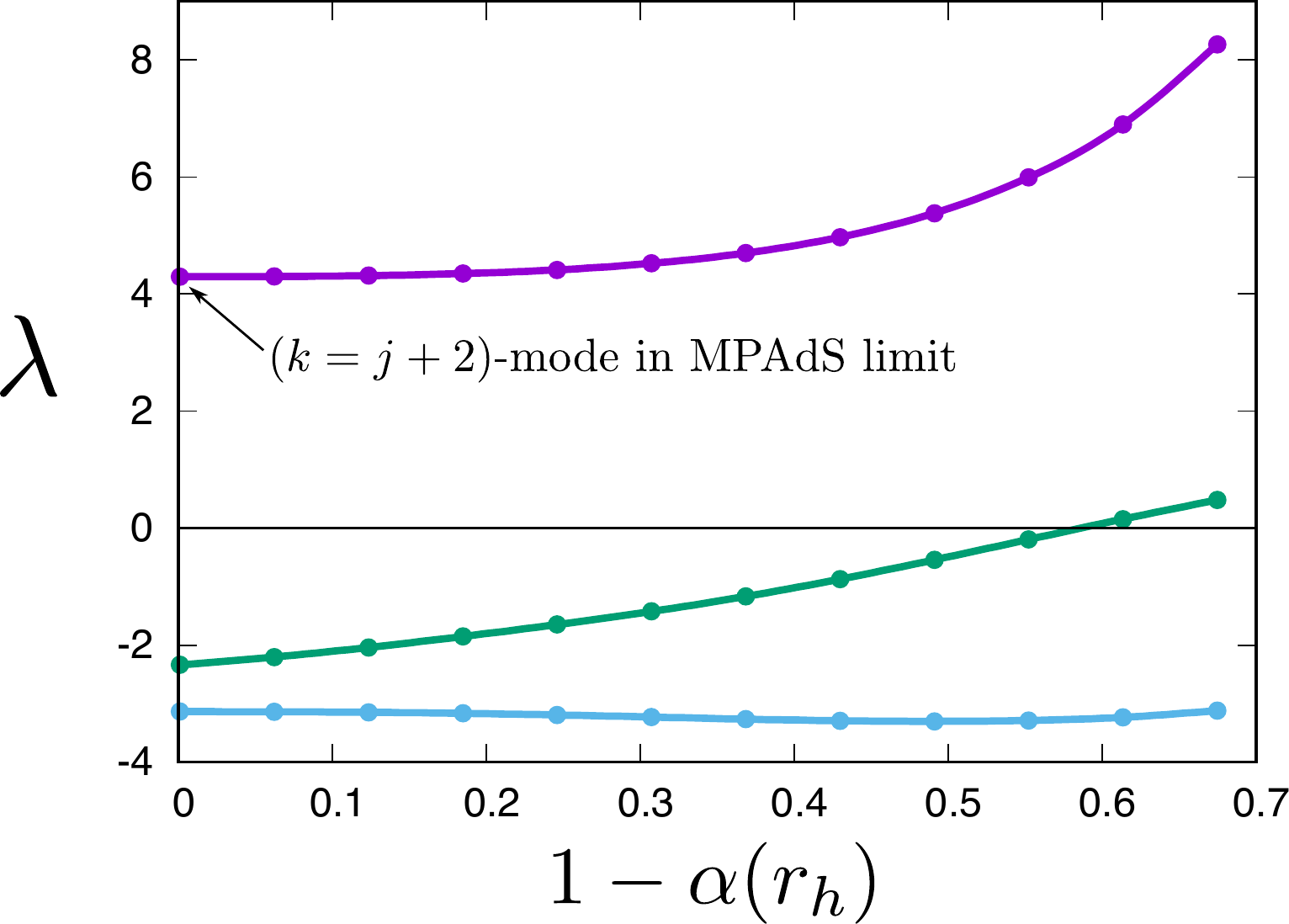}\label{lambda_grav_2j2_even}
  }
\subfigure[$j=1$ : odd parity]
 {\includegraphics[scale=0.32]{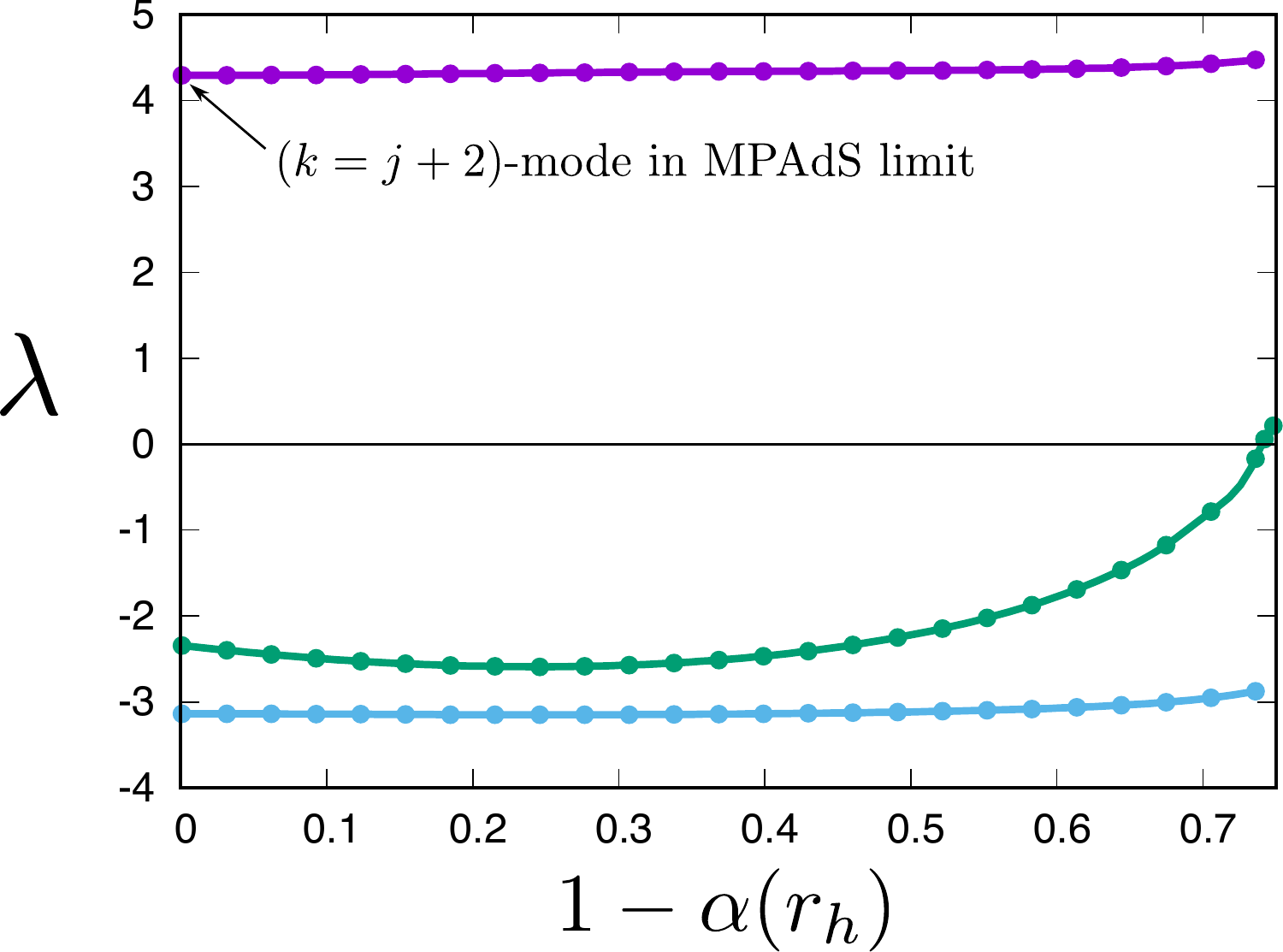}\label{lambda_grav_2j2_odd}
  }
 \caption{
Eigenvalue of the Lichnerowicz operator for $j=1/2,\,1$.
The horizon radius is fixed as $r_h=0.35$.
}
\label{lambda_grav}
\end{figure}

Even when a zero mode is found, we need to verify that it corresponds to the onset of a dynamical instability.  For all of these modes, we checked that the zero mode is not a gauge mode by using the technique which will be introduced in Appendix \ref{comgauge_grav}.

In Fig.\,\ref{lambda_grav}, we show the eigenvalue of the Lichnerowicz operator as a function of the horizon deformation parameter $1-\alpha(r_h)$ for $j=1/2,\,1$. The left edge of each figures is the MPAdS limit, where $\alpha(r_h)=1$ and modes with different $k$ decouple.
We chose the horizon radius as 
$r_h=0.35$ to clearly show the existence of the zero modes ($\lambda=0$).

For $j=1/2$, we find four eigenvalues as in  Fig.~\ref{lambda_grav_2j1}. One of them (the purple curve) approaches the
decoupled $h_{++}^{k=j+2=5/2}$ mode in the MPAdS limit. The eigenvalue crosses zero around $1-\alpha(r_h)\simeq 0.3$. 
The MPAdS at the branching point of the black resonator is unstable to this mode as studied in Appendix~\ref{grav_MPAdS}. Therefore, we deduce that the black resonators are unstable in $1-\alpha(r_h)\lesssim 0.3$ and stable in $1-\alpha(r_h)\gtrsim 0.3$ with respect to the ($j=1/2$)-perturbation.  In Section~\ref{sec:results}, we describe results for the $j=1/2$ mode in more detail, including the growth rates $\mathrm{Im}(\omega)$.

For $j=1$, we find three eigenvalues in each even and odd parity perturbation as in Figs.~\ref{lambda_grav_2j2_even} and \ref{lambda_grav_2j2_odd}. Purple curves approach the $k=j+2=3$ mode in the MPAdS limit. Unlike the case of $j=1/2$, we did not observe the purple curves reaching zero in the parameter region we examined. From the analysis in Appendix~\ref{grav_MPAdS}, we know that the black resonators are unstable to the $k=j+2=3$ perturbation near the branching points. Therefore, the black resonators are unstable to this perturbation at least in the region we explored.
We can also see that green curves cross zero.
This indicates that new unstable modes exist to the right of the zero points.

\subsection{Computing the frequency directly for resonators and geons}
\label{subsec:computeomegagrav}
As for the scalar field and Maxwell field perturbations of geons, we work with a new radial coordinate $\rho$ defined by $r=\rho\sqrt{2-\rho^2}/(1-\rho^2)$.  We also work with different rescaled variables than that of the black resonator.  For the geon, these are given by
\begin{align}
&\bigg(
h^k_{\tau\tau},\
h^k_{\tau\rho},\
h^k_{\tau 3},\
rh^k_{\rho\rho},\
h^k_{\rho3},\
h^k_{33},\
h^k_{+-},\
h^k_{\tau\pm},\
h^k_{\rho\pm},\
h^k_{\pm 3},\
h^k_{\pm\pm},\
\bigg) \nonumber\\
&=\rho^{2j}(2-\rho^2)^j(1-\rho^2)^2\bigg(
H^k_{\tau\tau},\
i\rho^{-1}H^k_{\tau\rho},\
H^k_{\tau3},\
\rho^{-2}(2-\rho^2)^{-1}H^k_{\rho\rho},\
i \rho^{-1}H^k_{\rho3},\
H^k_{33}, \nonumber\\
&\hspace{5cm}H^k_{+-},\
iH^k_{\tau\pm},\
\rho^{-1}H^k_{\rho\pm},\
iH^k_{\pm 3},\
H^k_{\pm\pm},\
\bigg)\ .
\end{align}
The boundary conditions at $\rho=1$ are similar to those for the black resonator at $r\to\infty$.  At the origin, we impose regularity.  We then set $\lambda=0$ and solve the resulting eigenvalue problem with $\omega$ as an eigenvalue.

For black resonators, we use a modified ansatz from before to compute the frequency.  Our background metric is given by
\begin{align}
\mathrm ds^2=\frac{1}{(1-\rho^2)^2}\bigg\{&-\rho^2(2-\rho^2)\hat f^{-2}\hat g \frac{\mathrm d\tau^2}{z_+^2}+\frac{4\mathrm d\rho^2}{(2-\rho^2)\hat g}\nonumber\\
&\qquad+\frac{1}{4z_+^2}\left[\hat\beta\left(\frac{1}{\hat\alpha}\sigma_1^2+\hat\alpha \sigma_2^2\right)+\frac{1}{\hat\beta^2}\left(\sigma_3+2\rho^2(2-\rho^2)\hat h\mathrm d\tau\right)^2\right]\bigg\}\;,
\label{BRgravansatz}
\end{align}
where the functions with hats depend on the coordinate $\rho$, and the solution is parametrised by $z_+$ and by $h(1)=\Omega$, the angular frequency.  Our metric perturbation ansatz is given by
\begin{align}
h_{\mu\nu}\mathrm dx^\mu\mathrm dx^\nu&=(1-\rho^2)^2\rho^{2i\kappa\omega}(2-\rho^2)^{i\kappa\omega}\bigg\{-\hat h_{\tau\tau}f^{-2}g\frac{\mathrm d\tau^2}{z_+^2}+\hat h_{\rho\rho}\frac{4\mathrm d\rho^2}{\rho^2(2-\rho^2)g}\nonumber\\
&\qquad +\frac{1}{4z_+^2}\bigg[\beta\left(\hat h_{--}\sigma_-^2+\hat h_{++}\sigma_+^2+2\hat h_{-+}\sigma_-\sigma_+\right)+\hat h_{33}\frac{1}{\beta^2}\left(\sigma_3+2\rho^2(2-\rho^2)\hat h\mathrm d\tau\right)^2\nonumber\\
&\qquad\qquad\quad+2\left(\hat h_{\tau3}\mathrm d\tau+i\hat h_{\rho3}\frac{\mathrm d\rho}{\rho}+ih_{3-}\sigma_-+ih_{3+}\sigma_+\right)\left(\sigma_3+2\rho^2(2-\rho^2)\hat h\mathrm d\tau\right)\bigg]\nonumber\\
&\qquad+2\bigg[i\left(\hat h_{\tau\rho}\frac{\mathrm d\rho}{\rho}+\hat h_{\tau-}\sigma_-+\hat h_{\tau+}\sigma_+\right)\mathrm d\tau+\left(\hat h_{\rho-}\sigma_-+\hat h_{\rho+}\sigma_+\right)\frac{\mathrm d\rho}{\rho}\bigg]\bigg\}\;,
\end{align}
where $\kappa=-\frac{z_+\hat f(0)}{2\hat g(0)}$.  We decompose the perturbation functions as
\begin{equation}
\begin{split}
\hat h_{AB}&=e^{-i\omega \tau} \sum_{|k|\leq j} \hat H^k_{AB}(\rho) D_k\ ,\quad
\hat h_{+-}=e^{-i\omega \tau} \sum_{|k|\leq j} \hat H^k_{+-}(\rho) D_k\ ,\\
\hat h_{A\pm}&=e^{-i\omega \tau} \sum_{|k\mp 1|\leq j} \hat H^k_{A\pm}(\rho) D_{k\mp 1}\ ,\quad
\hat h_{\pm\pm}=e^{-i\omega \tau} \sum_{|k\mp 2|\leq j} \hat H^k_{\pm\pm}(\rho) D_{k\mp 2}\ ,
\end{split}
\end{equation}
where $A,B=\tau,\rho,3$.

\subsection{Comment on gauge modes}
\label{comgauge_grav}
We conclude this section by showing that the normal modes found at the onset of instability cannot be gauge modes. The gauge transformation of the metric perturbation is given by $\delta h_{\mu\nu} = \nabla_\mu\zeta_\nu + \nabla_\nu\zeta_\mu$ where $\zeta_\mu$ denote the gauge parameters. These can be written in the $e^a$-basis \eqref{orthbasis} as $\zeta_\mu dx^\mu=\zeta_a e^a$ ($a=\tau,r,\pm,3$), and we expand $\zeta_a$ in the Wigner D-matrices as
\begin{equation}
\zeta_A= \sum_{|k|\leq j} \zeta^k_A(r) D_k\ ,\quad
\zeta_\pm=\sum_{|k\mp 1|\leq j} \zeta^k_\pm(r) D_{k\mp 1}\ ,
\end{equation}
where $A=\tau,r,3$. Then, the gauge transformation of $h_{\tau\tau}^k$, $h_{\tau 3}^k$, $h_{33}^k$ and $h_{\pm\pm}^k$ takes the form
\begin{equation}
\begin{split}
\delta h_{\tau\tau}^k &= 4ik h\zeta_\tau^k - (1+r^2)\{(1+r^2)f\}'g\zeta_r^k\ ,\\
\delta h_{\tau 3}^k &= -ik\zeta_\tau^k+\frac{r^2(1+r^2)}{2}g h' \beta \zeta_r^k+2ik h \zeta_3^k\ ,\\
\delta h_{3 3}^k &=\frac{1+r^2}{4}g(r^2\beta)' \zeta_r^k -2ik \zeta_3^k\ ,\\
\delta h_{++}^k &=-\frac{ir^2h}{(1+r^2)f}\left(\alpha-\frac{1}{\alpha}\right)\zeta_\tau^{k-2}
+\frac{1+r^2}{4}g\left\{r^2\left(\alpha-\frac{1}{\alpha}\right)\right\}'\zeta_r^{k-2}\\
&\hspace{6cm}
-\frac{2i}{\beta}\left(\alpha-\frac{1}{\alpha}\right)\zeta_3^{k-2}
+2\epsilon_{k-1}\zeta_+^k\ ,\\
\delta h_{--}^k&=\frac{ir^2h}{(1+r^2)f}\left(\alpha-\frac{1}{\alpha}\right)\zeta_\tau^{k+2}
+\frac{1+r^2}{4}g\left\{r^2\left(\alpha-\frac{1}{\alpha}\right)\right\}'\zeta_r^{k+2}\\
&\hspace{6cm}
+\frac{2i}{\beta}\left(\alpha-\frac{1}{\alpha}\right)\zeta_3^{k+2}
-2\epsilon_{k+2}\zeta_-^k\ .
\end{split}
\label{gauge1}
\end{equation}
Let us focus on the highest $k$. In the expressions of $\delta h_{\tau\tau}^{k=j}$, $\delta h_{\tau 3}^{k=j}$, $\delta h_{33}^{k=j}$, and $\delta h_{++}^{k=j+2}$, only three gauge parameters $\zeta^{k=j}_\tau$, $\zeta^{k=j}_r$, and $\zeta^{k=j}_3$ appear. (The term of $\zeta_+^{k=j+2}$ vanishes because $\epsilon_{j+1}=0$.) Eliminating these gauge parameters, we can construct gauge invariant variables. Similarly, for the lowest $k$, we can also construct the other gauge invariant variables from  $h_{\tau\tau}^{k=-j}$, $h_{\tau 3}^{k=-j}$, $h_{33}^{k=-j}$ and $h_{--}^{k=-j-2}$. Near the horizon, these gauge invariant variables approach
\begin{equation}
 g_\pm=h_{\pm\pm}^{k=\pm (j+2)}-\frac{\alpha-\alpha^{-1}}{j\beta }h_{33}^{k=\pm j}\bigg|_{r=r_h}\ .
\end{equation}
Therefore, if $g_+$ and $g_-$ are nonzero, the perturbations cannot be a gauge mode.
In our numerical calculations, we monitor this gauge invariant quantity and check that it does not reach zero at the onset of instability.

\section{Technical Details for Oscillating Geons}
\label{sec:techoscillgeons}
The equations of motion from the ansatz \eqref{eq:geon2} read
\begin{subequations}
\begin{multline}
B_{\tau\tau}-\frac{4 e^{2 \delta } f^2 (1-y) y B_{yy}}{\Omega ^2}-\frac{8 B^2 \left(1-B^3\right) e^{2 \delta } f}{3 \Omega ^2 (1-y)}-\frac{B_\tau^2}{B}-\frac{2 (1-y) y B_y B_\tau^2}{B^2}-B_\tau \delta_\tau+
\\
\frac{4 e^{2 \delta } f \left[6-\left(6-4 B+B^4\right) y-3f (1-2 y)\right] B_y}{3 \Omega ^2}+\frac{4 e^{2 \delta } f^2 (1-y) y B_y^2}{B \Omega ^2}=0\,,
\end{multline}
\begin{equation}
f_y+\frac{4 B}{3 (1-y)}-\frac{B^4}{3 (1-y)}+\frac{2}{y}-\frac{(2-y) f}{(1-y) y}-\frac{4 (1-y) y f^2 B_y^2}{4 B^2 f}-\frac{e^{-2 \delta } \Omega ^2 B_\tau^2}{4 B^2 f}=0\,,
\end{equation}
\begin{equation}
\delta_y+\frac{(1-y) y B_y^2}{B^2}+\frac{e^{-2 \delta} \Omega ^2 B_\tau^2}{4 f^2\,B^2}=0\,,
\end{equation}
\label{eq:dynamical}
\end{subequations}%
and
\begin{equation}
C\equiv f_\tau-2 y (1-y) f \frac{B_\tau B_y}{B^2}=0\,.
\label{eq:constraint}
\end{equation}

It is a relatively simple exercise to show that Eqs.~(\ref{eq:dynamical}) and their derivatives imply that
\begin{equation}
\left[\frac{1-y}{y^2}e^{-\delta}C\right]_y-\frac{2(1-y)^2 B_y^2}{y B^2} e^{-\delta} C=0\,.
\label{eq:hyper}
\end{equation}
The latter equality shows that $C$ should be regarded as a constraint equation. In particular, if $C$ vanishes on a particular hypersurface of constant $y$, Eq.~(\ref{eq:hyper}) shows that $C$ must vanish for all values of $y$. To see this more clearly, we note that Eq.~(\ref{eq:hyper}) can be formally integrated to give
\begin{equation}
C(\tau,y)=\frac{y^2}{1-y}e^\delta \beta (\tau)\,\exp \left[2\int_1^y \frac{(1-\tilde{y}) \tilde{y} B_{\tilde{y}}^2}{B^2} \, \mathrm{d}\tilde{y}\right]\,,
\end{equation}
where $\beta(\tau)$ is an integration function. Smoothness of the line element (\ref{eq:geon2}) at $y=1$ demands that $f$, $B$ and $\delta$ admit a regular Taylor series around $y=1$ with $f(\tau,1)=B(\tau,1)=1$. These conditions, together with Eqs.~(\ref{eq:dynamical}), imply that $C$ vanishes at $y=1$, and thus that $C=0$ everywhere in the integration domain $(\tau,y)\in[0,\pi]\times[0,1]$.

For numerical purposes it is convenient to perform the following function redefinitions
\begin{equation}
B=1+y^2(1-y)q_1\,,\qquad f=1+y^2(1-y)^2\,q_2\qquad\text{and}\qquad \delta = y^4 q_3\,.
\end{equation}

Solving the equations around $y=1$ and demanding regularity yields the following set of boundary conditions
\begin{subequations}
\begin{equation}
\left.q_{1\;y}+\frac{1}{16} \left[36 q_1+28 q_1^2+e^{-2 q_3} \Omega ^2 \left(q_{1\;\tau\tau}-q_{1\;\tau} q_{3\;\tau}\right)\right]\right|_{y=1}=0\,,
\end{equation}
\begin{equation}
\left.q_2+q_1^2\right|_{y=1}=0\,,
\end{equation}
and
\begin{equation}
\left.q_{3\;y}+4 q_3\right|_{y=1}=0\,.
\end{equation}
\end{subequations}

At the conformal boundary, located at $y=0$, we find
\begin{equation}
\left.q_{1\;y}-\frac{1}{12}(36 q_1^2+\Omega^2 q_{1\;\tau\tau})\right|_{y=0}=0\,,\quad \left. q_{2\;y}-3\,q_{2}\right|_{y=0}\,,\quad \text{and}\quad \left.q_3+q_1^2\right|_{y=0}=0\,,
\end{equation}
with the constraint equation Eq.~(\ref{eq:constraint}) further demanding $\left.q_{2\;\tau}\right|_{y=0}=0$. Note that the constraint is being explicitly enforced at $y=1$, but not at $y=0$. So we monitor $q_2(\tau,0)$ to get an idea of how well the constraints are being satisfied. Furthermore, one can show that there are no non-analytic pieces in the expansion of the conformal boundary to all orders in $y$.

To find our solutions we employed spectral collocation methods, with a uniform cosine-type grid along the $\tau$ direction and a Chebyshev grid along the holographic direction $y$. Our findings are consistent with exponential convergence, which seems to be backed up by the fact that we found no non-analytic behaviour at any of the edges of the integration domain with the gauge we used.

One can also go further and determine the holographic stress energy tensor using \cite{deHaro:2000vlm}
\begin{subequations}
\begin{equation}
\langle T_{\mu\nu}\rangle \mathrm{d}x^\mu\mathrm{d}x^\nu=\frac{1}{16 \pi}\left.\Bigg\{-3 q_2 \mathrm{d}t^2+\frac{1}{4} \left[\left(8 q_1-q_2\right) \sigma _3^2-\left(4 q_1+q_2\right) \left(\sigma _1^2+\sigma _2^2\right)\right]\Bigg\}\right|_{y=0}\,,
\end{equation}
from which one can read the total energy of the system\footnote{We will always measure the energy with respect to pure AdS, \emph{i.e.}~we neglect the Casimir energy of global AdS$_5$.}
\begin{equation}
E=-\frac{3\pi}{8} \left.q_2\right|_{y=0}\,.
\end{equation}
\end{subequations}

Besides solving for our solutions numerically, we can also use perturbation theory to construct these objects. This has been extensively used in the literature \cite{Bizon:2011gg,Dias:2011ss,Horowitz:2014hja,Dias:2016ewl,Martinon:2017uyo,Dias:2017tjg} in similar contexts and provides a good check of our numerical procedures. We expand all our functions in power series in a small parameter $\varepsilon$ as
\begin{subequations}
\begin{align}
&\Omega = \sum_{j=0}^{+\infty} \varepsilon^{2j}\Omega^{(2j)}\,,
\\
& q_1(\tau,y) = \sum_{j=0}^{+\infty} \varepsilon^{2j+1} \sum_{k=0}^{j}\hat{q}_1^{(j,k)}(y)\cos [(2k+1) \tau]+\sum_{j=1}^{+\infty} \varepsilon^{2j} \sum_{k=0}^{j}\tilde{q}_1^{(j,k)}(y)\cos (2\,k \tau)\,,
\\
& q_2(\tau,y) = \sum_{j=1}^{+\infty} \varepsilon^{2j+1} \sum_{k=0}^{j}\hat{q}_2^{(j,k)}(y)\cos [(2k+1) \tau]+\sum_{j=1}^{+\infty} \varepsilon^{2j} \sum_{k=0}^{j}\tilde{q}_2^{(j,k)}(y)\cos (2\,k \tau)\,,
\\
& q_3(\tau,y) = \sum_{j=1}^{+\infty} \varepsilon^{2j+1} \sum_{k=0}^{j}\hat{q}_3^{(j,k)}(y)\cos [(2k+1) \tau]+\sum_{j=1}^{+\infty} \varepsilon^{2j} \sum_{k=0}^{j}\tilde{q}_3^{(j,k)}(y)\cos (2\,k \tau)\,,
\end{align}
\end{subequations}%
and define $\varepsilon$ to be such that, to all orders in $\varepsilon$, the term proportional to $\cos \tau$ in the Fourier-Cosine expansion of $q_1(\tau,0)$ is $\varepsilon$.

To first order in $\varepsilon$, we find
\begin{equation}
\hat{q}_1^{(0,0)}(y)=\, _2F_1(-p,p+6;3;y)\,,\qquad \text{and}\qquad \Omega^{(0)} = 6+2\,p\,,
\end{equation}
where $\, _2F_1(a,b;c;z)$ is a Gaussian Hypergeometric function and $p$ indicates the radial overtone of the excitation in question.

Focusing on the fundamental mode with $p=0$, for the next few orders we find
\begin{subequations}
\begin{align}
&\hat{q}_1^{(0,0)}(y)=1\,, \quad \tilde{q}_{1}^{(1,0)}=\frac{1}{60} \left(9 y^2-1-3 y-15 y^3\right)\,,
\\
&\tilde{q}_2^{(1,0)}(y)=\frac{1}{20} \left(2 y^2-3-9 y\right)\,,\quad \tilde{q}_2^{(1,1)}(y)=\frac{1}{2} y^2 (2-3 y)\,,
\\
& \tilde{q}_3^{(1,0)}(y)=\frac{1}{20} \left(14 y-10-5 y^2\right)\,,\quad  \tilde{q}_3^{(1,1)}(y)=\frac{1}{28} \left(36 y^3-14+70 y-91 y^2\right)\,.
\end{align}
\end{subequations}
The equations can be used to infer the energy and $\Omega$ as a function of $\varepsilon$ to order $\varepsilon^2$, as in \eqref{eq:perturb}, which we reproduce here:
\begin{equation}
E=\frac{9 \pi  \varepsilon ^2}{160}\;,\qquad\Omega=6-\frac{11057 \varepsilon ^2}{90090}\;.
\label{eq:perturb2}
\end{equation}

\bibliography{bib_gp}
\end{document}